\documentclass[10pt,twocolumn,english,showpacs,preprintnumbers,amsmath,amssymb,floatfix,nofootinbib]{revtex4-1} 
\usepackage[normalem]{ulem}  
\usepackage{caption}
\usepackage{tikz,xcolor}
\usepackage[colorlinks = true,
linkcolor = blue,
urlcolor  = blue,
citecolor = blue,
anchorcolor = blue]{hyperref}
\definecolor{lime}{HTML}{A6CE39}
\DeclareRobustCommand{\orcidicon}{%
	\begin{tikzpicture}
		\draw[lime, fill=lime] (0,0)
		circle [radius=0.16]
		node[white] {{\fontfamily{qag}\selectfont \tiny ID}};
		\draw[white, fill=white] (-0.0625,0.095)
		circle [radius=0.007];
	\end{tikzpicture}
	\hspace{-2mm}
}
\foreach \x in {A, ..., Z}{%
\expandafter\xdef\csname orcid\x\endcsname{\noexpand\href{https://orcid.org/\csname orcidauthor\x\endcsname}{\noexpand\orcidicon}}
}
\usepackage[T1]{fontenc}
\usepackage[latin9]{inputenc}
\usepackage{color}
\usepackage{array}
\usepackage{amstext}
\usepackage{subfig}
\usepackage{graphicx}
\usepackage{esint}
\usepackage{rotating}
\usepackage{slashed}
\usepackage{siunitx}
\usepackage[normalem]{ulem}
\usepackage[font=small,labelfont=bf]{caption}
\usepackage{lipsum}
\usepackage{xcolor}
\usepackage{comment}
\usepackage{amssymb}
\newcommand{\xpom}{x_{\xpom}}

\makeatletter

\@ifundefined{textcolor}{}
{%
 \definecolor{BLACK}{gray}{0}
 \definecolor{WHITE}{gray}{1}
 \definecolor{RED}{rgb}{1,0,0}
 \definecolor{GREEN}{rgb}{0,1,0}
 \definecolor{BLUE}{rgb}{0,0,1}
 \definecolor{CYAN}{cmyk}{1,0,0,0}
\definecolor{MAGENTA}{cmyk}{0,1,0,0}
 \definecolor{YELLOW}{cmyk}{0,0,1,0}
 }

\@ifundefined{definecolor}
{\usepackage{color}}{}
\@ifundefined{definecolor}
{\usepackage{color}}{}
\makeatother
\usepackage{babel}

\def\Re{{\cal R \mskip-4mu \lower.1ex \hbox{\it e}\,}}
\def\Im{{\cal I \mskip-5mu \lower.1ex \hbox{\it m}\,}}

\def\tev{\,{\ifmmode\mathrm {TeV}\else TeV\fi}}
\def\gev{\,{\ifmmode\mathrm {GeV}\else GeV\fi}}
\def\mev{\,{\ifmmode\mathrm {MeV}\else MeV\fi}}
\def\to{\rightarrow}

\begin{document}

%
%%%%%%%%%%%%%%%%%%%%%%%%%%%%%%%%%%%%%%%%%%%%%%%%%%%%%%%%%%%%%%%%%%%%%%%%%%%%%%%%%%%%%%%%%%%%%%%%%%%%%%%%%%%%%%%%%%%%%%%%%%%%%%%%%%%%%%%
%
\title{Revisiting constraints on proton PDFs from HERA DIS, Drell-Yan, W/Z Boson production, and projected EIC measurements}
%
%%%%%%%%%%%%%%%%%%%%%%%%%%%%%%%%%%%%%%%%%%%%%%%%%%%%%%%%%%%%%%%%%%%%%%%%%%%%%%%%%%%%%%%%%%%%%%%%%%%%%%%%%%%%%%%%%%%%%%%%%%%%%%%%%%%%%%%
%

\author{Majid~Azizi$^{1}$\orcidD{}}
\email{Ma.Azizi@ipm.ir}

\author{Maryam~Soleymaninia$^{1}$\orcidB{}}
\email{Maryam\_Soleymaninia@ipm.ir}

\author{Hadi~Hashamipour$^{2}$\orcidA{}}
\email{Hadi.Hashamipour@lnf.infn.it}

\author{Maral~Salajegheh$^{3}$\orcidC{}} 
\email{Maral@hiskp.uni-bonn.de }

\author{Hamzeh~Khanpour$^{4,5,1}$\orcidE{}}
\email{Hamzeh.Khanpour@cern.ch}

\author{Ulf-G.~Mei{\ss}ner$^{3,6,7}$\orcidF{}}
\email{Meissner@hiskp.uni-bonn.de}

\affiliation {
$^{1}$School of Particles and Accelerators, Institute for Research in Fundamental Sciences (IPM), P.O.Box 19395-5531, Tehran, Iran.          \\  
$^{2}$Istituto Nazionale di Fisica Nucleare, Gruppo collegato di Cosenza, I-87036 Arcavacata di Rende, Cosenza, Italy.                       \\  
$^{3}$Helmholtz-Institut f\"ur Strahlen-und Kernphysik and Bethe Center for Theoretical Physics, Universit\"at Bonn, D-53115 Bonn, Germany.  \\  
$^{4}$AGH University, Faculty of Physics and Applied Computer Science, Al. Mickiewicza 30, 30-055 Krakow, Poland.                             \\  
$^{5}$Department of Physics, University of Science and Technology of Mazandaran, P.O.Box 48518-78195, Behshahr, Iran.                         \\   
$^{6}$Institute for Advanced Simulation (IAS-4), Forschungszentrum J\"ulich, D-52425 J\"ulich, Germany.                  \\
$^{7}$Tbilisi State University, 0186 Tbilisi, Georgia.\\
}

\date{\today}

%
%%%%%%%%%%%%%%%%%%%%%%%%%%%%%%%%%%%%%%%%%%%%%%%%%%%%%%%%%%%%%%%%%%%%%%%%%
\begin{abstract}

We present new parton distribution functions (PDFs) at next-to-leading order (NLO) and 
next-to-next-to-leading order (NNLO) in perturbative QCD, derived from a comprehensive global QCD analysis 
of high-precision data sets from combined HERA deep-inelastic scattering (DIS), 
the Tevatron, and the Large Hadron Collider (LHC). To improve constraints on quark 
flavor separation, we incorporate Drell-Yan pair production data, which provides 
critical sensitivity to the quark distributions. In addition, we include the latest 
W and Z boson production data from the CDF, D0, ATLAS, and CMS collaborations, 
further refining both quark and gluon distributions. Our nominal global QCD fit 
integrates these datasets and examines the resulting impact on the PDFs and their 
associated uncertainties. 
Uncertainties in the PDFs are quantified using the Hessian method, ensuring robust 
error estimates. Furthermore, we explore the sensitivity of the strong coupling 
constant, \(\alpha_s(M_Z^2)\), and proton PDFs in light of the projected measurements 
from the Electron-Ion Collider (EIC), where improvements in precision are expected.
The analysis also investigates the effects of inclusive 
jet and dijet production data, which provide enhanced constraints on the gluon PDF and \(\alpha_s(M_Z^2)\). 

\end{abstract}

\maketitle
\tableofcontents{}

%=================================================
\section{Introduction}
%=================================================

Deep Inelastic Scattering (DIS) is one of the most thoroughly studied processes in 
perturbative Quantum Chromodynamics (QCD), as reviewed in Refs.~\cite{South:2016cmx,Gross:2022hyw,Blumlein:2023aso,Blumlein:2012bf}. 
DIS plays a fundamental role in probing the internal structure of hadrons, particularly in 
determining parton distribution functions (PDFs), which describe the momentum distribution of quarks and gluons 
within a hadron~\cite{Gao:2017yyd}. 
Parton density functions, \(f_a(x, Q)\), are essential for accurate predictions in high-energy physics, especially at 
colliders such as the CERN Large Hadron Collider (LHC), and future facilities such as 
the Electron-Ion Collider (EIC)~\cite{AbdulKhalek:2021gbh,AbdulKhalek:2022hcn}, the Large 
Hadron Electron Collider (LHeC)~\cite{LHeC:2020van}, 
and the Future Circular Collider (FCC)~\cite{FCC:2018byv,FCC:2018vvp,FCC:2018bvk}. 

The upcoming EIC and LHeC are anticipated to enhance PDF determinations significantly  
by expanding the kinematic range accessible to current experiments. 
These facilities will enable a more precise extraction of PDFs by providing complementary 
insights into the proton structure. The EIC is expected to offer critical 
data in the large \(x\) region, which HERA was unable to fully explore, 
while the LHeC will achieve high precision in probing the small \(x\) region, 
thereby reducing uncertainties in the gluon and sea quark distributions. 
However, achieving such precision 
required for modern experiments presents two primary challenges: missing higher-order 
QCD corrections in theoretical calculations and inherent uncertainties in the PDFs themselves.  

Several global QCD analyses have been carried out by multiple collaborations, 
including CTEQ-TEA~\cite{Hou:2019efy,Ablat:2024muy}, NNPDF4.0~\cite{NNPDF:2021njg,Cruz-Martinez:2024cbz}, 
MSHT20~\cite{Bailey:2020ooq}, and other groups~\cite{H1:2021xxi,ATLAS:2021vod,Alekhin:2024bhs}, to address these 
challenges and improve the PDF determinations. In particular, the CTEQ-TEA collaboration 
has developed a new generation of general-purpose PDFs to supersede the CT18 set~\cite{Hou:2019efy}, 
aimed at a wide range of applications, including precision studies of electroweak processes, 
Higgs production, and searches for physics beyond the Standard Model (BSM).

Recent advancements in PDF determinations have been driven by the inclusion of high-precision 
data from the LHC. For instance, the NNPDF3.1 set~\cite{NNPDF:2017mvq} was the first to extensively 
incorporate LHC data, achieving a precision level of 3-5\% in PDF uncertainties. 
The latest NNPDF4.0 set~\cite{NNPDF:2021njg} builds on this by incorporating LHC Run~II 
data at \(\sqrt{s} = 13\) TeV and introducing machine learning techniques to optimize the 
PDF fitting process. This approach, along with rigorous validation through 
closure tests, represents a significant leap forward in PDF precision.

The MSHT collaboration introduced the MSHT20 set~\cite{Bailey:2020ooq}, determined through 
global analyses of available hard-scattering data up to NNLO accuracy. MSHT20 builds on 
the MMHT14 framework~\cite{Harland-Lang:2014zoa} with extended parameterizations and the 
inclusion of new datasets, ranging from the final HERA combined data on total and heavy-flavor 
structure functions to recent LHC measurements of vector boson production, 
inclusive jets, and top quark production at 7 and 8~TeV. Notable updates include 
improvements in the \(u - d\) valence quark difference and the strange quark PDFs, 
attributed to new data and updated parameterizations. Additionally, MSHT20 incorporates 
NNLO corrections for dimuon production in neutrino DIS, leading to reduced uncertainties 
in key processes such as Higgs and W/Z boson production at the LHC. A recent update 
from the MSHT collaboration is presented in Ref.~\cite{Harland-Lang:2024kvt}, where they 
discuss a global closure test of the fixed parameterization approach to PDF fitting.

In this work, we present new PDFs determined at NLO and NNLO accuracy through a 
comprehensive analysis of high-precision data from the LHC, Tevatron,  and the combined HERA DIS datasets. 
The focus of this study is to maximize the sensitivity of the PDFs by incorporating data 
from Drell-Yan pair production and W/Z boson production at the Tevatron and LHC, with 
particular emphasis on the strange quark density, which has been a topic of interest 
in recent QCD analyses~\cite{Bailey:2020ooq,Cooper-Sarkar:2018ufj,Sato:2019yez}. 
A combined PDF fit integrates all these datasets, providing a thorough examination of the 
resulting PDFs and their associated uncertainties. Uncertainty estimates are calculated 
using the Hessian method, ensuring robust uncertainty quantification.

Furthermore, we explore the impact of simulated inclusive DIS data from the EIC~\cite{AbdulKhalek:2021gbh,AbdulKhalek:2022hcn} 
on the determination of proton PDFs. The EIC is expected to provide complementary 
information in the large Bjorken-\(x\) region, which will be crucial for improving PDF 
precision at both NLO and NNLO accuracy~\cite{Armesto:2023hnw,Cerci:2023uhu}. We also 
estimate the expected experimental uncertainty in the strong coupling constant, \(\alpha_s(M_Z^2)\), when incorporating 
simulated EIC inclusive data into the QCD analysis.

Finally, we examine the significant role that jet and dijet production data play in constraining PDFs, 
particularly the gluon distribution. Jet production data are directly sensitive to the gluon content in 
the proton, and their inclusion is expected to reduce uncertainties in the gluon PDF and improve the 
determination of \(\alpha_s(M_Z^2)\). The impact of such datasets on 
the precision of \(\alpha_s\) is also examined and discussed in this work.

The remainder of this paper is organized as follows:  
In Sec.~\ref{Theoretical_Framework}, we outline the theoretical framework used in our analysis, 
focusing on the key aspects of perturbative QCD and the role of proton PDFs.  
The datasets employed in this analysis, including those from HERA DIS, Tevatron, and the LHC, 
are discussed comprehensively in Sec.~\ref{Data}.  
Sec.~\ref{method} describes the methodology utilized in the global PDF fit, highlighting the 
statistical tools and techniques used to quantify uncertainties.  
In Sec.~\ref{results}, we present our main findings, offering an in-depth examination of the 
impact of different datasets on the resulting PDFs and providing a comparison with previous determinations.  
This section also explores the influence of simulated inclusive DIS data from the future 
EIC experiments on the determination of proton PDFs and the precision of \(\alpha_s(M_Z^2)\).  
The role of simulated EIC DIS data in proton PDF determination is discussed in Sec.~\ref{EIC}.  
Sec.~\ref{jet} covers the impact of jet and dijet production data on proton PDF determination.  
A detailed examination of the strange-quark density is provided in Sec.~\ref{Strange-quark}.  
The effect of simulated EIC and jet production data on the determination of the 
strong coupling constant, \(\alpha_s\), is discussed in Sec.~\ref{Coupling}.  
Finally, Sec.~\ref{summary} offers concluding remarks and suggests future research 
directions for improving PDF precision and reducing theoretical uncertainties.

%=================================================
\section{Global QCD Analysis Framework}\label{Theoretical_Framework}
%=================================================

In this section, we outline the theoretical framework used in our global analysis to extract the proton PDFs 
from high-energy scattering data. Our analysis relies on QCD factorization theorems, 
which allow the separation of short-distance perturbative interactions, calculable within 
perturbative QCD, from long-distance non-perturbative effects encoded in the PDFs~\cite{Collins:1989gx}. 
This separation ensures that PDFs are universal across different hard-scattering processes, 
making them indispensable for predicting cross sections at high-energy colliders.

%-----------------------------------------------------
\subsection{Factorization in DIS} 
%-----------------------------------------------------

DIS is a cornerstone in the determination of PDFs. The DIS cross-section is expressed in terms of 
hadronic structure functions \( F_2(x, Q^2) \) and \( F_L(x, Q^2) \), which can be written as 
a convolution of PDFs and perturbatively calculable Wilson coefficients, \( C_i(x, \alpha_s(Q^2)) \), 
at a given scale \( Q^2 \). Specifically, the structure functions take the form~\cite{Gao:2017yyd,Martin:2009iq,Alekhin:2012ig}:
%
%++++++++++++++++
\begin{eqnarray}
F_{i}(x, Q^2) &=& \sum_{k=1}^{n_f} C_{i,k}(x, \alpha_s(Q^2)) \otimes x q^+_k(x, Q^2)  \\  \nonumber
&+&C_{i,g}(x, \alpha_s(Q^2)) \otimes x g(x, Q^2), \quad i = \{2, L\}, \\  \nonumber
\end{eqnarray}
%++++++++++++++++
%
where \( q^+_k(x, Q^2) \) are the quark and antiquark distributions, and \( g(x, Q^2) \) is the 
gluon distribution. The Wilson coefficients \( C_{i,k}(x, \alpha_s(Q^2)) \) are known up to NNLO for massless quarks, 
while treatments for massive quarks are implemented through the Variable Flavor Number Scheme (VFNS), 
allowing for a smooth transition across heavy quark mass thresholds~\cite{Martin:2009iq}.

%-----------------------------------------------------
\subsection{Drell-Yan Process}
%-----------------------------------------------------

The Drell-Yan process, in which a quark-antiquark pair annihilates into a virtual photon or \( Z \)-boson that subsequently decays into a 
lepton pair, is another critical process for constraining PDFs. 
The cross-section for the Drell-Yan process is factorized as~\cite{Gao:2017yyd,NuSea:2001idv,Hamberg:1990np}:
%
%++++++++++++++++
\begin{equation}
\sigma_{\text{DY}} = \sum_{q,\bar{q}} \int dx_1 \, dx_2 \, f_q(x_1, Q^2) \, f_{\bar{q}}(x_2, Q^2) 
\, \hat{\sigma}_{q\bar{q} \rightarrow \ell^+ \ell^-}(s),
\end{equation}
%++++++++++++++++
%
where \( f_q(x_1, Q^2) \) and \( f_{\bar{q}}(x_2, Q^2) \) are the PDFs for quarks and antiquarks, 
and \( \hat{\sigma}_{q\bar{q} \rightarrow \ell^+ \ell^-} \) is the hard-scattering 
cross-section for the pertinent subprocess. The Drell-Yan process is essential for constraining the 
sea quark and antiquark distributions, particularly at medium-to-large \( x \).

%-----------------------------------------------------
\subsection{Electroweak Boson Production}
%-----------------------------------------------------

The production of electroweak gauge bosons (W and Z) in hadronic collisions provides powerful 
constraints on the flavor decomposition of the quark PDFs. The inclusive cross-section for 
W or Z boson production can be expressed as~\cite{Gao:2017yyd,Nadolsky:2004vt}:
%
%++++++++++++++++
\begin{equation}
\sigma_{W/Z} = \sum_{q,\bar{q}} \int dx_1 \, dx_2 \, f_q(x_1, Q^2) \, f_{\bar{q}}(x_2, Q^2) 
\, \hat{\sigma}_{q\bar{q} \rightarrow W^\pm/Z}(s).
\end{equation}
%++++++++++++++++
%
These processes are sensitive to both quark and antiquark distributions, offering precise 
constraints on quark flavor separations. The inclusion of high-precision measurements 
from LHC experiments (e.g., ATLAS and CMS) allows for a more refined determination of the 
quark PDFs, particularly for the up, down, and strange quarks at low to moderate \(x\) values.

%-----------------------------------------------------
\subsection{Jet Production and Gluon PDF Constraints}
%-----------------------------------------------------

Jet production in hadronic collisions provides crucial information about the gluon content of the proton. 
The cross-section for inclusive jet production can be factorized as~\cite{Sapeta:2015gee,Catani:1996vz}:
%
%++++++++++++++++
\begin{equation}
\sigma_{\text{jet}} = \sum_{a,b=q,\bar{q},g} \int dx_1 \, dx_2 \, f_a(x_1, Q^2) \, f_b(x_2, Q^2) 
\, \hat{\sigma}_{ab \rightarrow \text{jets}}(s),
\end{equation}
%++++++++++++++++
%
where the partonic subprocess cross-sections \( \hat{\sigma}_{ab \rightarrow \text{jets}} \) 
include contributions from quark-quark, quark-gluon, and gluon-gluon scattering. Jet data from 
HERA and the LHC, particularly at high energies, provide important constraints on 
the gluon distribution. The recent inclusion of NNLO corrections to jet production 
processes has further reduced theoretical uncertainties and refined the determination 
of the gluon PDF, especially at low values of \(x\)~\cite{Ablat:2024muy}.

%-----------------------------------------------------
\subsection{Higher-Order Corrections in QCD}
%-----------------------------------------------------

The inclusion of higher-order QCD corrections, particularly at NNLO accuracy, is crucial for 
achieving precise determinations of PDFs. 
These corrections help to reduce the dependence of cross-sections on factorization and renormalization scales, 
thus providing more reliable theoretical predictions.
At NLO, the gluon PDF receives significant corrections from processes such as \( qg \rightarrow q \) 
and \( gg \rightarrow g \). 
Moving to NNLO, additional contributions from gluon-gluon and quark-antiquark 
interactions become relevant. NNLO corrections have been shown to significantly improve the 
agreement with experimental data, especially for processes dominated by gluons, 
such as Higgs production via gluon fusion~\cite{Harlander:2002wh} and inclusive jet production.
The calculation of NNLO splitting functions and Wilson coefficients allows for more 
precise DGLAP evolution of PDFs from the initial scale \( Q_0^2 \) to higher scales 
relevant for collider processes~\cite{Bertone:2024dpm}. 
These corrections are particularly impactful in the small \(x\) regime, where the 
gluon density rises steeply due to QCD evolution. Incorporating NNLO accuracy thus 
reduces theoretical uncertainties, especially in regions of phase space where gluon 
interactions dominate, contributing to improved predictions for hadron collider experiments.

%-----------------------------------------------------
\subsection{Heavy quark treatments in  QCD Analysis}
%-----------------------------------------------------

Heavy quark contributions, particularly from charm and bottom quarks, are treated using the 
Variable Flavor Number Scheme (VFNS)~\cite{Martin:2009iq,Thorne:2006qt,Alekhin:2012ig}. 
In this scheme, heavy quarks are produced perturbatively at energies above their mass thresholds. 
This approach allows for smooth matching of PDFs across these thresholds, ensuring that heavy 
quark PDFs are consistently included in both DIS and hadron collider cross-sections. 
Recent advancements, such as the inclusion of photon-gluon fusion for charm production, 
have provided valuable constraints on the gluon PDF at small \(x\), thereby improving the precision of the charm PDF. 
This is especially important for precision studies of the Higgs and electroweak sectors at the LHC.

The treatment of heavy quarks in perturbative QCD is crucial for determining PDFs in processes 
involving DIS and hadron collisions. Several schemes have been developed to properly 
account for heavy quark mass effects in different kinematic regions:
The Fixed Order plus Next-to-Leading Log (FONLL) approach~\cite{Forte:2002fg,Barontini:2024xgu} combines fixed-order (FO) 
calculations with next-to-leading log (NLL) resummations, enabling a smooth transition between the massive and 
massless cases. The FONLL scheme also allows for the simultaneous treatment of charm and bottom quarks, 
improving the flexibility and accuracy in PDF determinations, especially at scales where both heavy quarks contribute.
The ACOT scheme~\cite{Aivazis:1993kh}, along with its variants like S-ACOT and ACOT-\(\chi\)~\cite{Collins:1998rz}, 
retains mass dependence in the Wilson coefficients, ensuring an accurate treatment of heavy quarks near 
their production thresholds where \( Q^2 \sim m^2 \). This enables a precise 
description of heavy quark contributions in kinematic regions close to the threshold.
The ZMVFN scheme~\cite{Buza:1996wv} neglects heavy quark mass effects at high energy scales, 
simplifying calculations. However, it can be inaccurate near threshold regions where \( Q^2 \sim m^2 \), 
making it less suitable for precise studies involving heavy quark production near their mass thresholds.

The Thorne-Roberts (TR) method~\cite{Thorne:1997ga} is employed in the current analysis 
to ensure smooth transitions between the massive and massless regimes for heavy quark contributions to PDFs. 
This method provides a consistent framework for incorporating heavy quark effects 
across different kinematic regions, effectively bridging the transitions while maintaining the continuity of the PDFs. 
The TR scheme provided the proper handling of charm and bottom quarks, which is particularly 
important for accurate predictions at the LHC, including B-meson and Higgs production in association with heavy quarks.
The use of the TR scheme for heavy quark treatment also enables us to 
achieve consistent higher-order predictions and improved 
constraints on the gluon PDF, which is particularly significant for electroweak studies at the LHC.

%=================================================
\section{Experimental Measurements}\label{Data}
%=================================================

In this section, we provide a detailed overview of the data sets  employed in our QCD analysis. 
We outline the sources, selection criteria, and preprocessing steps applied to ensure the robustness and reliability of our results.
In global PDF analyses, DIS data sets, which span a 
wide range of Bjorken \(x\) and momentum transfer \(Q^2\), traditionally serve as the primary constraint. 
However, DIS data alone cannot sufficiently determine the gluon and sea quark distributions, particularly 
when it comes to distinguishing between individual quark flavors.
In addition to the data from HERA DIS, we utilize a broader array of hard-scattering 
cross sections, primarily from proton-proton (\(pp\)) collisions, to improve the precision 
of the PDF extractions. Recent studies have demonstrated that incorporating measurements from 
processes such as heavy flavor production, inclusive jet production, and top quark cross-sections 
significantly enhances the constraints on PDFs. These data sets provide improved coverage 
across a broad range of \(x\), \(Q^2\), and flavor combinations, ultimately leading to more 
accurate determinations of the parton distributions.

%--------------------------------
\begingroup
\squeezetable
\begin{table*}[htp]
\caption{\label{AllData} The list of all data sets: DIS HERA I+II, Drell-Yan and W/Z production data used in the present analysis. For each data set, we 
indicate process, measurement, reference and the ranges of their kinematic cuts such as $x$, $y$, $Q^2$ [GeV$^2$], $p_T$ [GeV], $E_T$ [GeV], $\sqrt{s}$ [TeV] and $\mathcal{L}$ [fb$^{-1}$].  }
\begin{center}
\begin{ruledtabular}
\begin{tabular}{l|c|l|c|lr}
~Data set  & Process & ~Experiment   & Ref.  & ~Kinematic ranges and details\\
\hline 
\multicolumn{1}{l}{\bf{HERA I+II}}\\ \hline	
\hline 			
& $e^{\pm}p \rightarrow \overset{(-)}{\nu} X$
&  ~HERA I+II CC $e^+p$ &\cite{Abramowicz:2015mha}&~$3\times 10^{2}\leq Q^2 \leq3\times 10^{4}$, &$8.0\times 10^{-3}\leq x \leq$0.4  \\ 
& & ~HERA I+II CC $e^-p$ &\cite{Abramowicz:2015mha}&~$3\times 10^{2}\leq Q^2 \leq3\times 10^{4}$, &$8.0\times 10^{-3}\leq x \leq$0.65\\ 
&$e^{\pm}p \rightarrow e^{\pm}X   $ & ~HERA I+II NC $e^-p$ &\cite{Abramowicz:2015mha}&~$60\leq Q^2 \leq5\times 10^{4}$, &$8.0\times 10^{-4}\leq x \leq$0.65\\ 
DIS $\sigma$	& & ~HERA I+II NC $e^-p$ 460~ &\cite{Abramowicz:2015mha} &~$6.5\leq Q^2 \leq8\times 10^{2}$, &$3.48\times 10^{-5}\leq x \leq$0.65\\ 
& & ~HERA I+II NC $e^-p$ 575~ &\cite{Abramowicz:2015mha}&~$6.5\leq Q^2 \leq8\times 10^{2}$, &$3.48\times 10^{-5}\leq x \leq$0.65\\ 
& & ~HERA I+II NC $e^+p$ 820~ &\cite{Abramowicz:2015mha}&~$6.5\leq Q^2 \leq3\times 10^{4}$, &$6.21\times 10^{-7}\leq x \leq$0.4\\ 
& & ~HERA I+II NC $e^+p$ 920~ &\cite{Abramowicz:2015mha}&~$6.5\leq Q^2 \leq3\times 10^{4}$, &$5.02\times 10^{-6}\leq x \leq$0.65 \\
\hline
\multicolumn{1}{l}{\bf{Drell-Yan}}\\\hline	
\hline 				
ATLAS 	& $Z/\gamma^*\rightarrow \ell \ell (\ell=e,\mu)$  & ~ATLAS low-mass DY 7 TeV & \cite{ATLAS:2014ape}&~$\mid \eta_\ell \mid \leq 2.4$, &   $12\leq m_{\ell \ell} \leq 66$ \\
&$Z/\gamma ^*\rightarrow e^\pm  $ & ~ATLAS high-mass DY 7 TeV& \cite{ATLAS:2013xny}&~$\mid \eta_\ell \mid \leq 2.1$, &   $116\leq m_{\ell \ell} \leq 1500$  \\
\hline
FNAL E866/NuSea 	&$pp(deut) \rightarrow \mu ^\pm X$ & ~E866 $\sigma ^d/ 2\sigma ^p$ (NuSea) &\cite{NuSea:2001idv} & ~$0.015 \leq x \leq 0.35$?, & $4.40 \leq m_{\ell\ell} \leq 12.9$
\\
\hline 
\multicolumn{1}{l}{\bf{W/Z production}}\\
\hline
\hline 
ATLAS	& $W^+ \rightarrow e^+ \nu$ &~ATLAS W 7 TeV ($\mathcal{L} = 4.6$ pb$^{-1}$) & \cite{ATLAS:2016nqi} & ~$\mid \eta_{\ell}\mid\leq 2.5$, & $\mid y_Z\mid\leq  3.6$ \\
& $W^+ \rightarrow \mu ^+ \nu $  & ~ATLAS W 7 TeV ($\mathcal{L} = 4.6$ pb$^{-1}$) &\cite{ATLAS:2016nqi}& ~$\mid \eta_{\ell}\mid\leq 2.5$, & $\mid y_Z\mid\leq  3.6$
\\& $Z \rightarrow e^+ e^- $  & ~ATLAS Z 7 TeV ($\mathcal{L} = 4.6$ pb$^{-1})$ &\cite{ATLAS:2016nqi}& ~$\mid \eta_{\ell}\mid\leq 2.5$, & $\mid y_Z\mid\leq  3.6$
\\& $Z \rightarrow \mu ^+ \mu ^- $  & ~ATLAS Z 7 TeV ($\mathcal{L} = 4.6$ pb$^{-1}$) &\cite{ATLAS:2016nqi}& ~$\mid \eta_{\ell}\mid\leq 2.5$, & $\mid y_Z\mid\leq  3.6$
\\& $W \rightarrow \ell \nu $  & ~ATLAS W 7 TeV ($\mathcal{L} = 35$ pb$^{-1}$)&\cite{ATLAS:2012sjl}&~$\mid \eta_{\ell}\mid\leq 2.4$, & $\mid y_Z\mid\leq  3.6$
\\& $Z \rightarrow \ell \ell $  & ~ATLAS Z 7 TeV ($\mathcal{L} = 35$ pb$^{-1}$) &\cite{ATLAS:2012sjl}&~$\mid \eta_{\ell}\mid\leq 2.5$, & $\mid y_Z\mid\leq  3.6$           \\ 			\hline 
CMS	&$ W\rightarrow e\nu $ &  ~CMS W electron asymmetry 7 TeV &\cite{CMS:2012ivw} & ~$\mid \eta_{\ell}\mid\leq 2.4$,&$P_T > 35$
\\ 
& $W\rightarrow \mu\nu$& ~CMS W muon asymmetry 7 TeV &\cite{CMS:2013pzl} & ~$\mid \eta_{\ell}\mid\leq 2.4$,&$P_T > 25$\\
& $W\rightarrow \mu\nu$ & ~CMS W muon asymmetry 8 TeV  &\cite{CMS:2016qqr} &~$\mid \eta_{\ell}\mid\leq 2.4$,&$P_T > 25$\\
&$ Z\rightarrow e^\pm(\mu ^\pm)$ &~CMS Z muon and electron 7 TeV  &\cite{CMS:2011wyd} & ~$\mid y \mid < 3.5$,&$60 \leq m_{\ell\ell} \leq 120$
\\
\hline 
CDF	&$ W \rightarrow e\nu \,$ &  ~CDF W charge asymmetry 1.96 TeV&\cite{CDF:2009cjw} & ~$2.1 \leq \mid \eta_{\ell}\mid\leq 3.5$, &$\mid y_W \mid \leq 3.0$\\
&$Z \rightarrow e^+e^-$ & CDF Z differential 1.96 TeV &\cite{CDF:2010vek} & $0.0 \leq y_{\ell\ell} \leq 2.9$, $66 \leq m_{\ell\ell} \leq 116$ \\

\hline 
D0	&$W \rightarrow e\nu \,$ &  ~D0 W electron asymmetry 1.96 TeV &\cite{D0:2014kma} & ~$\mid \eta ^e \mid \leq 2.9$,&$E^e_T > 25$
\\ 
& $W \rightarrow \mu \nu $\, & ~D0 W muon asymmetry 1.96 TeV &\cite{D0:2013xqc} & ~$\mid \eta ^\mu \mid \leq 2.0$,&$P^ \mu _T > 25$   \\
&$Z \rightarrow e^+ e^-$ & D0 Z differential 1.96 TeV &\cite{D0:2007djv} & $0.05 \leq y_{ee} \leq 2.75$, $71 \leq M_{ee} \leq 111$ \\

\end{tabular}
\end{ruledtabular}
%	}
\end{center}
\end{table*}
\endgroup
%--------------------------------

%=================================================
\subsection{DIS Data}
%=================================================

The HERA dataset, collected by the H1 and ZEUS collaborations between 1994 and 2007, 
represents a comprehensive set of DIS measurements of electrons and positrons scattered off protons. 
This dataset includes over 1 fb\(^{-1}\) of integrated luminosity, 
with proton beam energies ranging from 460 to 920 GeV and an electron beam 
energy fixed at 27.5 GeV~\cite{Abramowicz:2015mha}. The data spans six orders of 
magnitude in Bjorken \(x\) and the negative four-momentum-transfer squared, \(Q^2\) (from 0.045 GeV\(^2\) to 50000 GeV\(^2\)), 
making it the most precise dataset available for \(ep\) scattering. These measurements 
enabled the development of the {\tt HERAPDF2.0} PDFs, which were refined through QCD analyses 
at leading order (LO), NLO, and NNLO accuracy. 
Additionally, variants such as {\tt HERAPDF2.0Jets}, which incorporate jet production data, 
facilitated the simultaneous determination of PDFs and the strong coupling constant \(\alpha_s\), 
providing valuable insights into proton structure and fundamental interactions~\cite{Abramowicz:2015mha}.
The final combined HERA I+II dataset significantly improved constraints on small \(x\) sea 
quarks and gluon PDFs. This comprehensive dataset, covering both neutral current (NC) and 
charged current (CC) interactions at various proton beam energies (460, 575, 820, and 920 GeV), 
provided a consistent set of cross-section measurements across a wide kinematic range. 
The electron beam energy was fixed at 27.5 GeV, resulting in center-of-mass energies of 
approximately 225, 251, 300, and 320 GeV. The data are available as functions of \(Q^2\) and \(x\), 
covering the range \(0.045 \leq Q^2 \leq 50000\) GeV\(^2\) and \(6 \times 10^{-7} \leq x \leq 0.65\) 
for neutral current interactions, and \(200 \leq Q^2 \leq 50000\) GeV\(^2\) and \(1.3 \times 10^{-2} \leq x \leq 0.40\) 
for charged current interactions.
For the HERA I+II datasets included in our analysis, we follow the procedure in Ref.~\cite{ATLAS:2016nqi} 
and apply a kinematic cut at \(Q^2 > 10\) GeV\(^2\) to effectively remove higher-twist (HT) contributions.

%=================================================
\subsection{Drell-Yan Data Sets}
%=================================================

The precise measurement of inclusive \(W^+\), \(W^-\), and \(Z/\gamma^{*}\) production in \(pp\) collisions at 
the LHC provides a stringent test of perturbative QCD. The rapidity dependence of boson production 
in the Drell-Yan process offers critical constraints on the PDFs of the proton, as the boson rapidity 
is strongly correlated with the momentum fractions \(x_1\) and \(x_2\) carried by the partons 
participating in the hard-scattering subprocess. The weak and electromagnetic components of the 
neutral current (NC) process, \(Z/\gamma^{*} \rightarrow \ell\ell\), along with the weak charged 
current (CC) reactions, \(W^+ \rightarrow \ell^+\nu\) and \(W^- \rightarrow \ell^-\bar{\nu}\), 
probe the quark flavor structure of the proton, complementing the information obtained from DIS. 
In this section, we present the Drell-Yan datasets included in our analysis. \\

\textbf {LHC Drell-Yan Data}: 
The ATLAS collaboration provides a detailed measurement of the differential cross-section for 
the process \(Z/\gamma^{*} \rightarrow \ell \ell\) (with \(\ell = e, \mu\)) as a function of 
the dilepton invariant mass in proton-proton collisions at \(\sqrt{s} = 7~\text{TeV}\), 
using the ATLAS detector at the LHC~\cite{ATLAS:2014ape}. This analysis, based on data collected 
in 2011, corresponds to an integrated luminosity of \(1.6~\text{fb}^{-1}\), covering 
invariant masses between 26 GeV and 66 GeV in both the electron and muon channels. Additionally, 
data from 2010, with \(35~\text{pb}^{-1}\) of integrated luminosity, extend the measurement 
down to 12 GeV in the muon channel. These low-mass Drell-Yan measurements provide crucial constraints 
on PDFs in the low-\(x\) region, complementing higher-mass analyses, and have been included in our analysis.

Building on this, ATLAS also analyzed a dataset focused on the high-mass region of Drell-Yan production, 
offering precise measurements of the differential cross-section in proton-proton 
collisions at \(\sqrt{s} = 7\) TeV~\cite{ATLAS:2013xny}. This analysis, based on an 
integrated luminosity of \(4.9~\text{fb}^{-1}\), examines the \(Z/\gamma^{*} \rightarrow e^+e^-\) channel. 
The differential cross-section is measured as a function of the invariant mass \(m_{ee}\) in 
the range 116 GeV \(< m_{ee} <\) 1500 GeV, within a fiducial region where both the electron and 
positron have transverse momentum \(p_T > 25\) GeV and pseudorapidity \(|\eta| < 2.5\). 
These measurements extend the sensitivity of the analysis to higher \(x\) values, complementing the 
lower-mass data, and have been applied in our QCD fit.

\textbf {E866/NuSea (Fermilab)}: 
In addition to the LHC measurements, the E866/NuSea experiment at Fermilab, which is included in 
this analysis, has made significant contributions to our understanding of flavor asymmetry in 
the nucleon sea. This experiment measured the ratio of Drell-Yan yields from an 800 GeV/c proton 
beam incident on liquid hydrogen and deuterium targets~\cite{NuSea:2001idv}. The extensive dataset, 
comprising approximately 360,000 Drell-Yan muon pairs, enabled the determination of the 
ratio of \(\bar{d}\) to \(\bar{u}\) quark distributions in the proton sea across a broad 
range of Bjorken-\(x\). The results revealed a sharp downturn in the \(\bar{d}(x)/\bar{u}(x)\) 
ratio at large \(x\), leading to substantial revisions in global parameterizations of the 
nucleon sea and providing tighter constraints on valence PDFs. These findings have inspired 
further interest in extending measurements to higher \(x\) values 
using the 120 GeV/c proton beam from the Fermilab Main Injector.

%=================================================
\subsection{Precision W/Z Production Collider Data} 
%=================================================

In addition to DIS and Drell-Yan processes, recent collider data from the LHC and other 
sources provide complementary sensitivity for improving PDFs. In our global analysis, we 
incorporate W and Z boson production data from both the LHC (ATLAS and CMS collaborations) 
and the Tevatron (CDF and D0 collaborations), which offer crucial constraints on 
the quark flavor separation in PDFs, particularly for sea quarks.  \\

The ATLAS collaboration has provided two key datasets for this analysis:  \\

\textbf {Inclusive W/Z Production (ATLAS, 7 TeV)}: High-precision measurements of inclusive \(W^+ \rightarrow \ell^+ \nu\), 
\(W^- \rightarrow \ell^- \bar{\nu}\), and \(Z/\gamma^* \rightarrow \ell \ell\) (\(\ell = e, \mu\)) 
Drell-Yan production cross sections in proton-proton collisions at \(\sqrt{s} = 7\) TeV were 
published by ATLAS \cite{ATLAS:2016nqi}. 
These measurements are based on data corresponding to an integrated luminosity of 4.6 fb\(^{-1}\). 
The differential cross sections for \(W^+\) and \(W^-\) are measured within the 
lepton pseudorapidity range \(|\eta_\ell| < 2.5\), while the differential \(Z/\gamma^*\) cross sections 
are measured as a function of the absolute dilepton rapidity \(|y_{\ell\ell}| < 3.6\) across 
three dilepton mass intervals (46 GeV < \(m_{\ell\ell}\) < 150 GeV). 
These measurements, combined across the electron and muon channels, provide stringent 
constraints on the PDFs of the proton when compared to theoretical predictions using modern PDF sets.

\textbf {W/Z Production and Strange Quark Density}: Another ATLAS dataset, focused on inclusive \(W^\pm\) and \(Z\) boson 
production at \(\sqrt{s} = 7\) TeV, has been critical in improving the 
understanding of the light quark sea, particularly the strange quark density \cite{ATLAS:2012sjl}. 
This analysis includes a NNLO perturbative QCD fit and finds that the ratio of 
strange-to-down sea quark densities, \(r_s\), is approximately 1.00 at \(x = 0.023\) 
and \(Q^2_0 = 1.9\) GeV\(^2\), suggesting a flavor-symmetric light quark sea at low \(x\). 
This data plays a crucial role in refining the strange quark contribution to the proton sea quark content.   \\

The CMS collaboration has provided four key datasets that are included in our analysis:  \\

\textbf {Electron Charge Asymmetry (7 TeV)}: A precise measurement of the electron charge 
asymmetry in \(pp \rightarrow W + X \rightarrow e\nu + X\) production at \(\sqrt{s} = 7\) TeV, 
based on 840 pb\(^{-1}\) of data, was performed by CMS \cite{CMS:2012ivw}. 
This asymmetry measurement, as a function of electron pseudorapidity \(|\eta|\), imposes stringent 
constraints on PDFs by probing the differences in \(W^+\) and \(W^-\) production.

\textbf {Muon Charge Asymmetry (7 TeV)}: CMS also measured the muon charge asymmetry in 
inclusive \(pp \rightarrow W + X\) production, based on 4.7 fb\(^{-1}\) of data \cite{CMS:2013pzl}. 
The improved precision of this measurement, in 11 bins of muon pseudorapidity with different \(p_T\) thresholds, 
provides valuable input for determining valence and strange quark distributions.

\textbf {W Boson Production (8 TeV)}: A comprehensive measurement of the differential 
cross-section and charge asymmetry for inclusive \(pp \rightarrow W^{\pm} + X\) production 
at \(\sqrt{s} = 8\) TeV, based on 18.8 fb\(^{-1}\) of data, was carried out by CMS \cite{CMS:2016qqr}. 
These results, including differential cross-sections with respect to \(p_T\) and lepton pseudorapidity, 
are essential for constraining both the valence and sea quark distributions.

\textbf {Z Boson Production (7 TeV)}: CMS also measured the rapidity and transverse momentum 
distributions of \(Z\) bosons in \(pp\) collisions at \(\sqrt{s} = 7\) TeV, using 36 pb\(^{-1}\) 
of data \cite{CMS:2011wyd}. These measurements, with rapidity up to \(|y| < 3.5\) and transverse 
momentum up to 350 GeV, provide critical insights into the production dynamics of \(Z\) bosons and 
serve as an important input for refining the gluon and sea quark PDFs.

Finally, the Tevatron data sets from CDF and D0 Collaborations in which added to our data samples include: 

\textbf {CDF W Boson Production Charge Asymmetry}: The CDF collaboration measured the charge 
asymmetry in \(W\) boson production in \(p\bar{p}\) collisions at \(\sqrt{s} = 1.96\) TeV \cite{CDF:2009cjw}. 
This dataset provides key insights into the asymmetry between \(W^+\) and \(W^-\) production, 
helping to refine quark PDFs at intermediate \(x\).

\textbf {CDF Drell-Yan Production}: CDF also measured the differential cross-section \(d\sigma/dy\) 
for Drell-Yan \(e^+e^-\) pairs in the \(Z\) boson mass region at \(\sqrt{s} = 1.96\) TeV \cite{CDF:2010vek}, 
providing constraints on the quark-antiquark PDFs at intermediate \(x\).

\textbf {D0 Electron Charge Asymmetry}: D0 also measured the electron charge asymmetry in \(W \rightarrow e\nu\) 
events, providing complementary constraints to the muon charge asymmetry data \cite{D0:2014kma}.

\textbf {D0 Muon Charge Asymmetry}: D0 measured the muon charge asymmetry in \(W \rightarrow \mu\nu\) 
events at \(\sqrt{s} = 1.96\) TeV \cite{D0:2013xqc}, offering critical constraints on the \(u\) and \(d\) quark PDFs.

\textbf {Z Boson Rapidity Distribution}: Finally, D0 measured the rapidity distribution of \(Z/\gamma^* \rightarrow e^+ e^-\) 
in \(p\bar{p}\) collisions \cite{D0:2007djv}, providing detailed tests of QCD and electroweak 
theory through the shape of the rapidity distribution.

%=================================================
\section{Computational Setting and PDF Parametrization}\label{method}
%=================================================

In this section, we describe the computational framework and the parameterization strategy 
adopted for the determination of PDFs. The accurate extraction of the PDFs requires a 
robust computational environment that integrates various aspects of the QCD analysis, 
including the evolution of parton densities, the computation of physical observables, 
and the fitting of theoretical predictions to experimental data.
Our analysis is implemented within the {\tt xFitter} framework~\cite{xFitter:2022zjb,Alekhin:2014irh}, 
a versatile tool set designed for the global analysis of PDFs. This 
framework facilitates the numerical solution of the Dokshitzer-Gribov-Lipatov-Altarelli-Parisi (DGLAP) 
evolution equations~\cite{Altarelli:1977zs,Gribov:1972ri,Dokshitzer:1977sg}, 
ensuring consistency with perturbative QCD predictions up to NNLO accuracy. 
Moreover, {\tt xFitter} allows for the incorporation of various sources of experimental 
uncertainties, including statistical, systematic, and correlated uncertainties, into the PDF extraction process.
The PDF parameterization at the initial scale is a critical component of a QCD analysis, 
as it forms the basis for subsequent QCD evolution. We adopt a flexible functional 
form that captures the essential features of parton distributions while maintaining a 
balance between the number of free parameters and the stability of the fit.
The optimization of the parameterization involves fitting the experimental data to the 
theoretical predictions, which are computed by evolving the PDFs from the initial scale 
to the relevant experimental scales. This process is performed using the {\tt MINUIT} 
minimization package~\cite{James:1975dr,James:1994vla}, which provides both the central 
values of the parameters and their associated uncertainties.

%=================================================
\subsection{PDF Parameterization}\label{PDF_parametrisation}
%=================================================

In our QCD analysis, a flexible yet controlled functional form is employed to describe the wide range of 
experimental data without overfitting the PDFs. The initial scale is chosen to be \( Q_0^2 = 1.9 \, \text{GeV}^2 \), 
which lies below the charm mass threshold. This ensures that heavy-quark distributions are 
dynamically generated above their respective mass thresholds, 
which are set at \( m_c = 1.43 \, \text{GeV} \) for the charm quark and \( m_b = 4.5 \, \text{GeV} \) for the bottom quark. 
These thresholds are determined from the most recent heavy-quark 
differential cross-section measurements conducted at HERA.
As we mentioned before, our analysis follows the RT VFNS~\cite{Thorne:1997ga,Thorne:2006qt,Thorne:2012az}, 
where heavy quarks are dynamically included in the PDFs once the energy scale 
surpasses their respective mass thresholds~\cite{Martin:2009iq,Thorne:2006qt,Alekhin:2012ig}.

We use a HERAPDF-like parameterization for the valence and sea quark distributions, which has 
been well-validated in previous HERA global QCD analyses~\cite{H1:2015ubc}. 
The parameterizations for valence quarks \( x_{uv} \) and \( x_{dv} \) are constrained by 
the quark number sum rules, while the sea quark distributions are refined 
iteratively as new experimental datasets are included.

The gluon distribution poses significant challenges in the PDF parameterization, 
particularly at low \(x\). Previous studies, such as those in the 
MSTW08 analysis~\cite{Martin:2009iq, Martin:2001es}, observed that when the PDF evolution 
begins from a low scale \(Q_0^2 \sim 1 \, {\text{GeV}}^2\), the input gluon 
distribution can turn negative at very small values of \(x\), outside the kinematic 
region of the fit, causing issues for the PDFs at very low \(x\). To address this 
problem, and following the HERAPDF QCD analysis and similar work by MSTW08, 
we introduce an additional term in the gluon parameterization: \( A_{g'} x^{B_{g'}} (1 - x)^{C_{g'}} \). 
This term allows for greater flexibility in describing the gluon distribution, 
particularly in the small-\(x\) region. The inclusion of this correction is crucial for 
preventing unphysical negative gluon densities and stabilizing the fit across a wide range of \(x\) values.

The use of a two-component gluon distribution, 
represented by \( xg(x) = xg_1(x) + xg_2(x) \sim A_g x^{B_g} + A_{g'} x^{B_{g'}} \), 
also offers enhanced accuracy in describing both the small and large \(x\) behavior 
of the gluon. By capturing the steep rise at low \(x\) and the expected 
fall-off at high \(x\), this parameterization improves the overall precision of the proton PDFs. 
It also ensures that the gluon density fits smoothly with the constraints provided by experimental data, 
particularly in regions where data is sparse, such as for very small \(x\). This additional 
term ensures that the parameterization is sufficiently flexible at very low \(x\)~\cite{H1:2015ubc}.

A key aspect of understanding the proton structure is the determination of the strange-quark density. 
In our QCD analysis, we define the strange-quark fraction \(r_s\) as the ratio of the strange 
sea-quark distribution to the down sea-quark distribution~\cite{ATLAS:2016nqi}:
%
%--------------------------------
\begin{equation}
\label{PDF-rs}
r_s = \frac{s + \bar{s}}{2 \bar{d}}.  
\end{equation}
%--------------------------------
%
This parameter provides valuable insight into the relative contribution of strange quarks 
compared to down quarks in the proton sea. Recent ATLAS studies~\cite{ATLAS:2016nqi} have 
shown that \(r_s\) plays a crucial role in improving the precision of strange-quark distributions, 
which are critical for predictions of processes like W and Z boson production. Including \(r_s\) 
as a fit parameter allows us to extract precise information on the strange-quark content and 
its impact on other quark distributions. Our analysis refines this parameter to improve 
predictions for high-energy processes involving strange quarks.

In addition to the PDFs, the strong coupling constant \(\alpha_s\) at the Z boson mass (\(M_Z\)) 
is treated as a free parameter in our QCD fit. This ensures that the coupling evolves consistently 
with the energy scale, allowing us to account for uncertainties in both \(\alpha_s\) 
and the PDFs. The value of \(\alpha_s(M_Z)\) is critical for accurately predicting cross-sections 
in processes like jet and top quark production at the LHC, where gluon interactions dominate.

Finally, we adopt the following functional forms for the parton distributions at the initial scale \(Q_0^2\):
%
%--------------------------------
\begin{align}
\label{PDF-Q0}
xu_{v}(x) &= A_{uv} x^{B_{uv}} (1 - x)^{C_{uv}} (1 + E_{uv} x^2), \nonumber\\
xd_{v}(x) &= A_{dv} x^{B_{dv}} (1 - x)^{C_{dv}},  \nonumber\\
x{\bar{u}}(x) &= A_{\bar{u}} x^{B_{\bar{u}}} (1 - x)^{C_{\bar{u}}}, \nonumber\\
x{\bar{d}}(x) &= A_{\bar{d}} x^{B_{\bar{d}}} (1 - x)^{C_{\bar{d}}}, \nonumber\\
xg(x) &= A_{g} x^{B_{g}} (1 - x)^{C_{g}} - A_{g'} x^{B_{g'}} (1 - x)^{C_{g'}}, \nonumber\\
x{s}(x)  &= A_{\bar{s}} \,  r_s \,  x^{B_{\bar{s}}} (1-x)^{C_{s}}.
\end{align}
%--------------------------------
%

The above functional forms for \( xu_{v} \), \( xd_{v} \), \( x{\bar{u}} \), \( x{\bar{d}} \), and \( x{\bar{s}} \) 
are tailored to ensure accurate predictions at both low and high values of \(x\). 
In this analysis, we assume similar behavior of the up and down sea-quark distributions, 
setting \(A_{\bar{d}} = A_{\bar{u}}\) and \( B_{\bar{d}} = B_{\bar{u}} \). These assumptions are 
consistent with the HERAPDF-like~\cite{H1:2015ubc} and ATLAS-epWZ16~\cite{ATLAS:2016nqi} parameterizations 
for the sea quark distributions as well.  
For the strange distributions, we consider \( xs(x) = x\bar{s}(x) \), and also assume \(A_{\bar{s}} = A_{\bar{d}}\) 
and \(B_{\bar{s}} = B_{\bar{d}}\). 
Given the enhanced sensitivity to the strange-quark distribution through the
CMS and ATLAS W/Z data sets, the ${C_{s}}$ appear as a free parameter in the fit. 
The above assumptions also simplify the parameter space, especially 
given the limited experimental data constraints on these distributions.  
Following the HERAPDF analysis, on which our current work is closely based, as 
well as the MSTW2008 analysis, we have fixed the parameter \(C'_{g}\) to \(C'_{g} = 25\). 
As discussed in Refs.~\cite{H1:2015ubc,Martin:2009iq}, both analyses have demonstrated that the QCD fits 
are generally not sensitive to the exact value of \(C'_{g}\), provided that \(C'_{g} \gg C_{g}\), ensuring 
that the additional term does not contribute significantly at large \(x\).

%=================================================
\subsection{Definition and Minimization of \(\chi^2\) Function}\label{chi2}
%=================================================

In this section, we focus on two essential aspects of our QCD analysis: 
the optimization procedure for determining the PDFs and the incorporation of the experimental uncertainties. 
The free parameters of the PDFs are estimated using experimental data through a 
likelihood maximization method. Assuming that the data points are normally distributed, 
the likelihood maximization reduces to minimizing the \(\chi^2\) 
function, which quantifies the agreement between theory and experiment.

The \(\chi^2\) function used in our analysis follows the formalism of 
the {\tt xFitter} framework~\cite{xFitter:2022zjb,Alekhin:2014irh} and is expressed as follows:
%
%-------------------------------
\begin{eqnarray}\label{Chi}
\label{eq:chi2}
\chi^{2} (\{\zeta_k\}) & = & 
\sum_{i} 
\frac{\left[ \mathcal{E}_{i} - 
\mathcal{T}_{i} \left(\{\zeta\}\right) 
\left(1 - \sum_{j} 
\gamma_{j}^{i} {b}_{j}\right) 
\right]^{2}  }
{\delta^{2}_{{i}, \mathrm{unc}} 
\mathcal{T}_{i}^{2}(\{\zeta\})+ 
{\delta}^{2}_{i, \mathrm{stat}} 
\mathcal{E}_{i} \mathcal{T}_{i} 
(\{\zeta\}) 
\left(1 - \sum_{j} \gamma^{i}_{j} 
{b}_{j}\right)} \nonumber  \\ 
& + & \sum_{i} \ln \frac{\delta^{2}_{{i}, 
\mathrm{unc}} 
\mathcal{T}_{i}^{2} (\{\zeta\}) + 
\delta^{2}_{{i}, \mathrm{stat}} 
\mathcal{E}_{i} \mathcal{T}_{i}  
(\{\zeta\}) } 
{ \delta^{2}_{{i}, \mathrm{unc}} 
\mathcal{E}_{i}^{2} + 
\delta^{2}_{{i}, \mathrm{stat}} 
\mathcal{E}^{2}_{i}}  + 
\sum_{j} b_{j}^{2}\,,
\end{eqnarray}
%-------------------------------
%
where \(\mathcal{E}_{i}\) represents the experimental measurements, 
and \(\mathcal{T}(\{\zeta_k\})\) are the theoretical predictions 
that depend on a set of parameters \(\{\zeta_k\}\). The uncertainties in 
each measurement are separated into two components: \(\delta_{i,\text{stat}}\), 
the statistical uncertainty, and \(\delta_{i,\text{unc}}\), 
the uncorrelated systematic uncertainty, while \(\gamma_{j}^{i}\) denotes the 
correlated systematic uncertainties. The nuisance parameters \(b_{j}\) 
quantify the strength of the correlated systematic uncertainties across the data set.

This formulation allows us to correctly incorporate both statistical and 
systematic uncertainties into the minimization process. 
The nuisance parameters \(b_{\alpha}\) are introduced to control deviations in the 
correlated systematic uncertainties, with a penalty term \(\sum_{\alpha} b_{\alpha}^2\) 
added to the \(\chi^2\) function to constrain these deviations. The minimization of this 
function is performed using the {\tt MINUIT}~\cite{James:1975dr,James:1994vla} package 
via the {\tt MIGRAD} algorithm~\cite{James:1975dr}, as implemented in {\tt xFitter}. 
Once the minimum is found, the uncertainties in the fitted parameters are evaluated using 
the Hessian matrix, which is central to uncertainty propagation in the context of global QCD fits.

In a limited number of experimental studies, an alternative approach is employed wherein 
the covariance matrix, \(\text{cov}_{ij}\), is used in place of Eq.~\ref{Chi} to account for correlations between data points:
%
%-------------------------------
\begin{equation}
\chi^2_E(a, \lambda) = \sum_{k=1}^{N_\text{pt}} \frac{(D_k - T_k(a) - 
\sum_{\alpha} \lambda_\alpha \beta_{k\alpha})^2}{s_k^2} + 
\sum_{\alpha} \lambda_\alpha^2,
\end{equation}
%-------------------------------
%
where \(D_k\) represents the experimental data, \(T_k(a)\) represents the theoretical 
predictions for the \(k\)$^{\rm th}$ data point, and \(\lambda_\alpha\) are nuisance parameters 
that account for the correlated systematic uncertainties from source \(\alpha\). 
The total uncertainty \(s_k\) is the combination of statistical and uncorrelated 
systematic uncertainties. The \(\lambda_\alpha\) parameters are optimized 
analytically during the minimization process. In this formulation, the covariance 
matrix \(\text{cov}_{ij}\) captures both uncorrelated and correlated uncertainties and is given by:
%
%-------------------------------
\begin{equation}
\text{cov}_{ij} = s_i^2 \delta_{ij} + \sum_{\alpha} \beta_{i\alpha} \beta_{j\alpha},
\end{equation}
%-------------------------------
%
where \(\beta_{i\alpha}\) characterizes the sensitivity of the \(i\)$^{\rm th}$ and \(j\)$^{\rm th}$ data 
points to the correlated systematic uncertainty \(\alpha\). This decomposition allows for 
a proper treatment of experimental uncertainties, ensuring that correlated systematic effects are propagated consistently.

The minimization of the \(\chi^2\) function is followed by the propagation of uncertainties 
from the fit parameters to the physical observables. This is accomplished by generating a set of 
eigenvector PDF sets, which represent variations of the PDF parameters along the directions 
defined by the eigenvectors of the Hessian matrix. For \(N\) free parameters, a total of \(2N+1\) 
PDF sets are generated: one central fit and two variations along each eigenvector 
direction. Finally, the calculated proton PDF sets are made available in the standard {\tt LHAPDF} format~\cite{Buckley:2014ana}.

%=================================================
\subsection{Determination of PDF uncertainties}\label{uncertainty}
%=================================================

The Hessian formalism is widely used in the global analysis of PDFs, such as in the CT18 framework, 
to study the uncertainties associated with the fitted PDFs and their physical predictions~\cite{Pumplin:2001ct,Giele:2001mr}.   
The method is based on a quadratic approximation to the \(\chi^2\) function around its minimum. 
The eigenvectors of the Hessian matrix characterize the directions in the parameter space in 
which the \(\chi^2\) increases most rapidly, corresponding to directions where the fit 
is most sensitive to the data. This allows us to generate a set of orthonormal basis PDFs, 
known as eigenvector PDF sets, which can be used to calculate uncertainties in physical observables by 
varying the fit parameters along each eigenvector direction~\cite{Martin:2009iq}.

In the Hessian formalism, the \(\chi^2\) function near its minimum can 
be expressed as a quadratic form:
%
%-------------------------------
\begin{equation}
\Delta \chi^2 = \chi^2(\boldsymbol{\zeta}) - \chi^2(\boldsymbol{\zeta}_0) \approx \sum_{i,j} 
\frac{\partial^2 \chi^2}{\partial \zeta_i \partial \zeta_j} (\zeta_i - \zeta_i^0)(\zeta_j - \zeta_j^0),
\end{equation}
%-------------------------------
%
where \(\boldsymbol{\zeta}_0\) represents the best-fit parameter vector, \(\boldsymbol{\zeta}\) are the 
fit parameters, and \(\frac{\partial^2 \chi^2}{\partial \zeta_i \partial \zeta_j}\) represents 
the second-order derivatives of \(\chi^2\) with respect to the parameters \(\zeta_i\). 
This second-order derivative matrix is called the Hessian matrix, denoted as \(H_{ij}\):
%
%-------------------------------
\begin{equation}
H_{ij} = \frac{\partial^2 \chi^2}{\partial \zeta_i \partial \zeta_j}.
\end{equation}
%-------------------------------
%
The Hessian matrix captures the shape and curvature of the \(\chi^2\) surface near its minimum, 
reflecting how sensitive the fit is to variations in the parameters.
Diagonalizing this matrix yields the eigenvectors \(\mathbf{v}_k\) and corresponding eigenvalues \(\lambda_k\), 
which provide a natural basis for exploring the parameter space. 
The eigenvectors define directions in the parameter space along which the \(\chi^2\) increases most rapidly, 
indicating where the fit is most sensitive to variations in the parameters.

The parameter variations along the eigenvector directions are determined by solving the following eigenvalue problem:
%
%-------------------------------
\begin{equation}
H \mathbf{v}_k = \lambda_k \mathbf{v}_k.
\end{equation}
%-------------------------------
%
In practice, these eigenvectors are used to construct a set of orthonormal basis PDFs, known as eigenvector PDF sets. 
These eigenvector sets allow for the systematic variation of PDF parameters, 
and hence the estimation of uncertainties on physical observables.
Once the eigenvectors of the Hessian matrix are known, the uncertainty in any physical 
observable \(\mathcal{O}\) that depends on the PDFs can be calculated~\cite{Martin:2009iq}. 

The eigenvector basis is constructed by transforming the parameters of the fit into a new set of 
coordinates where the Hessian matrix is diagonal. In this new coordinate system, 
the parameters are normalized such that displacements from the minimum in each 
direction correspond to fixed increases in \(\chi^2\). The uncertainty in any observable \(\mathcal{O}\) 
is then calculated by evaluating the variation of \(\mathcal{O}\) along each eigenvector direction. 
This method provides a robust framework for propagating uncertainties in the PDFs to 
predictions for physical observables.

The uncertainty in any observable \(\mathcal{O}\) that depends on the PDFs is calculated as~\cite{Pumplin:2001ct,Martin:2009iq}, 
%
%-------------------------------
\begin{equation}
\label{eq:Hessian}
\Delta \mathcal{O} = \sqrt{\sum_{i=1}^{N} \left( \frac{\mathcal{O}_i^+ - \mathcal{O}_i^-}{2} \right)^2},
\end{equation}
%-------------------------------
%
where \(\mathcal{O}_i^+\) and \(\mathcal{O}_i^-\) are the observable values computed using the 
positive and negative variations of the \(i\)$^{th}$ eigenvector. 
This method assumes that the variations of the observables \(\mathcal{O}\) can be 
approximated as linear within the region where the \(\chi^2\) function behaves quadratically. 
Such an assumption holds well near the minimum, ensuring reliable propagation of 
uncertainties from the PDFs to the final physical observables.

To ensure that the fit remains well-behaved and does not overfit the data, especially in regions where 
the data are sparse or have large uncertainties, regularization methods are applied. 
In the {\tt xFitter} framework, regularization is implemented through the inclusion of 
length penalty terms in the \(\chi^2\) function. These penalty terms discourage overly complex PDF shapes, 
promoting smoother distributions that are more physically reasonable. Regularization is particularly 
important in global PDF fits, where the data cover a wide range of kinematic regions, 
and the parameterization of the PDFs must be flexible enough to 
accommodate this diversity while avoiding overfitting.

Regularization also plays a crucial role in determining the appropriate degree of flexibility 
in the parameterization of the PDFs. Too much flexibility can lead to poorly constrained parameters 
and instabilities in the fit, while too little flexibility may prevent the fit from adequately describing the data. 
By introducing regularization terms, we can control the complexity of the PDF parameterization and 
ensure that the fit remains stable and reliable across the full kinematic range.

%=================================================  
\section{Fit results}\label{results}  
%=================================================  

This section presents the main results and findings of this work. As outlined below, 
different QCD fits have been performed, considering various data sets. 
We first examine the quality of our QCD fits, then present the extracted PDFs, 
and finally compare the fit quality across a selection of fitted data for different scenarios.

To explore the impact of Drell-Yan and W/Z boson production data from the CMS, 
ATLAS, D0, and CDF collaborations on the shape of different parton species and 
their associated uncertainties, we performed four different QCD fits at NLO and NNLO accuracy. 
The details of these approaches are described below:

%-----------------------
\begin{itemize}  
%-----------------------

\item \textbf{Fit~A}:  
In the first fit, we incorporate only the HERA I+II~\cite{Abramowicz:2015mha} experimental data, 
which serves as a crucial baseline for any QCD analysis. Utilizing 1016 data points, as 
per the \(Q^2\) cut mentioned earlier, the results from Fit~A provide a robust foundation for 
assessing the impact of additional experimental measurements on the PDFs.

\item \textbf{Fit~B}:  
In this fit, we build on Fit~A by including measurements related to the Drell-Yan process 
from the ATLAS collaboration~\cite{ATLAS:2013xny,ATLAS:2014ape} and 
the E866 experiment~\cite{NuSea:2001idv}. The configuration of this fit remains 
identical to Fit~A, with the only difference being the inclusion of Drell-Yan data. 
This fit utilizes 1082 data points, as indicated in Table~\ref{tab:Chi2}, allowing 
us to evaluate how Drell-Yan measurements affect the shape of different parton species 
and their associated uncertainties.

\item \textbf{Fit~C}:  
Fit~C is similar to Fit~B, but instead of Drell-Yan data, it includes CMS and 
ATLAS measurements of W/Z boson cross-sections~\cite{ATLAS:2012sjl,ATLAS:2016nqi,CMS:2016qqr,CMS:2012ivw,CMS:2013pzl,CMS:2011wyd}, 
as well as the corresponding measurements from the D0 
and CDF collaborations at Tevatron~\cite{D0:2014kma,D0:2013xqc,D0:2007djv,CDF:2010vek}. 
This fit is performed using 1289 data points, enabling us 
to study the specific impact of W/Z boson production data on the extracted PDFs.

\item \textbf{Fit~D}:  
The final, or nominal, fit is performed using the combination of all 
the data sets discussed above, providing the most comprehensive QCD analysis. 
This fit integrates the HERA I+II, Drell-Yan, and W/Z boson datasets, yielding a 
complete picture of the PDFs and their associated uncertainties. The PDFs in 
this fit are extracted from 1355 experimental data points.

%-----------------------
\end{itemize}
%-----------------------

%------------------chi2-table---------------------
\begin{table*}
\caption{\label{tab:Chi2}
The extracted numerical values for the correlated \(\chi^2\), 
log penalty \(\chi^2\), and the total \(\chi^2\) per degree of 
freedom (dof) associated with each experimental measurement. 
The table also lists the input data sets included in the analysis of our PDFs at NNLO accuracy. 
For each data set, the name of the experiment, and the corresponding reference are also provided. } 
\begin{center}
%  \rowcolors{2}{lightgray}{}
%  \resizebox*{\textwidth}{!}{
\begin{tabular}{p{2.02cm}lp{2.02cm}p{2.02cm}p{2.02cm}p{2.02cm}p{2.02cm}}
\toprule
Observable &  Experiment   & Reference  & {\tt Fit A}   & {\tt Fit B}   & {\tt Fit C}   & {\tt Fit D} \\
\hline
%      \midrule
%-----------------------HERA------------------------
& HERA1+2 CC $e ^+ p$ & \cite{Abramowicz:2015mha} & 46 / 39& 42 / 39& 44 / 39& 43 / 39  \\
& HERA1+2 CC $e ^- p$ & \cite{Abramowicz:2015mha} & 55 / 42& 60 / 42& 69 / 42& 68 / 42  \\ 
& HERA1+2 NC $e ^- p$ & \cite{Abramowicz:2015mha} & 222 / 159& 224 / 159& 220 / 159& 222 / 159  \\ 
DIS & HERA1+2 NC $e ^- p$ 460 & \cite{Abramowicz:2015mha} & 195 / 177& 195 / 177& 195 / 177& 197 / 177  \\
& HERA1+2 NC $e ^- p$ 575 & \cite{Abramowicz:2015mha} & 187 / 221& 188 / 221& 187 / 221& 189 / 221  \\  
& HERA1+2 NC $e ^+ p$ 820 & \cite{Abramowicz:2015mha} & 53 / 61& 55 / 61& 53 / 61& 53 / 61  \\
& HERA1+2 NC $e ^+ p$ 920 & \cite{Abramowicz:2015mha} & 351 / 317& 361 / 317& 359 / 317& 379 / 317  \\
\hline
%-----------------------DY------------------------
& ATLAS high mass & \cite{ATLAS:2013xny} & - & 13 / 13& - & 9.2 / 13  \\
& ATLAS low mass 1.6 fb$^{-1}$ & \cite{ATLAS:2014ape} & - & 7.5 / 8& - & 8.7 / 8  \\ 
Drell-Yan & ATLAS low mass extended 35 pb$^{-1}$ & \cite{ATLAS:2014ape} & - & 6.4 / 6& - & 7.4 / 6  \\ 
& E866 low mass & \cite{NuSea:2001idv} & - & 11 / 10& - & 14 / 10  \\ 
& E866 mid mass & \cite{NuSea:2001idv} & - & 14 / 14& - & 13 / 14  \\
& E866 high mass & \cite{NuSea:2001idv} & - & 13 / 15& - & 23 / 15  \\
\hline
%-----------------------W------------------------
& ATLAS $W^+$  & \cite{ATLAS:2012sjl} & - & - & 15 / 11& 15 / 11  \\              
& ATLAS $W^+$  & \cite{ATLAS:2012sjl} & - & - & 15 / 11& 15 / 11  \\ 
& ATLAS $W^-$  & \cite{ATLAS:2012sjl} & - & - & 8.9 / 11& 8.9 / 11  \\
& ATLAS $W^+$  & \cite{ATLAS:2016nqi} & - & - & 13 / 11& 13 / 11  \\ 
& ATLAS $W^-$  & \cite{ATLAS:2016nqi} & - & - & 10 / 11& 9.6 / 11  \\ 
& CMS $W^+$  & \cite{CMS:2016qqr} & - & - & 1.6 / 11& 3.3 / 11  \\
& CMS $W^-$ & \cite{CMS:2016qqr} & - & - & 1.2 / 11& 2.3 / 11  \\
& CMS $W$ asymmetry & \cite{CMS:2016qqr} & - & - & 7.5 / 11& 11 / 11  \\
& CMS $W$ asymmetry & \cite{CMS:2012ivw} & - & - & 10.0 / 11& 9.2 / 11  \\            
& CMS $W$ asymmetry  & \cite{CMS:2013pzl} & - & - & 13 / 11& 14 / 11  \\  
$W/Z$ Bosons &   CDF $W$ asymmetry & \cite{CDF:2009cjw} & - & - & 38 / 13& 47 / 13  \\
& D0 $W \rightarrow e\nu$ asymmetry & \cite{D0:2014kma} & - & - & 31 / 13& 31 / 13  \\     
& D0 $W \rightarrow \mu\nu$  asymmetry & \cite{D0:2013xqc} & - & - & 15 / 10& 14 / 10  \\   
%  \hline
%-----------------------Z------------------------
&   ATLAS high mass CF $Z$  & \cite{ATLAS:2016nqi} & - & - & 3.9 / 6& 3.9 / 6  \\ 
&   ATLAS high mass CC $Z$  & \cite{ATLAS:2016nqi} & - & - & 5.6 / 6& 5.5 / 6  \\
&   ATLAS low mass $Z$  & \cite{ATLAS:2016nqi} & - & - & 22 / 6& 23 / 6  \\
&   ATLAS peak CF $Z$ & \cite{ATLAS:2016nqi} & - & - & 5.5 / 9& 8.4 / 9  \\ 
&   ATLAS peak CC $Z$ & \cite{ATLAS:2016nqi} & - & - & 13 / 12& 15 / 12  \\
&   ATLAS $Z$ & \cite{ATLAS:2012sjl} & - & - & 2.9 / 8& 2.5 / 8  \\  
&   CMS $Z$ & \cite{CMS:2011wyd} & - & - & 65 / 35& 67 / 35  \\
&   CDF $Z$  & \cite{CDF:2010vek} & - & - & 28 / 28& 28 / 28  \\
&   D0 $Z$ & \cite{D0:2007djv} & - & - & 22 / 28& 22 / 28  \\ 
  \hline
  \hline
%-----------------------chi2------------------------
  Correlated $\chi^2$ & & & 49& 61& 95& 113  \\ 
  Log~penalty~$\chi^2$~& & & -10.87& -17.10& -5.05& -13.22  \\ 
  %\rowcolor{white}
%      \midrule
  Total $\chi^2$ / dof & & & 1149 / 1000 & 1236 / 1066& 1551 / 1273& 1678 / 1339  \\
 & & & = 1.149 & = 1.159 & = 1.218 & = 1.253  \\
\hline
\hline
 % \rowcolor{white}
%      \midrule
%  $\chi^2$ p-value  & 0.00 & 0.00 & 0.00 & 0.00   \\ 
%      \bottomrule
    \end{tabular}
%  }
  \end{center}
\end{table*}
%------------chi2-table---------------------------
%--------------------------------

%=================================================  
\subsection{Fit quality}  
%=================================================  

The \(\chi^2\) values characterizing our NNLO nominal QCD fit to the H1/ZEUS combined datasets, 
as well as the Drell-Yan and W/Z production data from the LHC and Tevatron, 
are listed in Table~\ref{tab:Chi2}. As shown, our NNLO QCD fits provide a good 
description of the data sets both in terms of individual and total data sets 
from HERA, LHC, and Tevatron. Several observations can also be made based 
on the \(\chi^2\) values for both the individual and total datasets.

For \textbf{Fit~A}, where we utilize only the HERA data, the 
total \(\chi^2/\text{dof}\) is 1.149, indicating a good quality QCD fit 
within the chosen PDF parameterization of Eq.~\eqref{PDF-Q0}. This baseline 
fit sets the standard against which the inclusion of additional data sets is evaluated.

The inclusion of Drell-Yan data in \textbf{Fit~B} slightly increases the 
total \(\chi^2\) value to 1.159. Despite this slight increase, the 
overall fit quality remains very good, and the description of the 
individual data sets is still satisfactory. This modest change suggests 
that the additional Drell-Yan data are largely compatible 
with the existing HERA data, with minimal tension between the datasets.

When combining the HERA data with the W/Z boson production 
cross-section data in \textbf{Fit~C}, we observe a further increase in 
the total \(\chi^2\) to 1.218. While the overall fit quality 
remains acceptable, some individual data sets show slight discrepancies. 
Specifically, the CDF \(W\) asymmetry data~\cite{CDF:2009cjw} and the ATLAS 
low-mass \(Z\) data~\cite{ATLAS:2016nqi} show less agreement with our NNLO theory 
predictions, contributing to the increase in the total \(\chi^2\).

In our nominal fit, \textbf{Fit~D}, which incorporates all 
data sets discussed, the total \(\chi^2\) increases slightly to 1.253. This fit 
includes the effects of all combined data, and the observed increase is 
primarily driven by the tension between certain datasets. In addition to 
the CDF \(W\) asymmetry~\cite{CDF:2009cjw} and ATLAS low-mass \(Z\) 
data~\cite{ATLAS:2016nqi}, a noticeable increase in the \(\chi^2\) value is 
also seen for the HERA I+II NC \(e^+ p\) 920 dataset~\cite{Abramowicz:2015mha}. 

Overall, while there are slight increases in the total \(\chi^2\) values 
with the addition of new data sets, the fits remain well within acceptable ranges, 
confirming that our approach provides a consistent and comprehensive 
description across a wide range of experimental data.

We should also highlight here that the overall fit quality is improved at 
NNLO compared to NLO accuracy. For our nominal fit, \texttt{Fit~D}, at NLO accuracy, 
the total \(\chi^2\) value is found to be higher (\(\chi^2 = 1.270\)), indicating a 
less precise description of the data. In contrast, the NNLO fit provides a much 
closer agreement with the experimental measurements, as reflected in the 
lower \(\chi^2\) values for both the individual and combined datasets. 

This demonstrates the importance of incorporating higher-order QCD corrections 
to achieve a more accurate determination of the PDFs and their associated uncertainties. 
The difference between NNLO and NLO is mainly due to the fact that perturbative 
corrections are generally more significant for hadron-collider processes compared to 
HERA DIS, emphasizing the necessity of including higher-order QCD corrections, 
especially for high-precision LHC measurements. Additionally, the quantity and precision of 
the current LHC data are now sufficient to clearly demonstrate the superiority of NNLO calculations.

%=================================================  
\subsection{Parton distributions}  
%=================================================  

We now turn to the examination of our extracted PDFs. First, we present the complete set of four PDFs and 
compare them to the baseline {\tt Fit~A} results. Following this, we provide a 
detailed discussion on the impact of different data set selections on the shapes 
of various parton species and their associated uncertainties.

In Table~\ref{tab:par}, we list the numerical values of the best-fit parameters and 
their associated uncertainties, obtained from our QCD fits at the input scale \(Q_0^2 = 1.9\) GeV\(^2\). 
The parameter values are provided for all four QCD fit sets discussed in the text. 
The strong coupling constant, \(\alpha_s(M_Z)\), is treated as a free parameter in 
our QCD analyses. The value of \(\alpha_s(M_Z)\) extracted from our QCD fits is 
also presented   in Table~\ref{tab:par}, reflecting the impact of including different 
data sets on the determination of the strong coupling constant. 
Additionally, as discussed earlier, the parameter \(r_s\), defined as the 
relative strange-to-down sea quark fraction \(r_s = \frac{s + \bar{s}}{2 \bar{d}}\), 
is also treated as a free parameter in our QCD fits. The values of \(r_s\) extracted 
from the fits are presented in Table~\ref{tab:par}.  This parameter is essential for 
understanding the relative behavior of the strange and down sea quark distributions, 
and its inclusion allows for a more flexible parameterization of the PDFs.

Several observations can be made based on the numbers presented in this table. 
We parameterize six different PDFs: the valence distributions \(xu_v\) and \(xd_v\); 
the gluon density \(xg\); the sea quark densities \(x\bar{d}\) and \(x\bar{u}\); and finally, 
a symmetric strange quark distribution \(xs = x\bar{s}\). 
The parameters \(A_{u_v}\), \(A_{d_v}\), and \(A_{g}\) are determined by the number and momentum sum rules. 

As observed, nearly all parameters are well-determined from the fit to the data. 
However, some parameters, such as \(C_s\), \(C_{\bar{d}}\), and \(C_g\), exhibit relatively 
large uncertainties, particularly when only the HERA data are used. This indicates that the 
HERA combined DIS datasets alone do not sufficiently constrain these parameters. 
Notably, as shown in Table~\ref{tab:par}, the inclusion of additional data sets, such as Drell-Yan 
and W/Z production data, significantly improves the determination of these parameters, 
reducing their associated uncertainties.

%--------------------------------
\begin{table*}
%--------------------------------
\caption{\label{tab:par}
The numerical values of the best-fit parameters and their associated 
uncertainties obtained from our QCD fits at the input scale \(Q_0^2 = 1.9\) GeV\(^2\). 
The parameter values are provided for all four different QCD fit sets discussed in the text.} 
\begin{center}
%%	\scalebox{0.70}{%
\begin{tabular}{l | c c c c}
\hline 	\hline 
Parameter   ~&~ {\tt Fit A} ~&~ {\tt Fit B} ~&~ {\tt Fit C} ~&~ {\tt Fit D}        \\ 	\hline 
%      \midrule
$A_{\bar{d}}$ & $0.152 \pm 0.017$& $0.190 \pm 0.020$& $0.1542 \pm 0.0077$& $0.1672 \pm 0.0085$  \\ 
% $A_{dv}$ & $\textcolor{blue}{ 1.0000 }$& $\textcolor{blue}{ 1.0000 }$& $\textcolor{blue}{ 1.0000 }$& $\textcolor{blue}{ 1.0000 }$  \\ 
%$A_g$ & $\textcolor{blue}{ 1.0000 }$& $\textcolor{blue}{ 1.0000 }$& $\textcolor{blue}{ 1.0000 }$& $\textcolor{blue}{ 1.0000 }$  \\ 
$A_{g'}$ & $0.376 \pm 0.089$& $0.0042 \pm 0.0029$& $0.353 \pm 0.056$& $0.187 \pm 0.013$  \\ 
%$A_{uv}$ & $\textcolor{blue}{ 1.0000 }$& $\textcolor{blue}{ 1.0000 }$& $\textcolor{blue}{ 1.0000 }$& $\textcolor{blue}{ 1.0000 }$  \\ 
$B_{\bar{d}}$ & $-0.127 \pm 0.015$& $-0.095 \pm 0.018$& $-0.1167 \pm 0.0083$& $-0.1043 \pm 0.0089$  \\ 
$B_{dv}$ & $0.903 \pm 0.094$& $1.019 \pm 0.095$& $0.711 \pm 0.019$& $0.775 \pm 0.015$  \\ 
$B_g$ & $-0.48 \pm 0.18$& $1.08 \pm 0.50$& $-0.789 \pm 0.021$& $-0.854 \pm 0.016$  \\ 
$B_{g'}$ & $-0.57 \pm 0.12$& $-0.839 \pm 0.088$& $-0.817 \pm 0.017$& $-0.877 \pm 0.010$  \\ 
$B_{uv}$ & $0.752 \pm 0.034$& $0.872 \pm 0.039$& $0.736 \pm 0.011$& $0.7650 \pm 0.0085$  \\ 
$C_{\bar{d}}$ & $8.7 \pm 5.7$& $6.42 \pm 0.48$& $3.66 \pm 0.25$& $5.68 \pm 0.32$  \\ 
$C_{dv}$ & $4.53 \pm 0.43$& $4.08 \pm 0.41$& $4.239 \pm 0.085$& $4.364 \pm 0.083$  \\ 
$C_g$ & $3.5 \pm 1.2$& $3.9 \pm 1.4$& $3.8 \pm 1.0$& $0.57 \pm 0.20$  \\ 
$C_{g'}$ & 25 (Fixed) & 25 (Fixed)& 25 (Fixed)& 25 (Fixed)  \\ 
$C_{s}$ & $8.6 \pm 7.5$& $25.7 \pm 7.8$& $13.1 \pm 1.6$& $16.4 \pm 1.5$  \\ 
$C_{\bar{u}}$ & $5.04 \pm 0.82$& $10.14 \pm 0.68$& $5.83 \pm 0.29$& $8.34 \pm 0.46$  \\ 
$C_{uv}$ & $4.840 \pm 0.097$& $4.561 \pm 0.082$& $4.954 \pm 0.078$& $4.943 \pm 0.067$  \\ 
$E_{uv}$ & $11.3 \pm 1.5$& $11.3 \pm 2.2$& $9.51 \pm 0.55$& $9.86 \pm 0.59$  \\   	\hline  
$\alpha_s(M_Z)$ & $0.1170 \pm 0.0038$& $0.13005 \pm 0.001$& $0.1075 \pm 0.0030$& $0.1128 \pm 0.0014$  \\ 
$r_s$ & $0.72 \pm 0.47$& $1.06 \pm 0.41$& $1.025 \pm 0.071$& $1.069 \pm 0.053$  \\
\hline 	\hline   
\end{tabular}
%	}
\end{center}
%--------------------------------
\end{table*}
%--------------------------------

\begin{figure*}[!htb]
\vspace{0.5cm}
\begin{center}
\includegraphics[scale = 0.55]{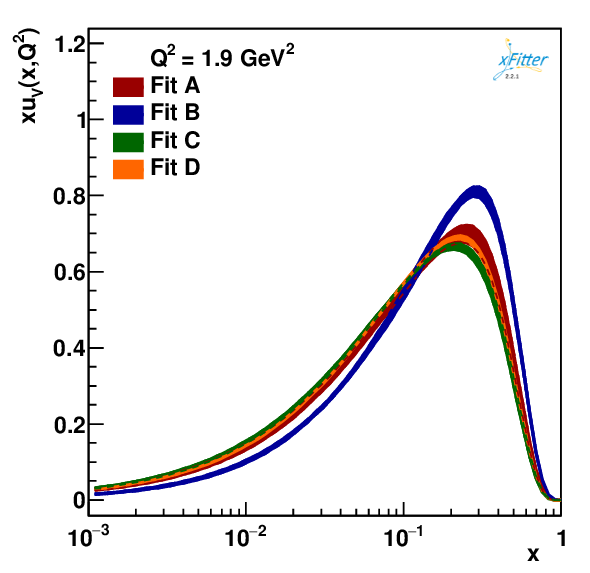}					    
\includegraphics[scale = 0.55]{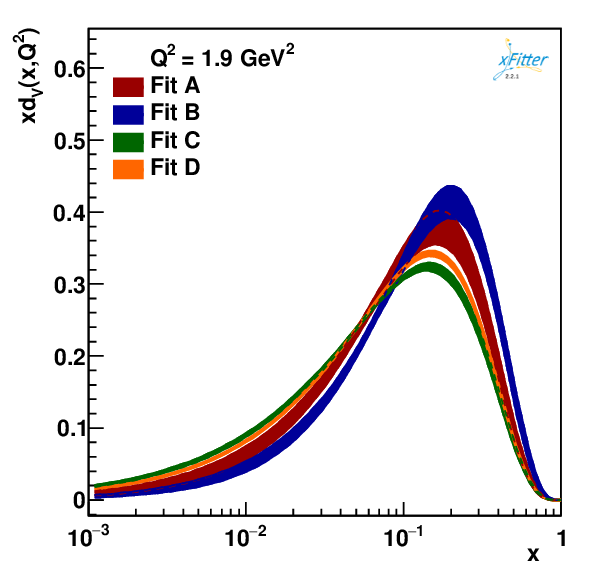}		
\includegraphics[scale = 0.55]{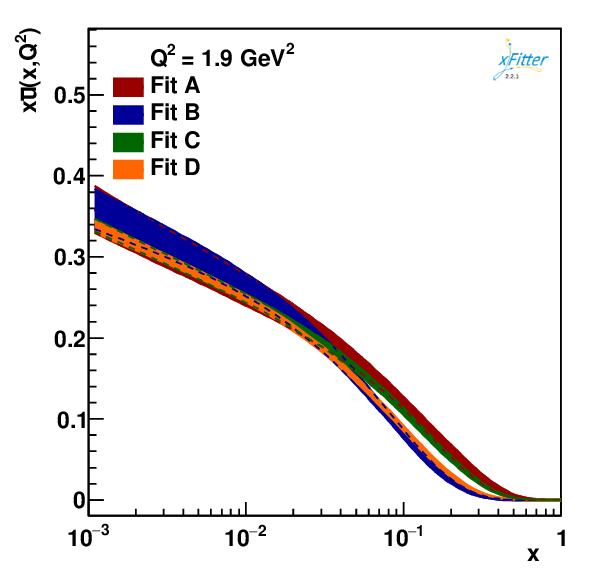}		    	
\includegraphics[scale = 0.55]{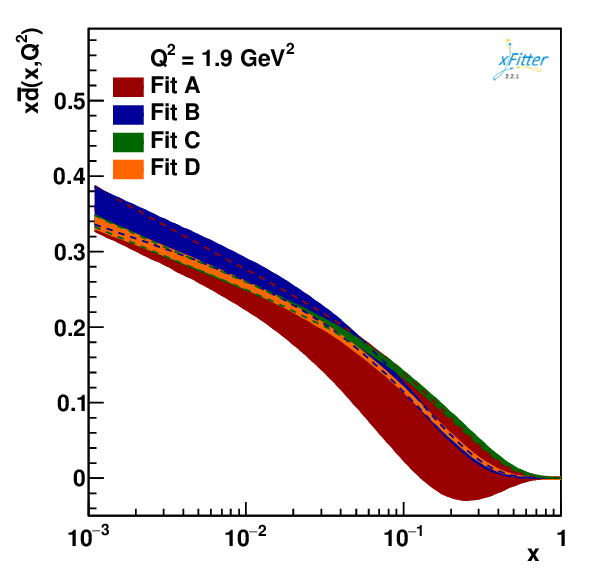}		
\includegraphics[scale = 0.55]{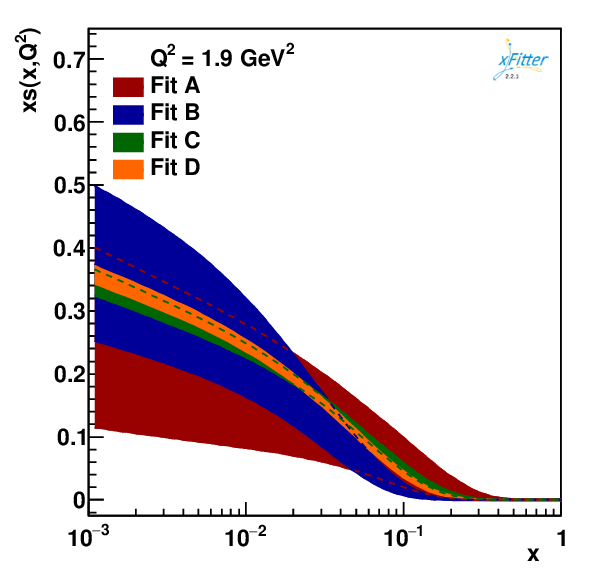}
\includegraphics[scale = 0.55]{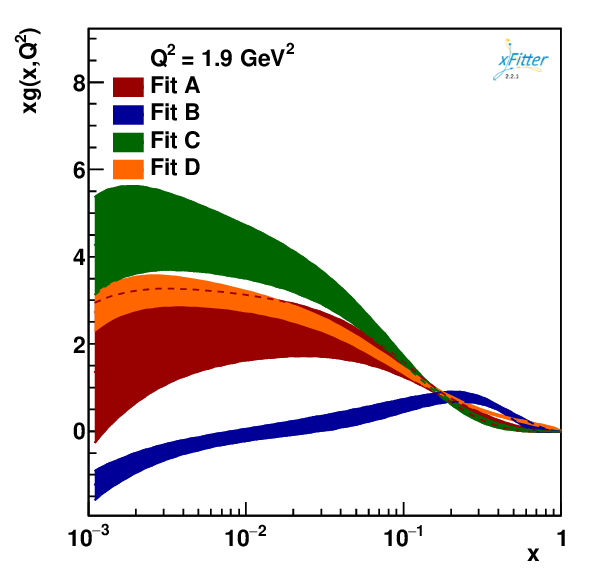}      
\caption{The parton distribution of $xu_v$, $xd_v$, $x\bar{u}$, $x\bar{d}$, $xs$ 
and $xg$ for Fit~A, Fit~B, Fit~C and Fit~D as a function of $x$ and at the input scale $Q_0^2=$1.9 GeV$^2$.}
\label{fig:PDF-1.9}
\end{center}
\end{figure*}

\begin{figure*}[!htb]
\vspace{0.5cm}
\begin{center}
\includegraphics[scale = 0.55]{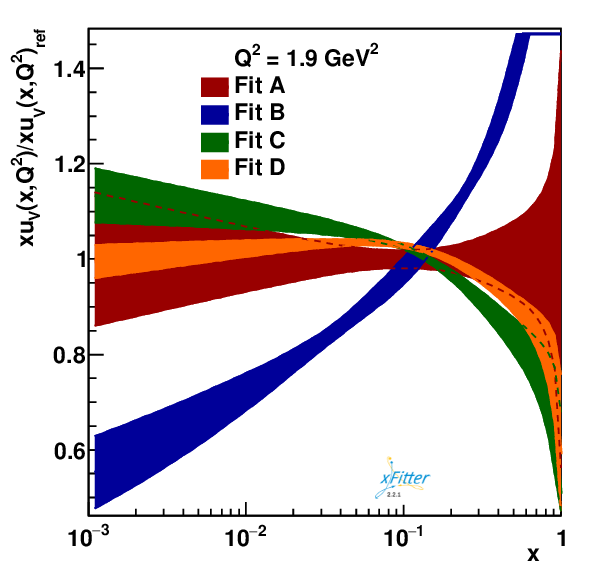}
\includegraphics[scale = 0.55]{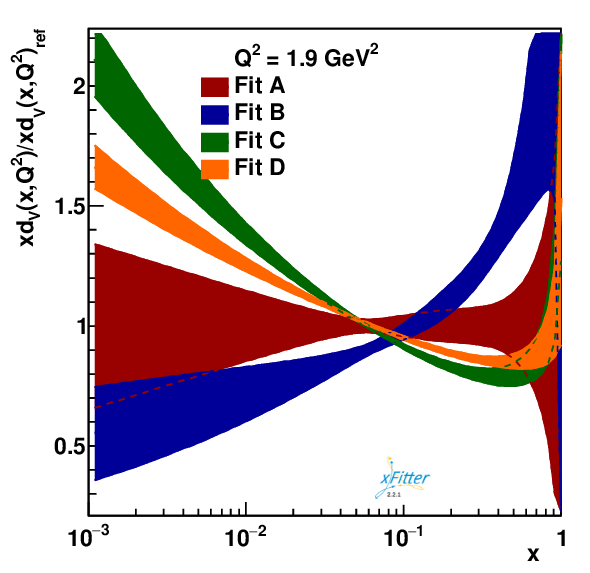}
\includegraphics[scale = 0.55]{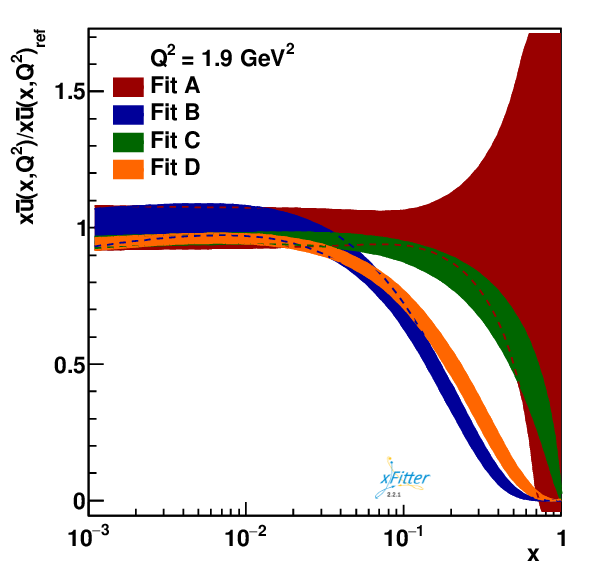}		    	
\includegraphics[scale = 0.55]{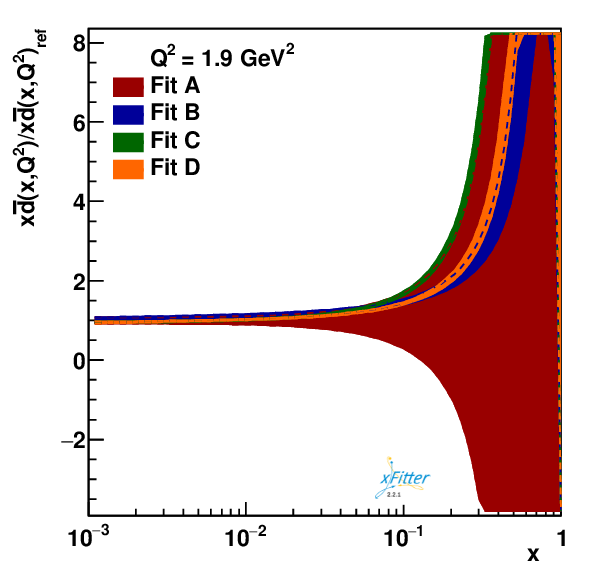}
\includegraphics[scale = 0.55]{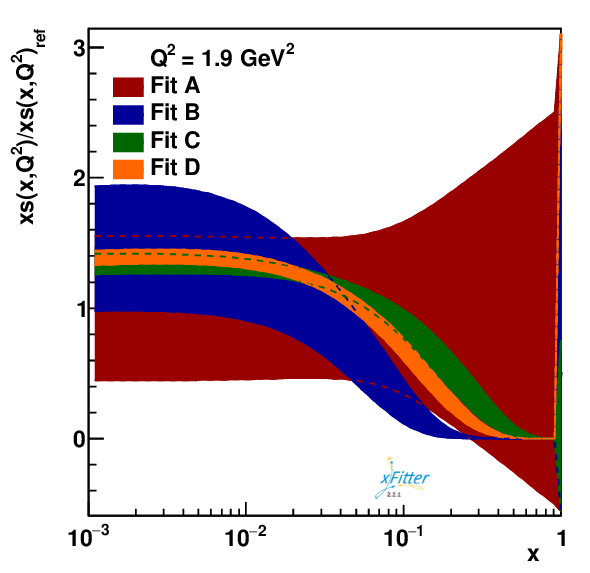}
\includegraphics[scale = 0.55]{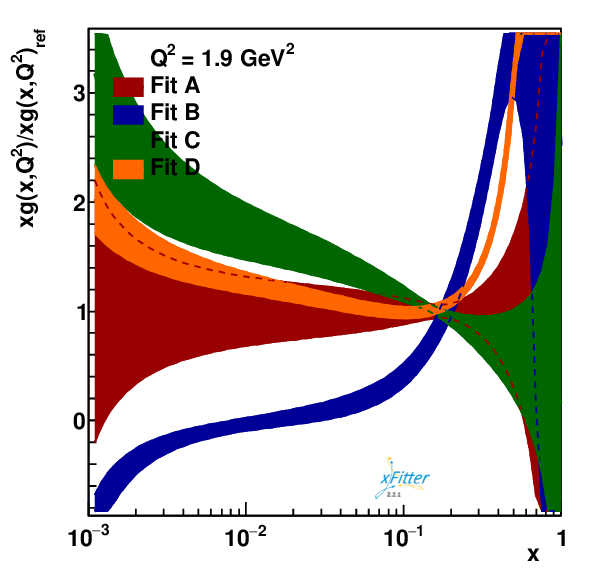}     
\caption{The ratios of parton distribution $(xf(x,Q^2)/xf(x,Q^2)_{ref})$ for 
$xu_v$, $xd_v$, $x\bar{u}$, $x\bar{d}$, $xs$ and $xg$ as a function of $x$ and 
at the input scale $Q_0^2=$1.9 GeV$^2$. 
The comparisons are displayed as ratios $xf(x,Q^2)/xf(x,Q^2)_{\rm {ref}}$ relative to our baseline QCD analysis, \texttt{Fit~A}}.
\label{fig:ref-1p9}
\end{center}
\end{figure*}

In the following discussion, we turn our attention to the obtained PDFs and their uncertainties. 
Detailed comparisons of the NNLO quark and gluon PDFs at the input scale \(Q_0^2 = 1.9\) GeV\(^2\) 
are presented in Figs.~\ref{fig:PDF-1.9} and \ref{fig:ref-1p9}. To investigate the impact of 
including different data selections, we compare all four PDF determinations with each other. 
Specifically, in Fig.~\ref{fig:ref-1p9}, the comparisons are displayed as ratios \(xf(x,Q^2)/xf(x,Q^2)_{\rm {ref}}\) 
relative to our baseline QCD analysis, \texttt{Fit~A}. 
The uncertainty bands of the PDFs are calculated using the Hessian method, as discussed earlier.

Notable differences in the shapes of the PDFs are observed, especially at small values of \(x\). 
These variations indicate that different data sets impose different types of constraints on the PDFs. 
For example, the inclusion of Drell-Yan and W/Z boson production data tends to 
impact the strange quark and gluon distributions more prominently at lower \(x\), 
demonstrating the sensitivity of the PDF shapes to the specific experimental inputs.

As can be seen, the inclusion of the Drell-Yan data in \texttt{Fit~B} significantly 
affects the shape of the valence quark distributions, particularly over medium to 
large values of \(x\). These differences are even more pronounced for the strange 
quark and gluon distributions, where notable variations are observed in both the 
shape and the uncertainty bands across the entire range of \(x\).

Focusing first on the strange quark PDF, it is evident that there are relatively 
poor constraints in \texttt{Fit~A}, which relies solely on the HERA data. Even with 
the addition of Drell-Yan data sets in \texttt{Fit~B}, the constraints on the strange 
quark remain limited, as reflected by the large associated uncertainty bands. 
However, the inclusion of LHC and Tevatron W/Z boson production cross-section 
data in \texttt{Fit~C} leads to much better constraints and significantly reduced 
error bands for the strange quark PDF across all \(x\) values.

For the gluon PDFs, the differences are even more pronounced. The inclusion of additional 
data sets, particularly those from W/Z boson production, affects the central values; 
however, it does not lead to a significant reduction in the uncertainty bands. This highlights the 
critical role of jet production data sets from collider DIS and hadron colliders 
in providing stronger constraints on the gluon distribution across both small and large values of \(x\).

In our QCD fits discussed above, we chose not to include jet production data sets, as 
the main motivation of this work is to specifically examine the constraints on individual 
PDFs from HERA DIS, Drell-Yan, and W/Z boson production data sets. This approach allows 
us to isolate and better understand the individual contributions and the impact of these 
particular data sets on the determination of PDFs. The inclusion of jet and dijet production 
data in our QCD fits will be discussed in Sec.~\ref{jet}, where we will explore their 
effects on the extracted PDFs and the strong coupling constant.

%--------------------------------
%--------------------------------

%--------------------------------
%--------------------------------

\begin{figure*}[!htb]
\vspace{0.5cm}
\begin{center}
\includegraphics[scale = 0.55]{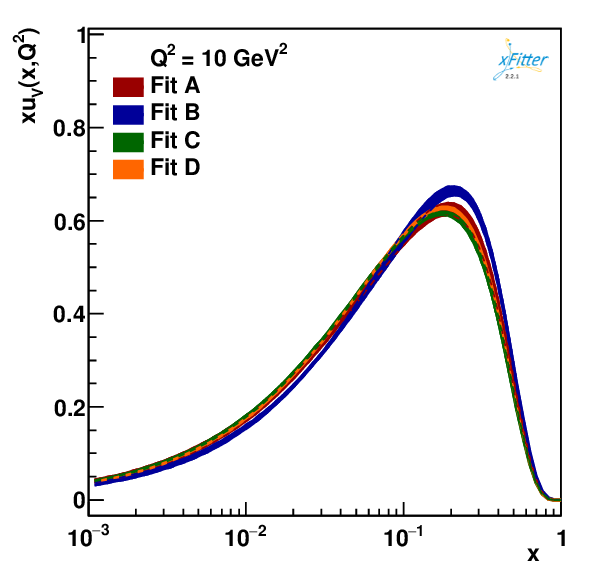}					    
\includegraphics[scale = 0.55]{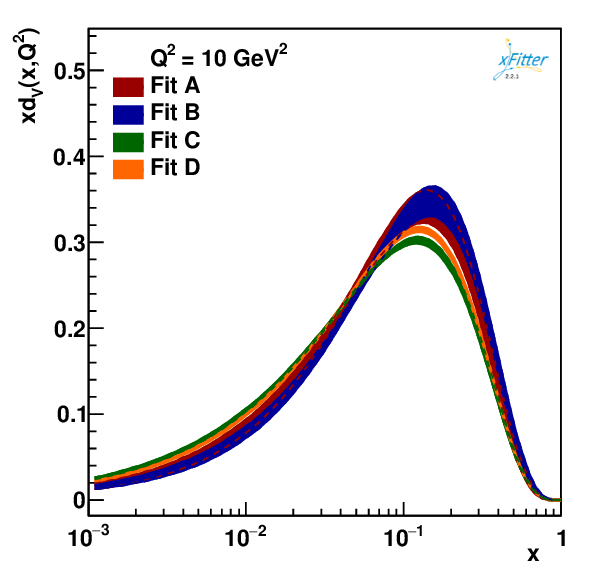}		
\includegraphics[scale = 0.55]{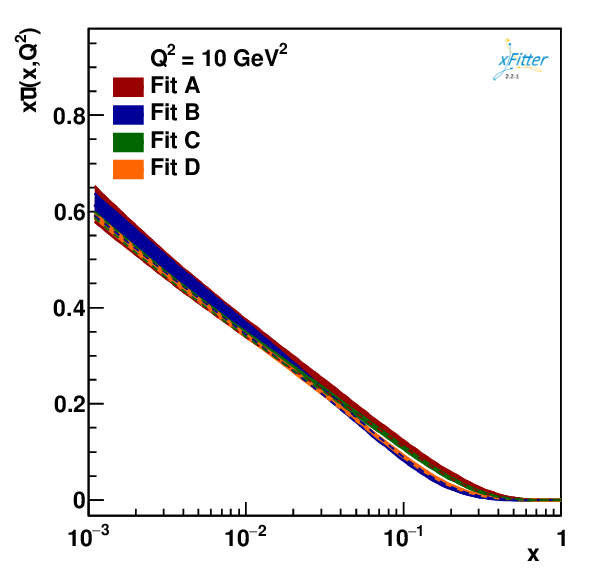}		    	
\includegraphics[scale = 0.55]{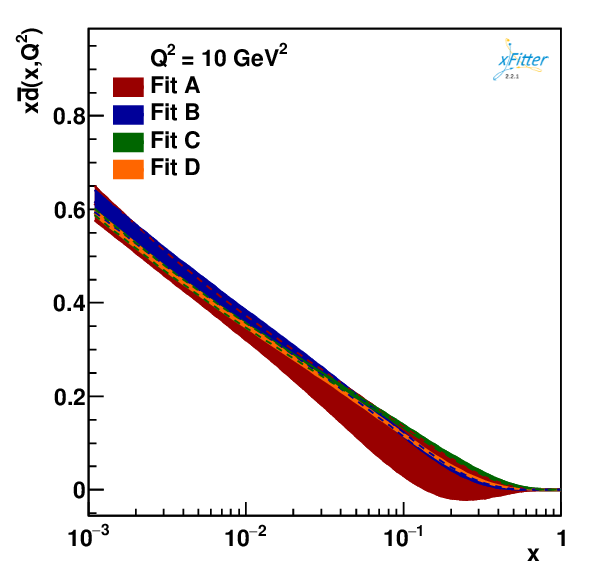}		
\includegraphics[scale = 0.55]{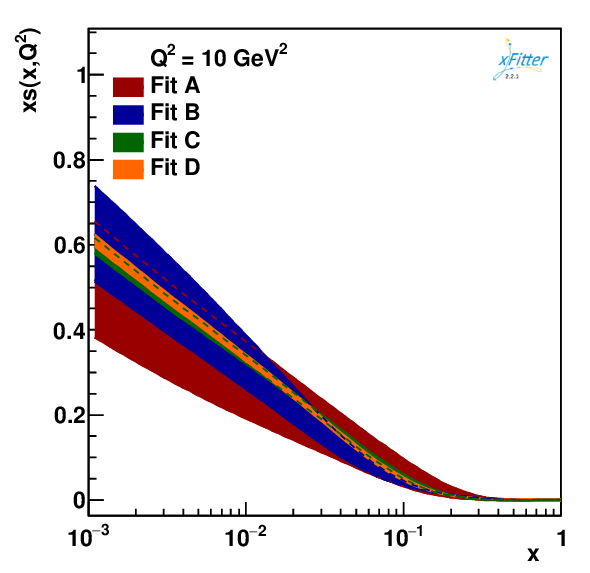}
\includegraphics[scale = 0.55]{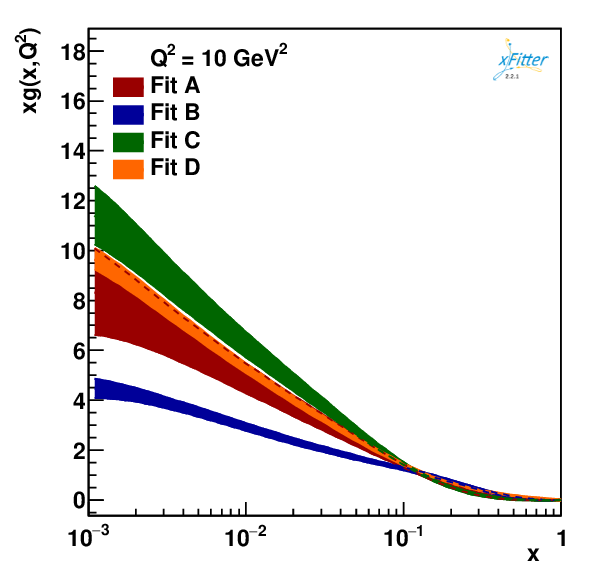}      
\caption{Same as Fig.~\ref{fig:PDF-1.9} but this time for $Q^2=$10~GeV$^2$.}
\label{fig:PDF-10}
\end{center}
\end{figure*}

\begin{figure*}[!htb]
\vspace{0.5cm}
\begin{center}
\includegraphics[scale = 0.55]{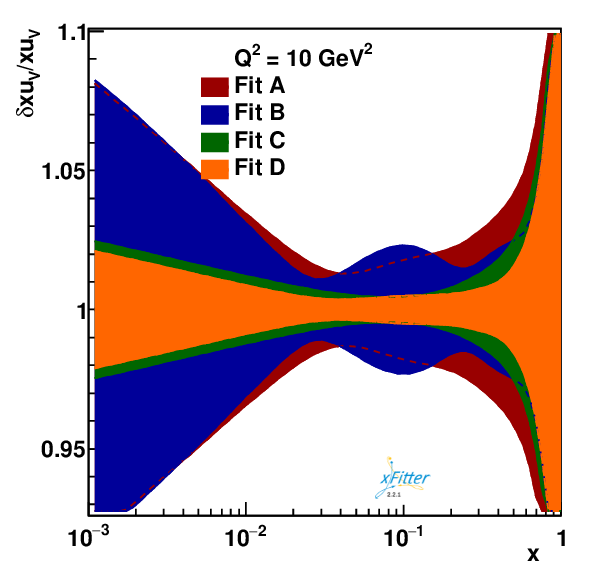}
\includegraphics[scale = 0.55]{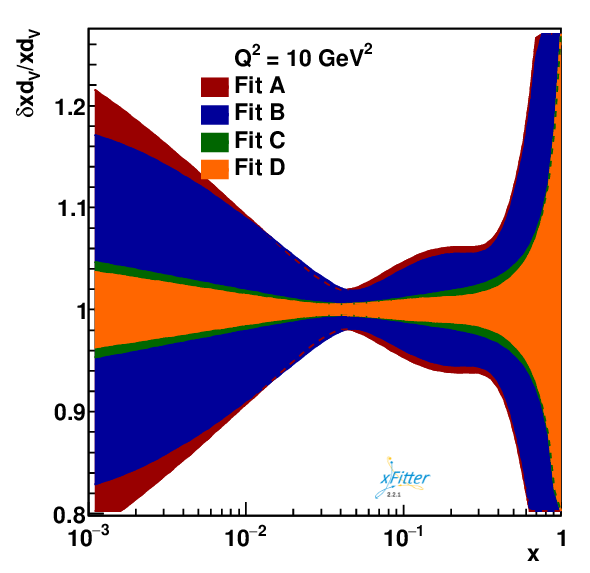} 
\includegraphics[scale = 0.55]{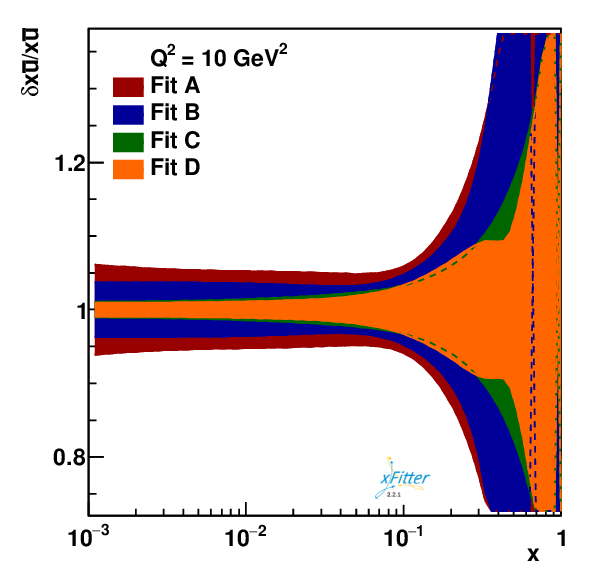}		    	
\includegraphics[scale = 0.55]{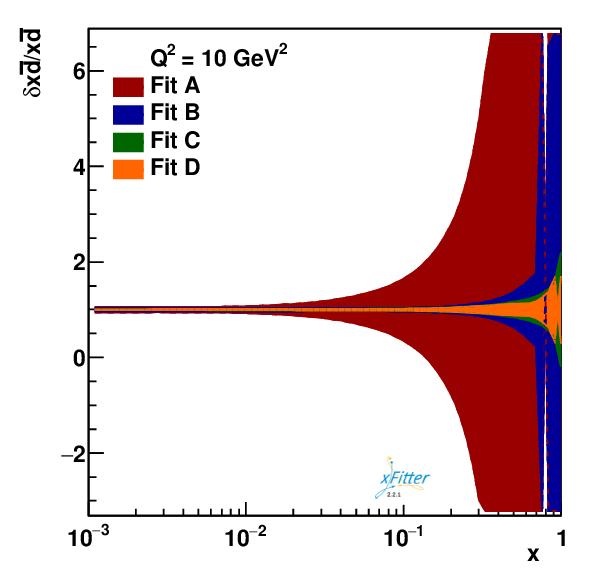}
\includegraphics[scale = 0.55]{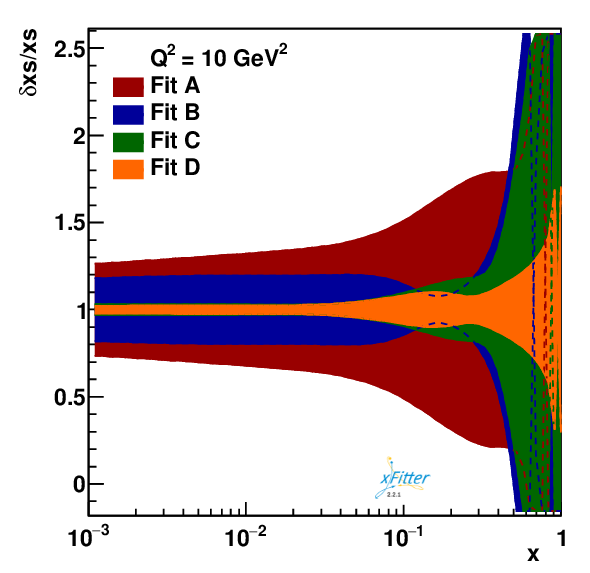} 
\includegraphics[scale = 0.55]{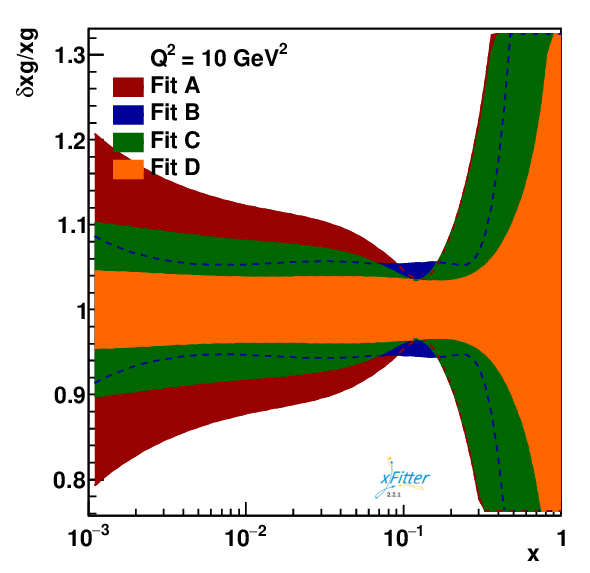}     
\caption{Same as Fig.~\ref{fig:ref-1p9} but this time for $Q^2=$10~GeV$^2$. 
The relative uncertainties $\delta xq(x,Q^2)/xq(x,Q^2)$  are also shown.}
\label{fig:ref-10}
\end{center}
\end{figure*}

\begin{figure*}[!htb]
\vspace{0.5cm}
\begin{center}
\includegraphics[scale = 0.55]{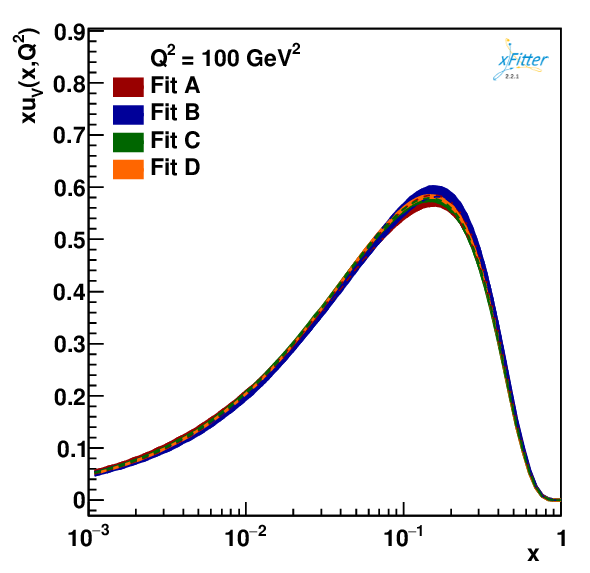}					    
\includegraphics[scale = 0.55]{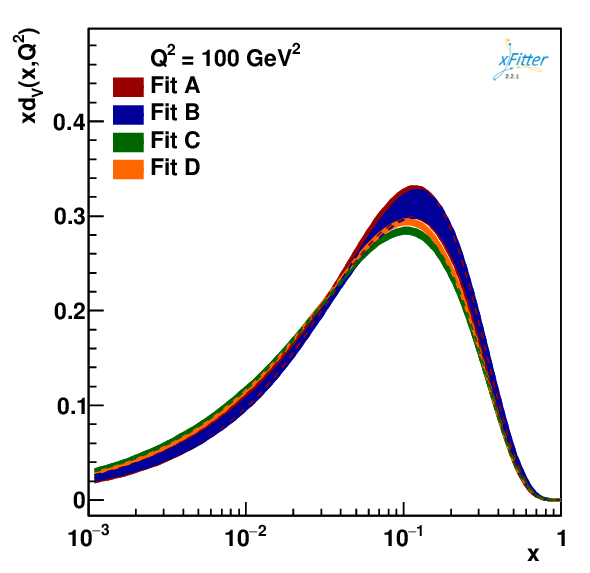}		
\includegraphics[scale = 0.55]{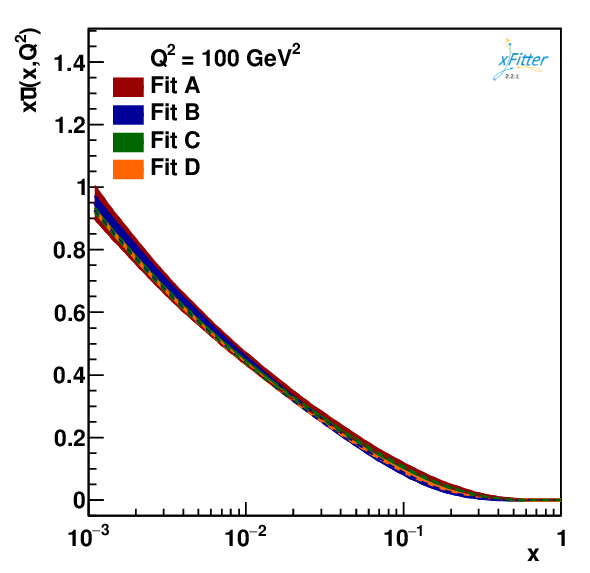}		    	
\includegraphics[scale = 0.55]{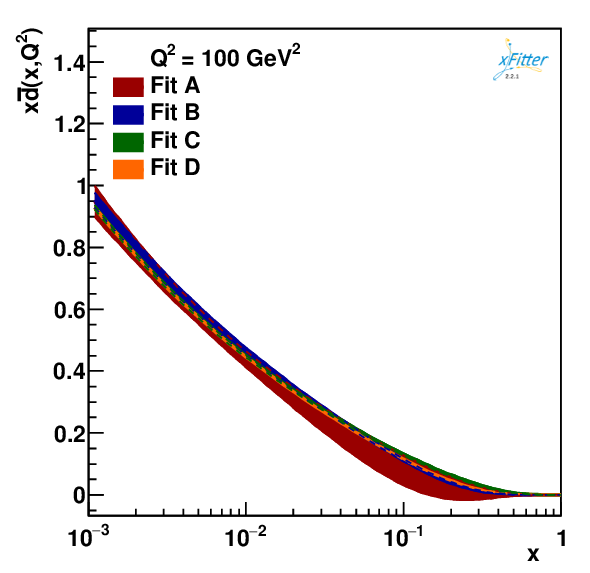}		
\includegraphics[scale = 0.55]{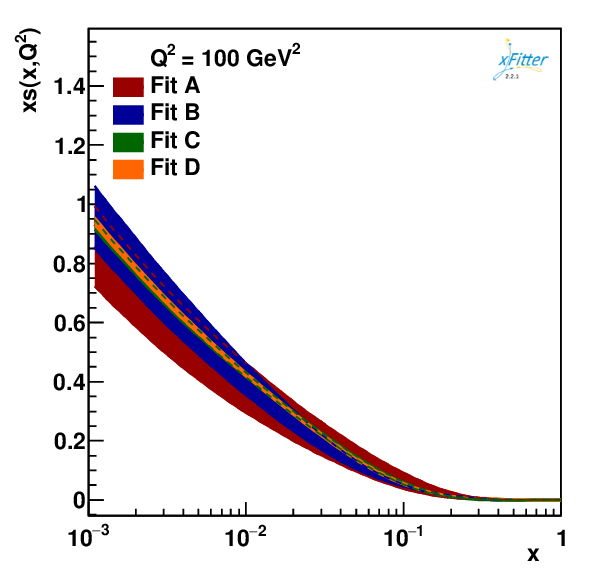}
\includegraphics[scale = 0.55]{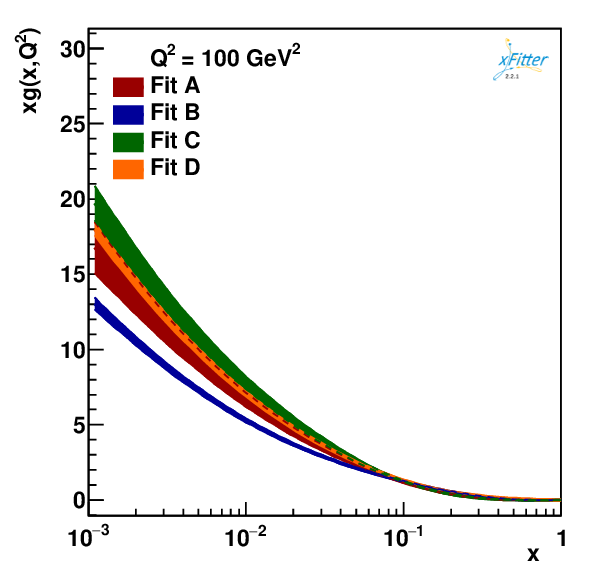}      
\caption{Same as Fig.~\ref{fig:PDF-1.9} but this time for $Q^2=$100~GeV$^2$.}
\label{fig:PDF-100}
\end{center}
\end{figure*}

\begin{figure*}[!htb]
\vspace{0.5cm}
\begin{center}
\includegraphics[scale = 0.55]{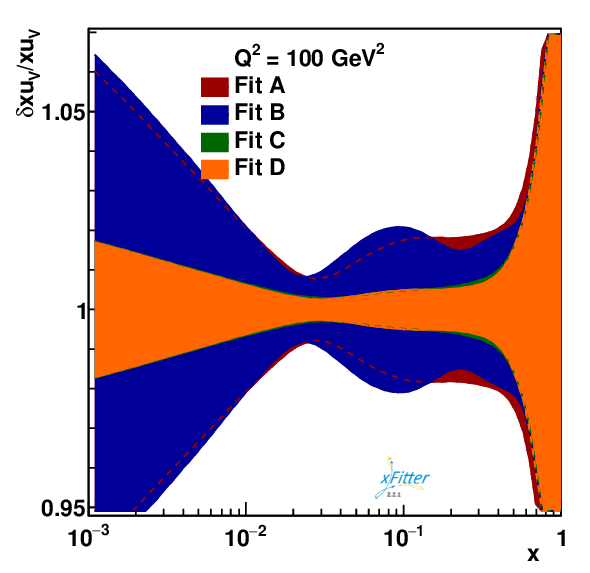}
\includegraphics[scale = 0.55]{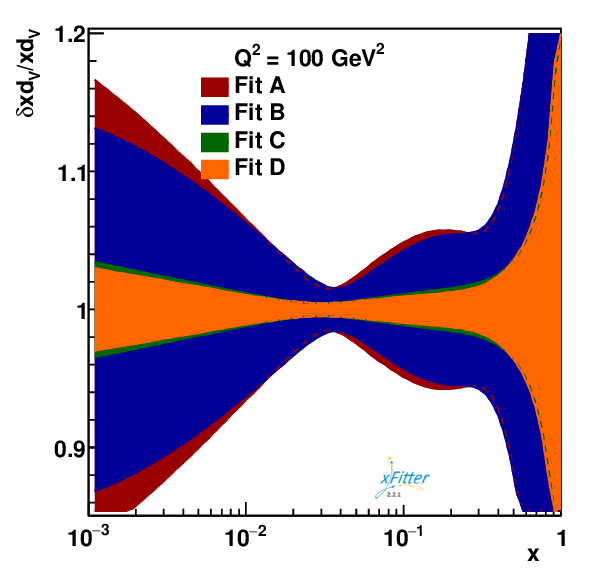} 
\includegraphics[scale = 0.55]{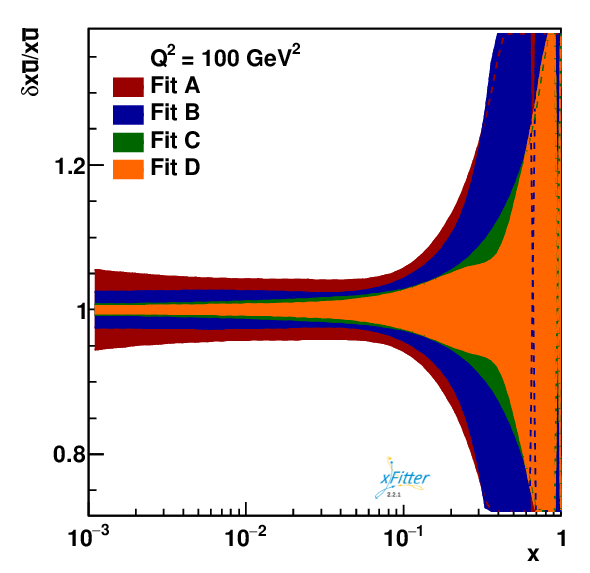}		    	
\includegraphics[scale = 0.55]{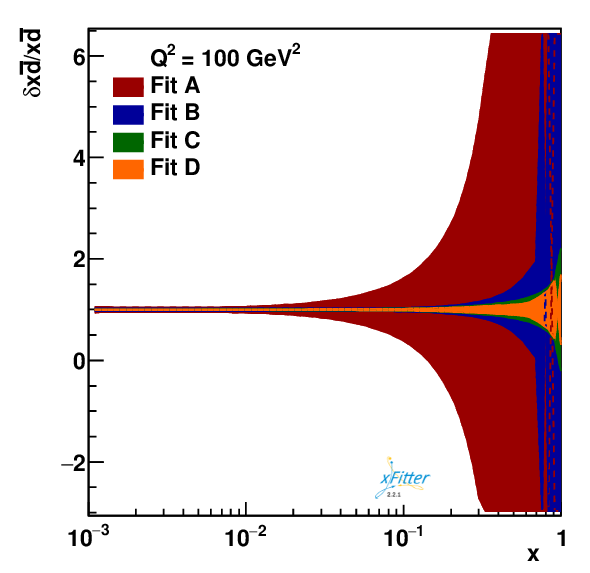}
\includegraphics[scale = 0.55]{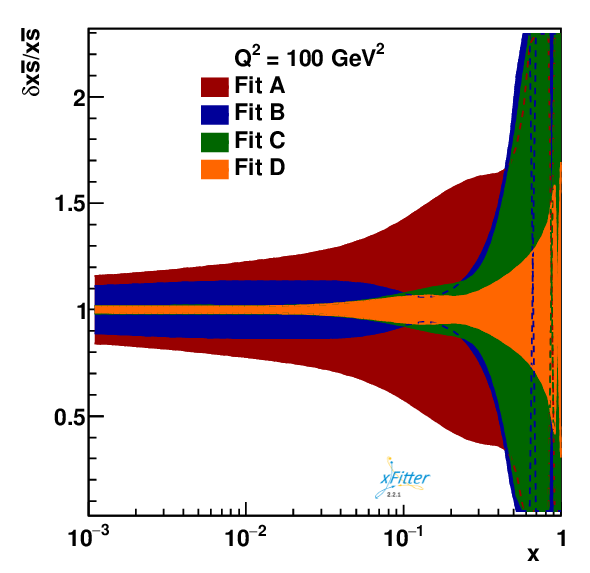} 
\includegraphics[scale = 0.55]{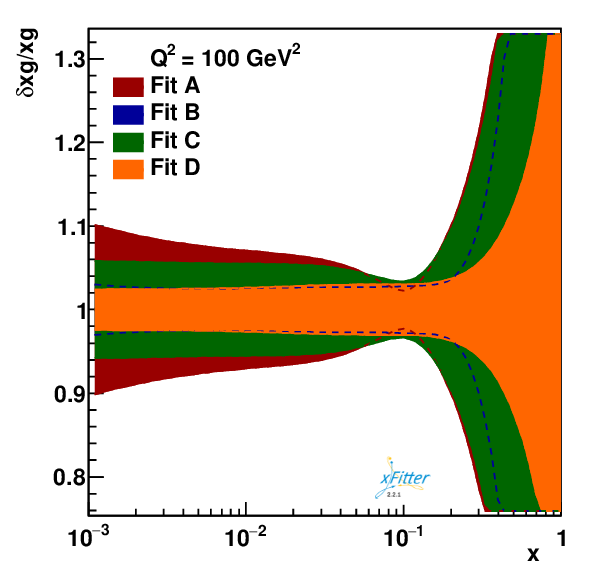}    
\caption{Same as Fig.~\ref{fig:ref-1p9} but this time for $Q^2=$100~GeV$^2$.
The relative uncertainties $\delta xq(x,Q^2)/xq(x,Q^2)$  are also shown.}
\label{fig:ref-100}
\end{center}
\end{figure*}

Detailed comparisons of the NNLO quark and gluon PDFs are presented in Figs.~\ref{fig:PDF-10}, 
\ref{fig:ref-10}, \ref{fig:PDF-100}, and \ref{fig:ref-100}, this time 
at higher values of \(Q^2 = 10\) and 100~GeV\(^2\), respectively. The general observations 
discussed earlier remain valid for the PDFs at these higher scales as well. 
However, it is important to highlight a key finding apparent in these plots 
regarding the combined inclusion of all data sets. As shown, incorporating 
all the data sets together results in more significant constraints on the PDFs, 
which in turn leads to noticeably smaller uncertainty bands for all parton species, 
as illustrated in the ratio plots of Figs.~\ref{fig:ref-10} and \ref{fig:ref-100}.

%--------------------------------
%--------------------------------

%--------------------------------
%--------------------------------

%--------------------------------
%--------------------------------

%--------------------------------
%--------------------------------

%=================================================  
\subsection{Impact of higher-order QCD corrections}\label{NLO_NNLO} 
%=================================================  

In this section, we discuss the effects of higher-order QCD corrections on the extracted PDFs and their 
associated uncertainties. As presented in Table~\ref{tab:Chi2}, the overall fit quality remains 
excellent for both individual and global data sets for NNLO accuracy, 
indicating that the NNLO corrections result in a good description of the 
experimental data across a wide range of observables.

The inclusion of NNLO corrections is vital for achieving high-precision results, 
especially for processes such as Drell-Yan production, W/Z boson production, and 
jet production, which are sensitive to higher-order effects. The CT18, MSHT20, 
and NNPDF4.0 analyses also emphasize the importance of NNLO corrections in reducing 
theoretical uncertainties and improving the overall agreement between theory and 
experimental data. For example, the MSHT20 analysis shows improvements in the 
description of LHC data, particularly for gluon-sensitive processes, while NNPDF4.0 
utilizes advanced machine learning techniques to incorporate NNLO 
corrections and reduce biases in PDF parameterizations.

\begin{figure*}[!htb]
%\vspace{0.5cm}
\begin{center}
\includegraphics[scale = 0.48]{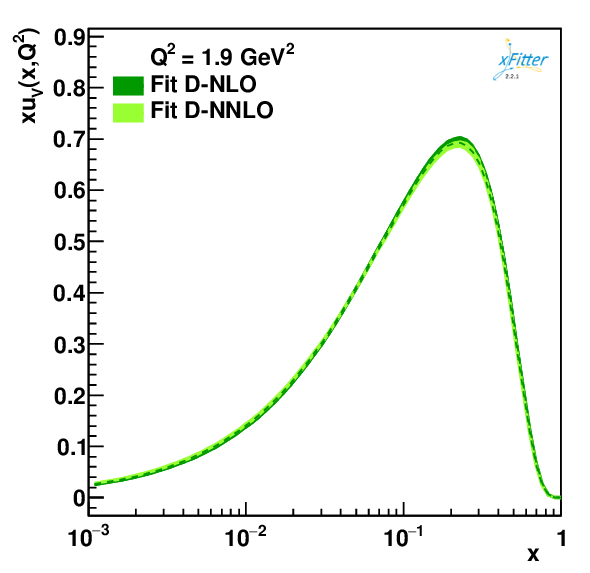}	
\includegraphics[scale = 0.48]{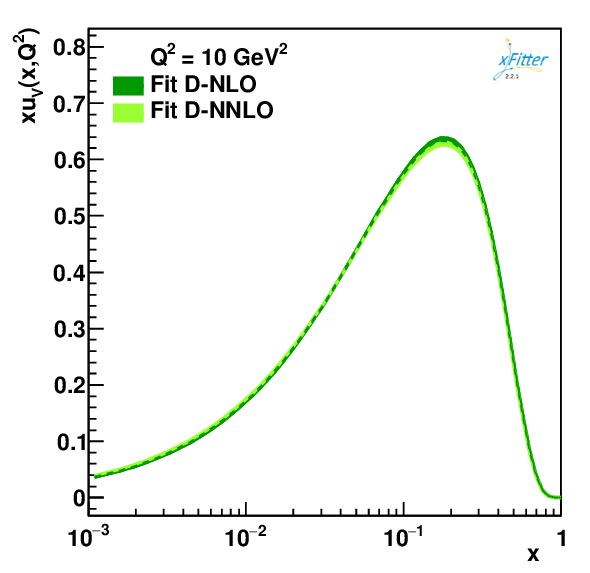}        
\includegraphics[scale = 0.48]{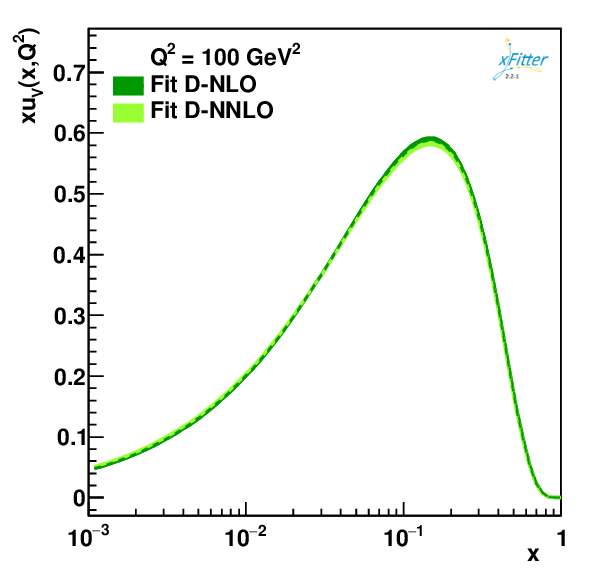}  \\  
\includegraphics[scale = 0.48]{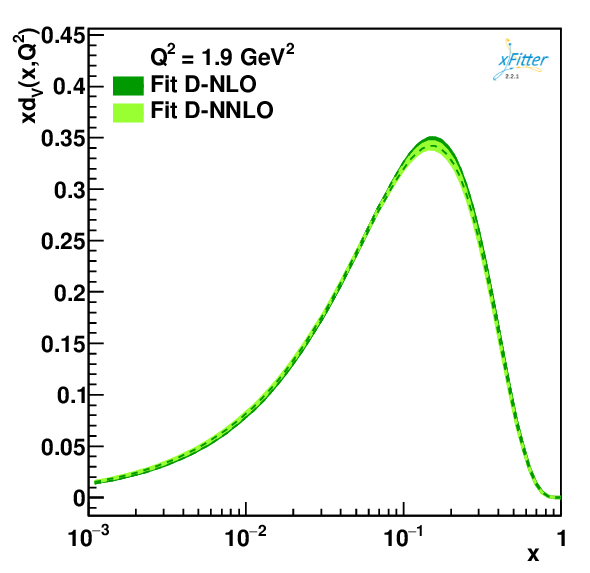}	
\includegraphics[scale = 0.48]{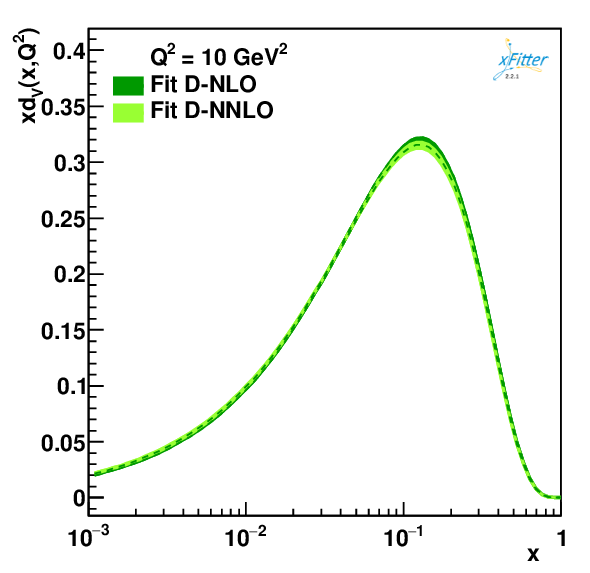}        
\includegraphics[scale = 0.48]{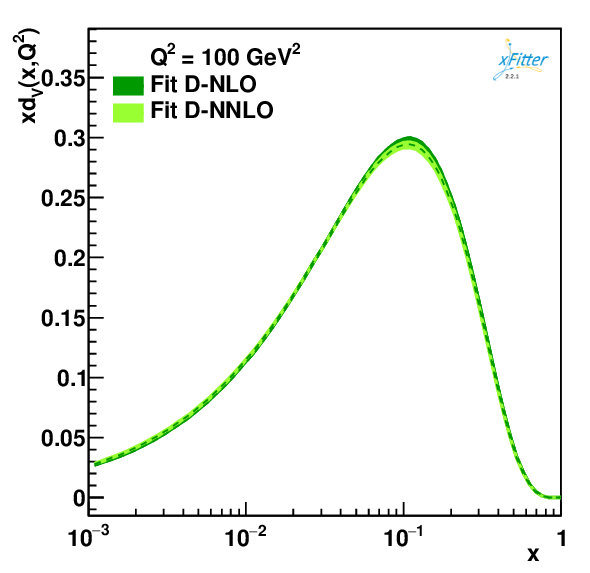}  \\
\includegraphics[scale = 0.48]{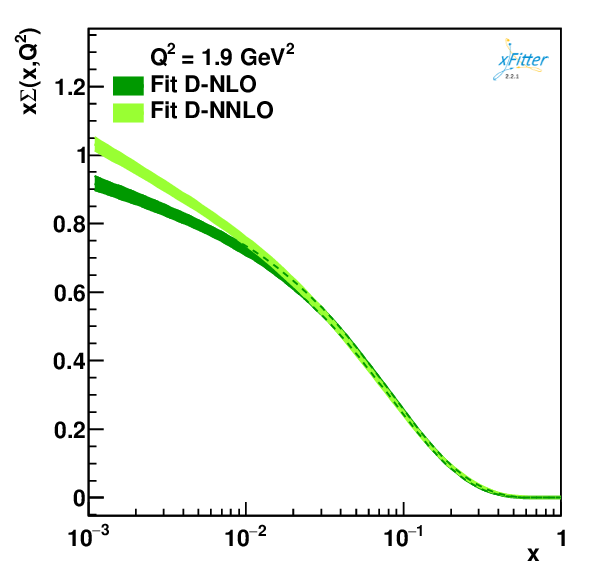}	
\includegraphics[scale = 0.48]{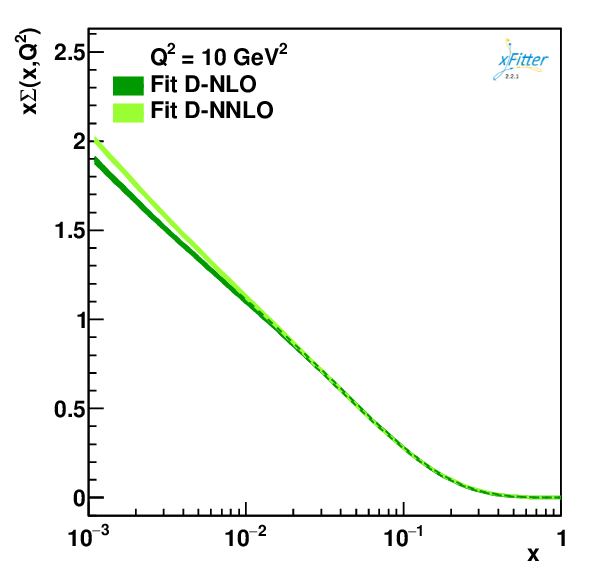}        
\includegraphics[scale = 0.48]{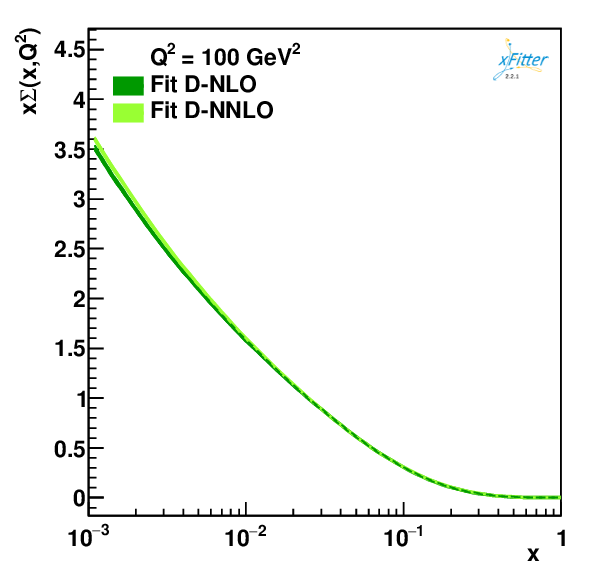}  \\
\includegraphics[scale = 0.48]{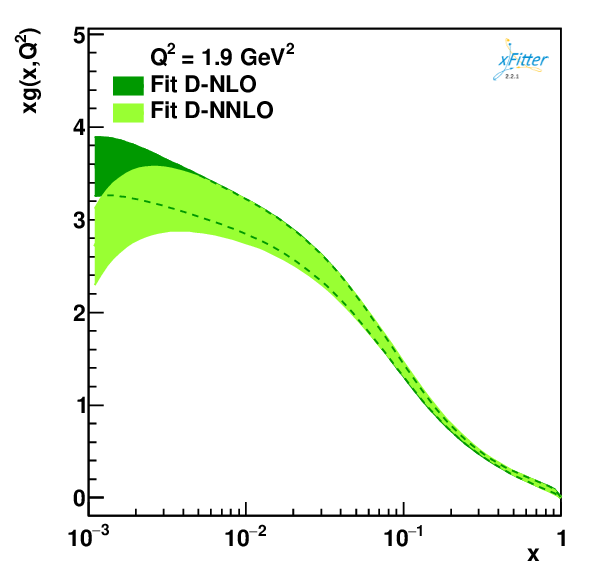}	
\includegraphics[scale = 0.48]{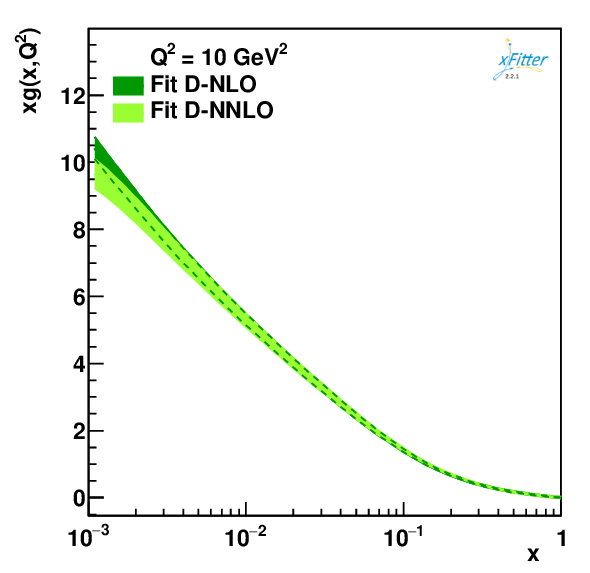}        
\includegraphics[scale = 0.48]{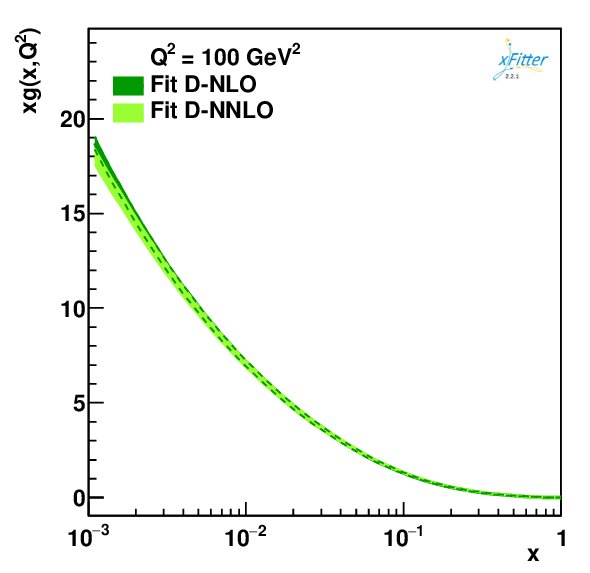}  \\
\caption{The parton distribution of $xu_v$, $xd_v$, $x\Sigma$,  
and $xg$ for our NLO and NNLO QCD fits as a 
function of $x$ and at $Q^2=$1.9, 10 and 100~GeV$^2$.} 
\label{fig:NLO-NNLO}
\end{center}
\end{figure*}

In Fig.~\ref{fig:NLO-NNLO}, we show the PDFs at both NNLO and NLO as 
functions of \(x\) for three different \(Q^2\) values: 1.9, 10, and 100~GeV\(^2\). 
The uncertainty bands, calculated using the Hessian method, 
illustrate the effect of higher-order corrections on the PDFs. The impact of 
NNLO QCD corrections on the central values of the PDFs is generally modest relative 
to the uncertainties; however, differences can be observed in specific parton 
distributions. For instance, the total singlet distribution is larger at NNLO than at NLO, 
particularly in the small to medium \(x\) range. The gluon PDF shows a slight reduction at NNLO 
in the small to medium \(x\) region, reflecting the sensitivity of 
gluon-dominated processes to higher-order effects.

%---------------Log-------- NLO-NNLO-------------------------
%--------------------------------
%--------------------------------

\begin{figure*}[!htb]  
%\vspace{0.5cm}
\begin{center}
\includegraphics[scale = 0.48]{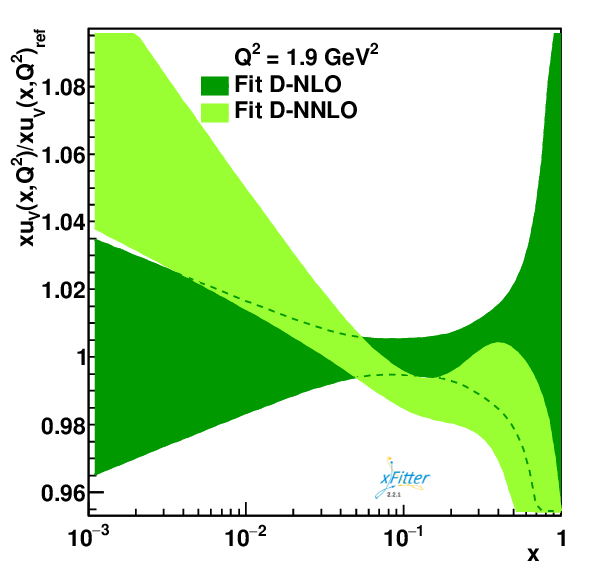}	
\includegraphics[scale = 0.48]{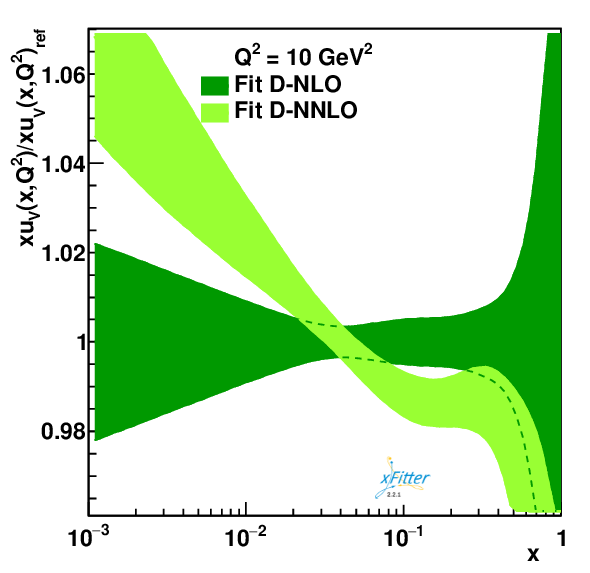}        
\includegraphics[scale = 0.48]{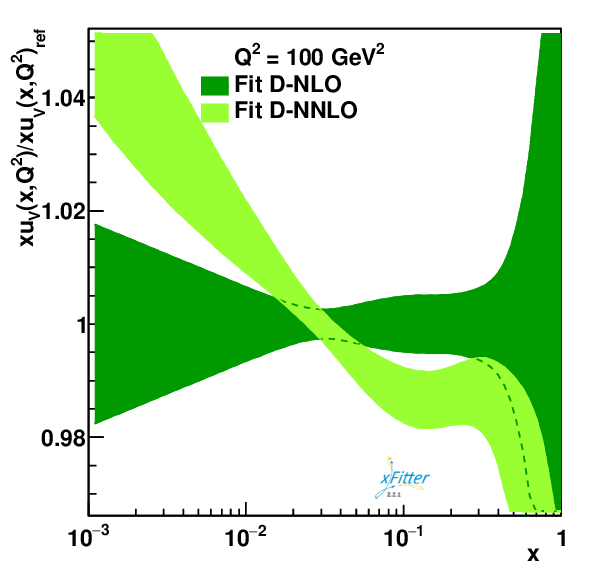}   \\  
\includegraphics[scale = 0.48]{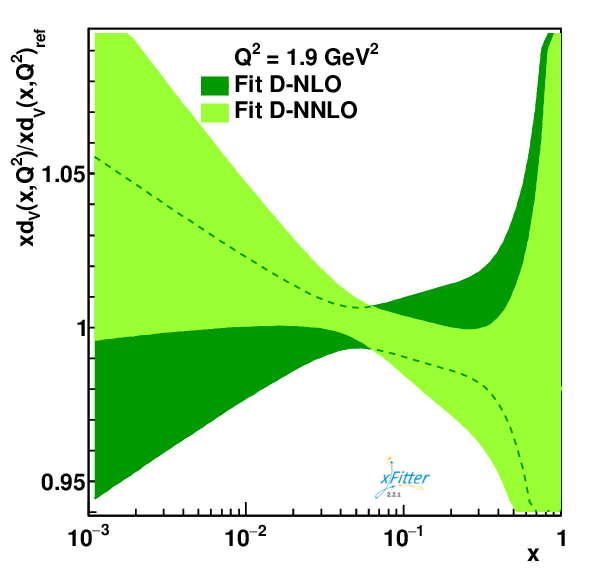}	
\includegraphics[scale = 0.48]{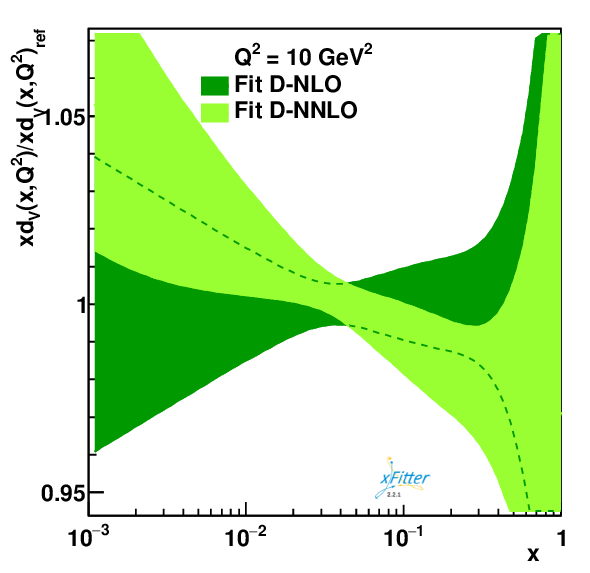}        
\includegraphics[scale = 0.48]{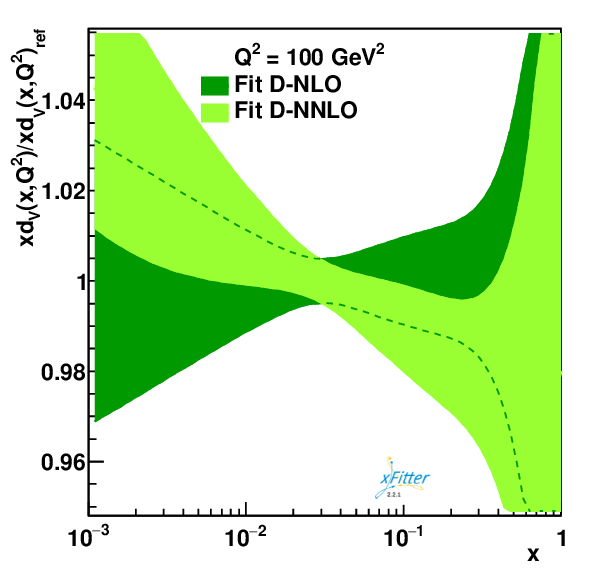}   \\
\includegraphics[scale = 0.48]{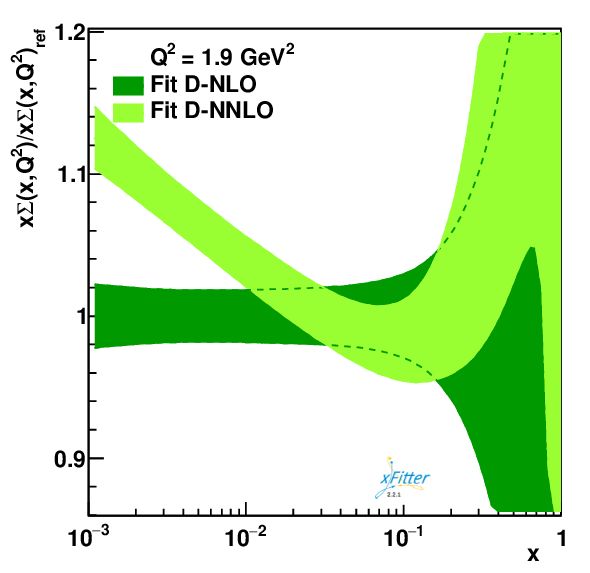}	
\includegraphics[scale = 0.48]{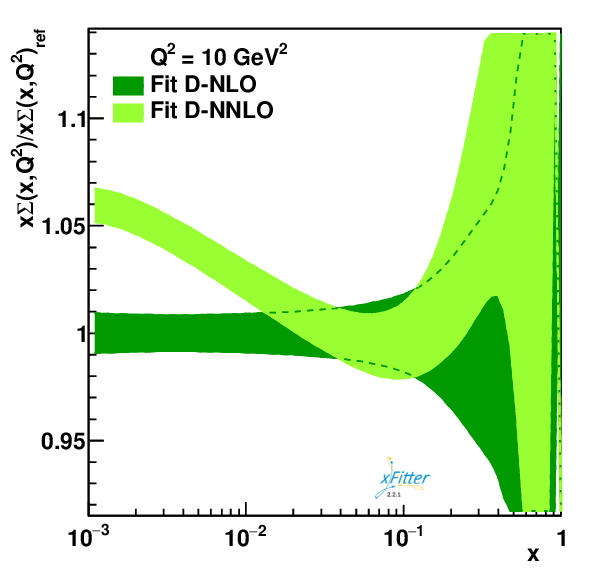}        
\includegraphics[scale = 0.48]{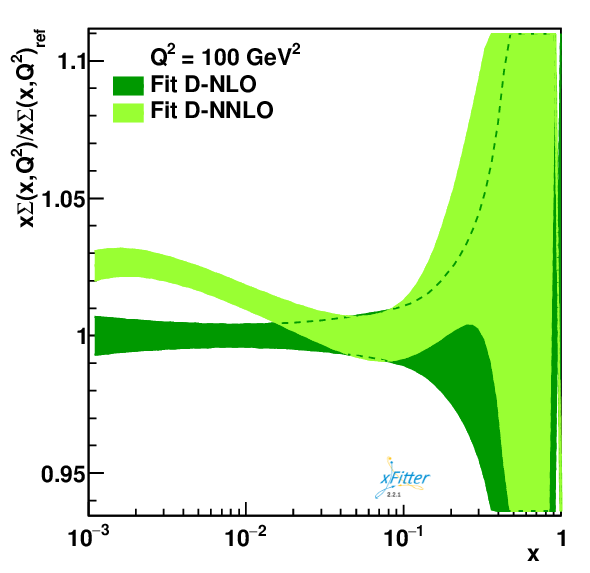}   \\
\includegraphics[scale = 0.48]{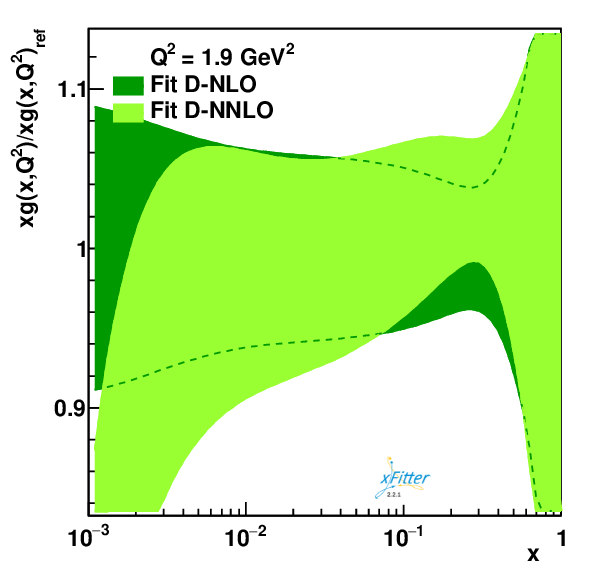}	
\includegraphics[scale = 0.48]{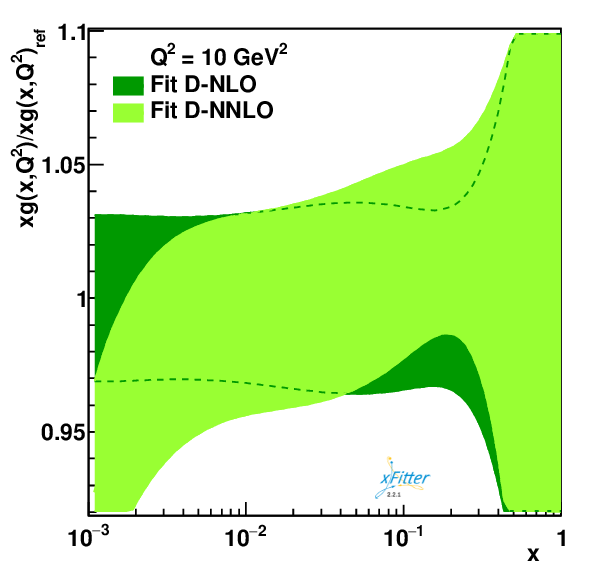}        
\includegraphics[scale = 0.48]{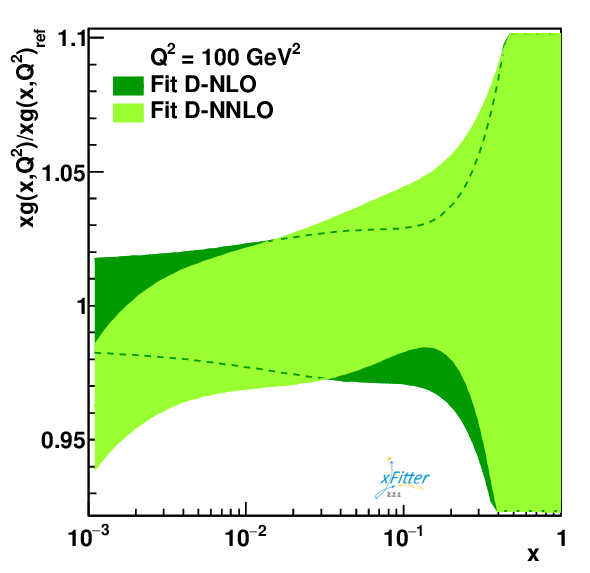}   \\  
\caption{The ratio of parton distribution $xf(x,Q^2)/xf(x,Q^2)_{\rm {ref}}$  
as a function of $x$ and at $Q^2=$1.9, 10 and 100~GeV$^2$.}
\label{fig:NLO-NNLO-ratio}
\end{center}
\end{figure*}

Additionally, as Fig.~\ref{fig:NLO-NNLO-ratio} demonstrates, the changes in the central 
values of the PDFs due to NNLO corrections are most pronounced at lower values of \(x\), 
where higher-order effects are more substantial. Both the total singlet and gluon 
distributions exhibit the largest deviations from their NLO counterparts.

%---------------Log-------- NLO-NNLO-ratio------------------------
%--------------------------------
%--------------------------------   

While the central values of the PDFs shift slightly due to NNLO corrections, the 
associated uncertainties remain largely unaffected. This stability is evident for all 
parton species, as the uncertainty bands at NLO and NNLO are of comparable size. 
The \(\chi^2\) values in Table~\ref{tab:nlo-nnlo} confirm that the NNLO fits 
provide only a modest improvement in the overall uncertainty reduction. This suggests 
that the inclusion of NNLO corrections does not drastically reduce uncertainties 
but primarily leads to shifts in the central values, particularly for the singlet and gluon PDFs.

Table~\ref{tab:nlo-nnlo} also compares the \(\chi^2\) values obtained 
from the NLO and NNLO fits for the different data sets analyzed in this work. 
As shown, the total \(\chi^2\) is slightly lower at NNLO, with a 
total \(\chi^2/\text{dof}\) of 1678/1339 compared to 1699/1339 at NLO. 
This improvement is consistent with the results from the CT18 and MSHT20 studies, 
where NNLO corrections lead to better overall fit quality and improved agreement with experimental data.

%--------------------------------
\begin{table}[h]
\begin{ruledtabular}
\begin{center}	
\caption{The values of $\chi^2$ obtained from our NLO and NNLO fits for the 
different data sets analyzed in this work. } 
\begin{tabular}{c|r|c}
Experiment & NLO ({\tt Fit D}) & NNLO ({\tt Fit D})   \\       \hline         \hline	
%            \multicolumn{4}{l}{{\bf CMS Experiment}}\\
 HERA I+II & 1155 / 1016& 1151/ 1016  \\
 %\hline
  CMS $W /Z$ & 106.4 / 90 & 106.8 / 90  \\  
 %\hline
  ATLAS $W /Z$ & 108.6 / 91 & 104.8 / 91  \\  
%\hline
  ATLAS Drell-Yan & 26.3 / 27 & 25.3 / 27  \\  
%\hline
  D0 $W /Z$ & 68 / 51 & 67 / 51  \\ 
%\hline
  CDF $W /Z$ & 73 / 41 & 75 / 41  \\
%\hline
  E866 Drell-Yan & 50 / 39 & 50 / 39  \\ 
\hline  
\hline
Correlated $\chi^2$  & 114& 113  \\ 
Log penalty $\chi^2$  & -1.26& -13.22  \\ 
Total $\chi^2$ / dof  & 1699 / 1339& 1678 / 1339  \\ 
%\hline			 
\end{tabular}	
\label{tab:nlo-nnlo}
\end{center}
\end{ruledtabular}
\end{table}
%--------------------------------

Notably, the fit to the ATLAS \(W/Z\) data shows a more pronounced improvement at NNLO, 
with a \(\chi^2/\text{dof}\) of 104.8/91 compared to 108.6/91 at NLO. This indicates that 
NNLO corrections are particularly important for precision measurements of electroweak 
processes, where small corrections can significantly improve the fit. Similarly, 
the \(\chi^2\) for the HERA I+II dataset shows a slight improvement at NNLO (1151/1016 vs. 1155/1016 at NLO).

Overall, the \(\chi^2\) values demonstrate that the inclusion of NNLO 
corrections leads to modest improvements in fit quality across most data sets, 
confirming the necessity of higher-order corrections for precise determinations of 
PDFs, particularly in processes dominated by gluon-gluon or quark-gluon interactions. 
While NLO fits remain reasonably accurate for many applications, NNLO fits are essential 
for precision studies at high-energy colliders such as the LHC and 
future experiments like the EIC, LHeC, and FCC (he, hh).

%=================================================  
\subsection{Comparison to other PDF sets}  
%=================================================  

In this section, we compare our nominal NNLO PDFs (\texttt{Fit D}) with other recent global sets, 
specifically NNPDF4.0~\cite{NNPDF:2021njg}, CT18~\cite{Hou:2019efy}, and MSHT20~\cite{Bailey:2020ooq}. 
All results are presented at \(Q^2 = 1.9\), 10, and 100~GeV\(^2\), normalized to our nominal (\texttt{Fit~D}) NNLO PDFs. 
The PDF uncertainties are consistently shown at the 68\% confidence level (CL) for all sets, 
including those from the comparison groups, ensuring a fair basis for comparison. 
This consistency in uncertainty presentation is intended to provide a 
clear understanding of the relative uncertainties across the different PDF sets.

As discussed in Sec.~\ref{PDF_parametrisation}, our analysis includes six independently 
parameterized distributions: \( xu_{v} \), \( xd_{v} \), \( x{\bar{u}} \), \( x{\bar{d}} \), \( x{\bar{s}} \), and \( xg \). 
Similar to CT18 and MSHT20, we rely on a dynamically generated charm distribution through the VFNS, rather than 
independently parameterizing the charm quark. 
In contrast, NNPDF4.0 independently parameterizes eight PDFs, which include both the strange and charm distributions. 
It is also important to note that there are substantial differences in the underlying data sets used in these QCD analyses.

Overall, the four parton sets shown in these plots are generally in good agreement 
within their respective uncertainties, though some differences in shape are observed. 
These differences are more pronounced at the input scale, \(Q^2_0 = 1.9\)~GeV\(^2\), 
particularly for the total singlet \(x\Sigma\) and gluon \(xg\) distributions. For the 
valence densities, \(xu_v\) and \(xd_v\), NNPDF4.0 is slightly suppressed at 
intermediate values of \(x\) compared to the others. Notably, our gluon density is larger than the others, 
especially at small values of \(x\). Our valence and total singlet densities are 
in good agreement and are consistently within the uncertainty envelopes of CT18 and MSHT20.

More pronounced differences are observed for the gluon distribution. However, it is important to 
note that all other PDF sets include jet data, which provide additional constraints 
on the gluon PDF. In contrast, our \texttt{Fit~D} does not include any jet production 
data sets, as our primary goal is to investigate the impact of Drell-Yan and W/Z boson 
production data on the shape of different parton species and their associated uncertainties. 
The gluon distribution in our analysis is in fair agreement in the small \(x\) region, 
which is relevant for dominant Higgs boson production at the LHC.

\begin{figure*}[!htb]
\vspace{0.5cm}
\begin{center}
\includegraphics[scale = 0.4]{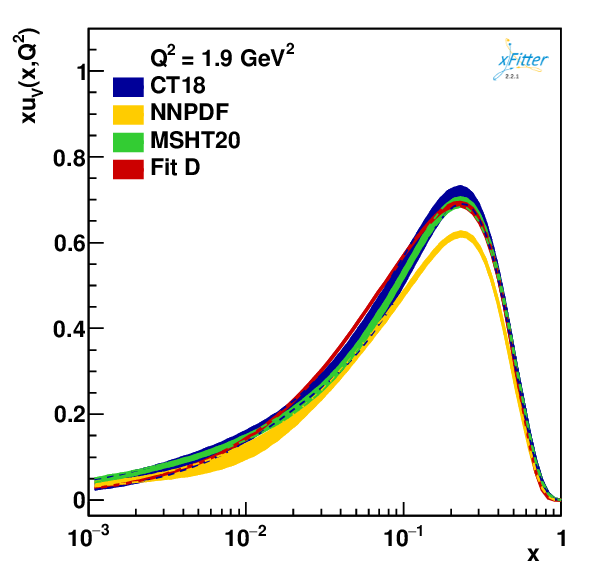}	
\includegraphics[scale = 0.4]{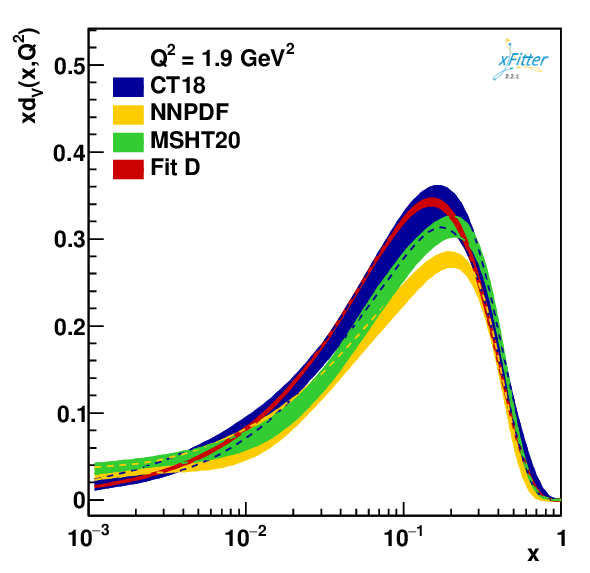}		
\includegraphics[scale = 0.4]{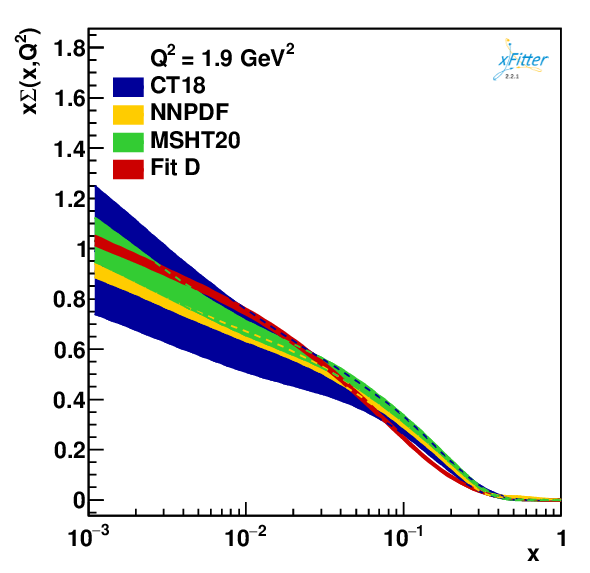}        
\includegraphics[scale = 0.4]{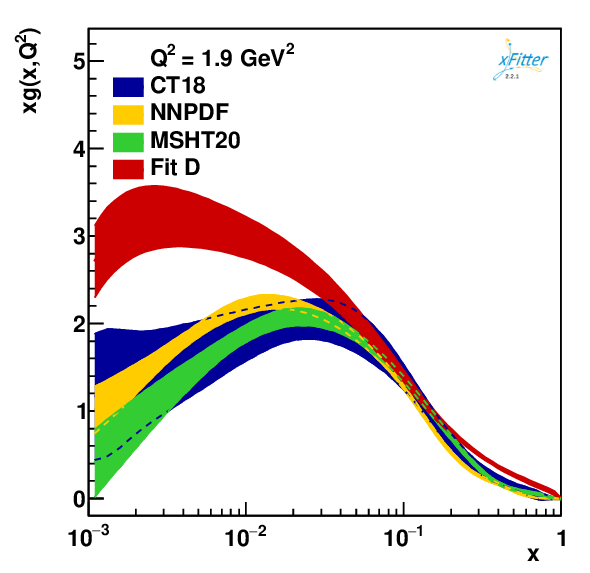}  
\includegraphics[scale = 0.4]{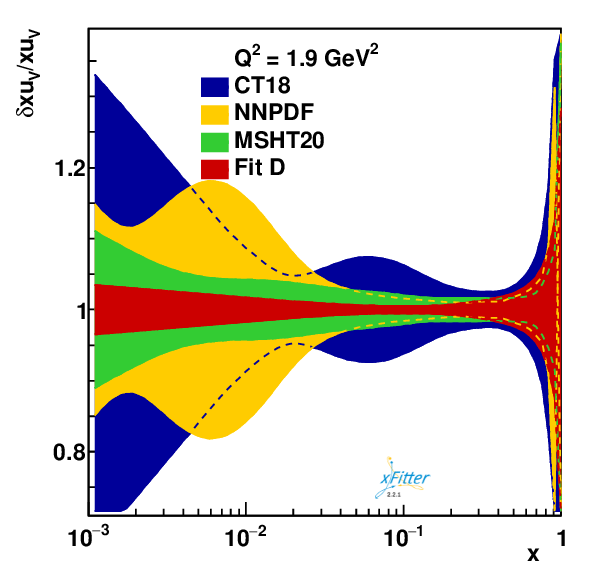}	
\includegraphics[scale = 0.4]{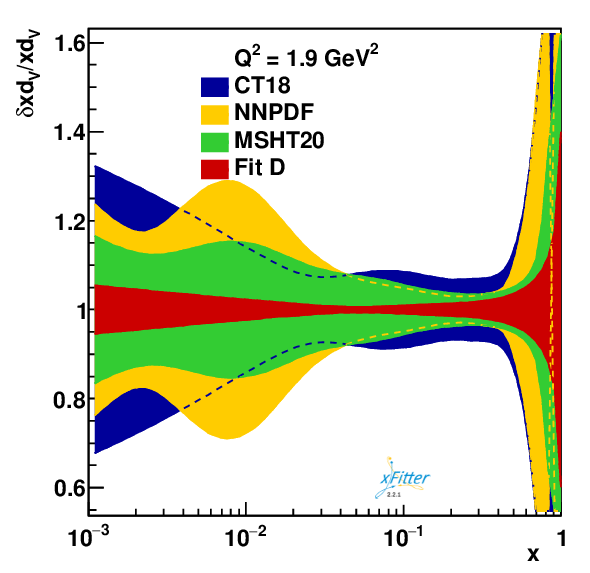}		
\includegraphics[scale = 0.4]{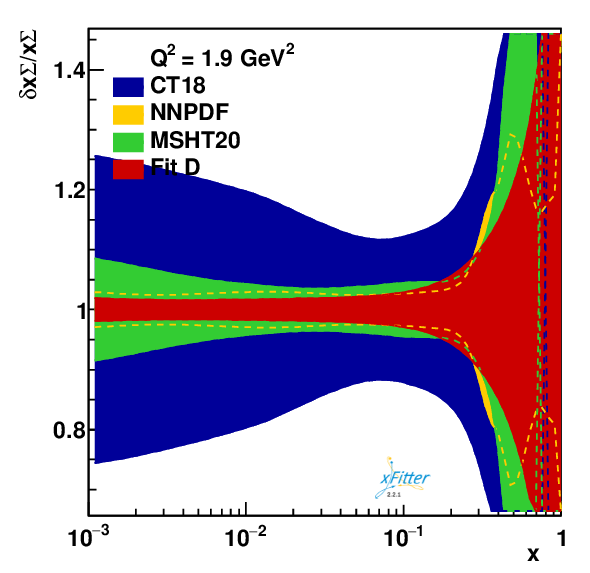}        
\includegraphics[scale = 0.4]{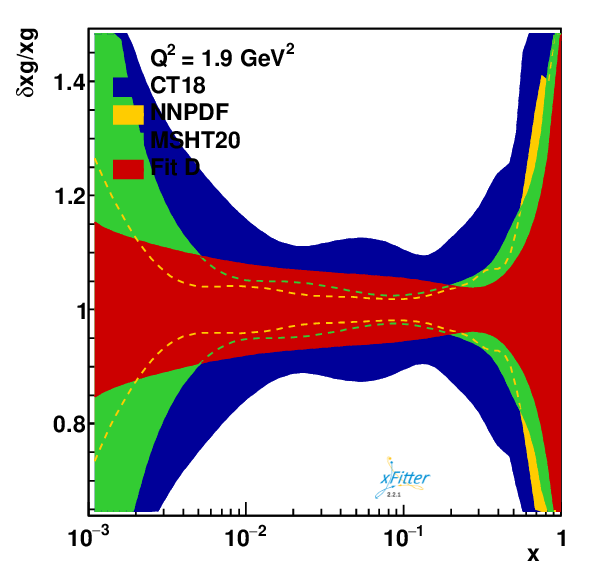}  \\
\caption{Comparison of our nominal fit ({\tt Fit~D})  at NNLO with other very recent determinations of PDFs, including 
NNPDF4.0~\cite{NNPDF:2021njg}, CT18~\cite{Hou:2019efy} and MSHT20~\cite{Bailey:2020ooq} 
for $xu_v$, $xd_v$, $x\Sigma$, and $xg$ distribution as a function of $x$ and at $Q^2 =1.9$ GeV$^2$. 
The relative uncertainties $\delta xq(x,Q^2)/xq(x,Q^2)$  are also shown. }
\label{fig:ModernPDF-Q0}
\end{center}
\end{figure*}

More interesting findings emerge from Fig.~\ref{fig:ModernPDF-Q0} when comparing PDF 
uncertainties. Our uncertainties for the valence and total singlet distributions are smaller 
than those of all other sets across the entire \(x\) region. However, 
the associated uncertainty for our gluon PDF is generally larger compared to the uncertainties 
of NNPDF4.0 and MSHT20 for \(x < 0.1\). The CT18 analysis, on the other hand, 
exhibits generally larger uncertainties for all parton densities shown in Fig.~\ref{fig:ModernPDF-Q0}.

%---------------Log-------- NNPDF and CTEQ gluon PDF NNLO at Q_0^2 = Figure 1.9-------------------------
%--------------------------------
%--------------------------------

\begin{figure*}[!htb]
\vspace{0.5cm}
\begin{center}
\includegraphics[scale = 0.4]{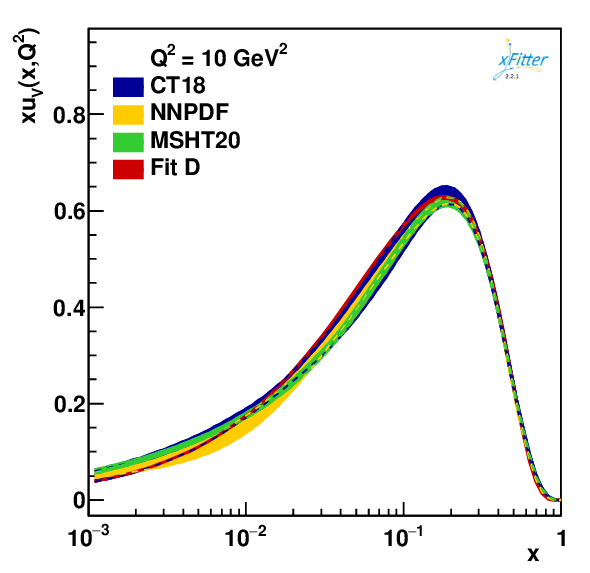}	
\includegraphics[scale = 0.4]{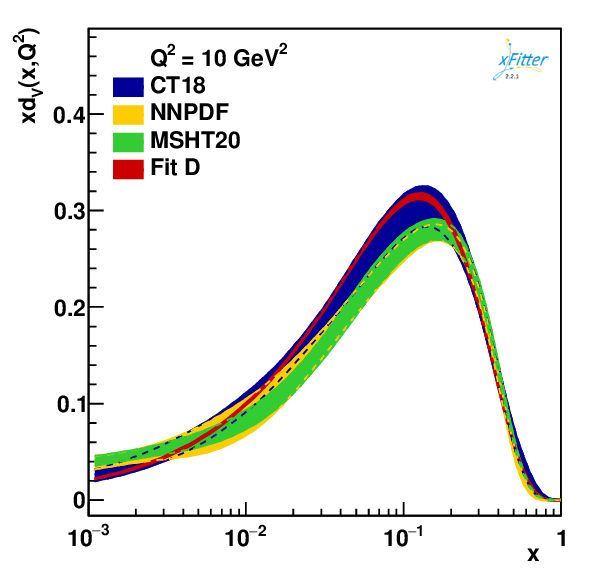}		
\includegraphics[scale = 0.4]{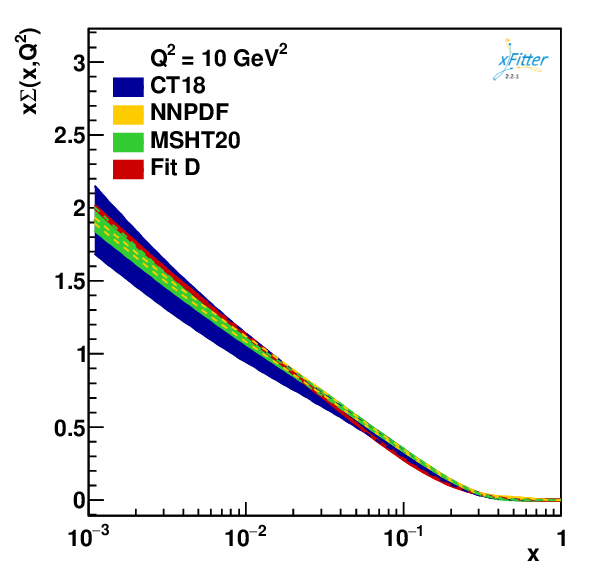}        
\includegraphics[scale = 0.4]{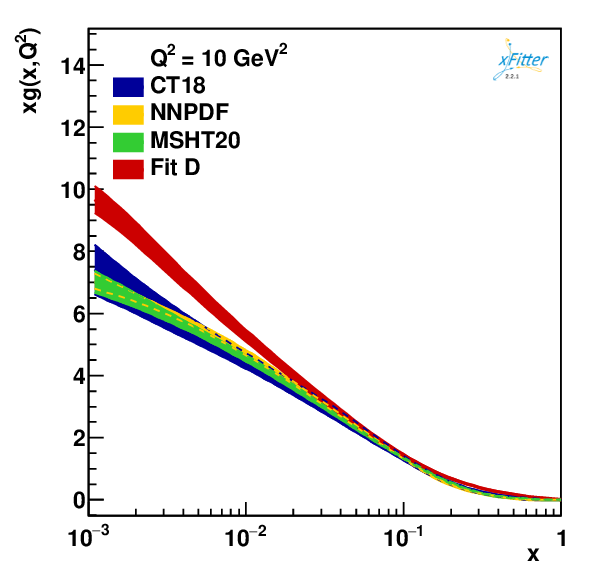}  
\includegraphics[scale = 0.4]{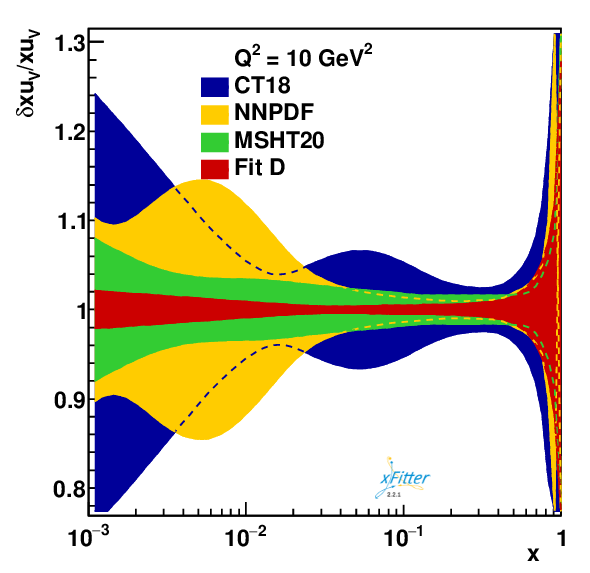}	
\includegraphics[scale = 0.4]{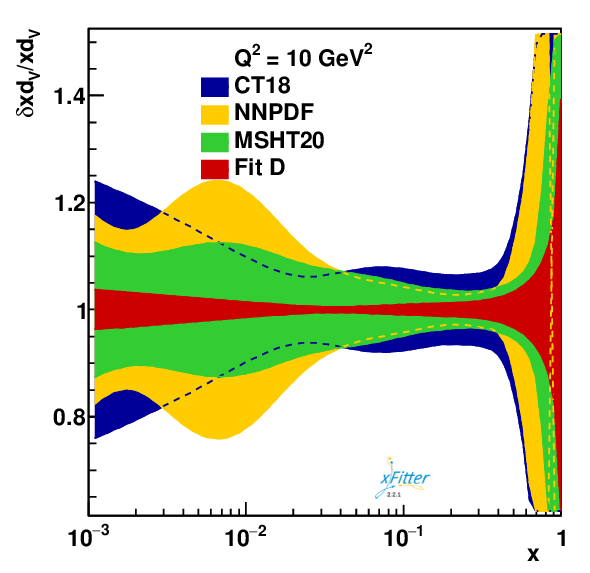}		
\includegraphics[scale = 0.4]{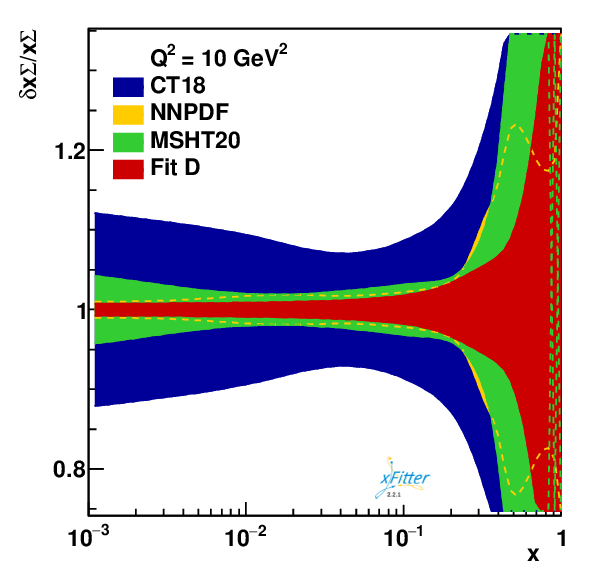}        
\includegraphics[scale = 0.4]{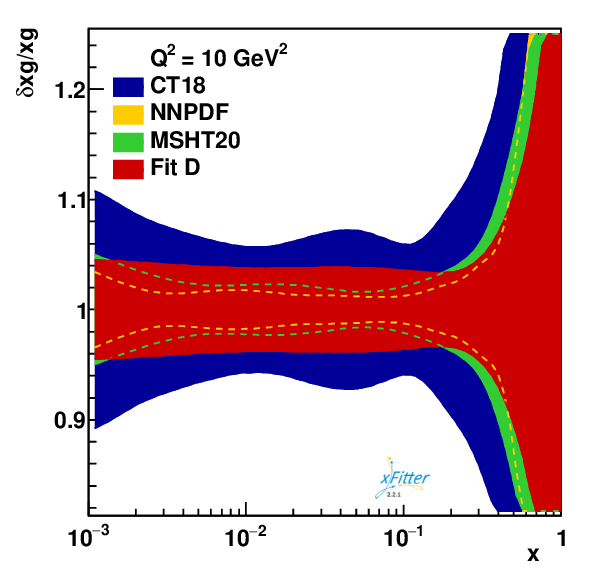}  \\
\caption{Same as Fig.~\ref{fig:ModernPDF-Q0} but this time for higher  $Q^2$ value of  10~GeV$^2$.}
\label{fig:ModernPDF-10}
\end{center}
\end{figure*}

\begin{figure*}[!htb]
\vspace{0.5cm}
\begin{center}
\includegraphics[scale = 0.4]{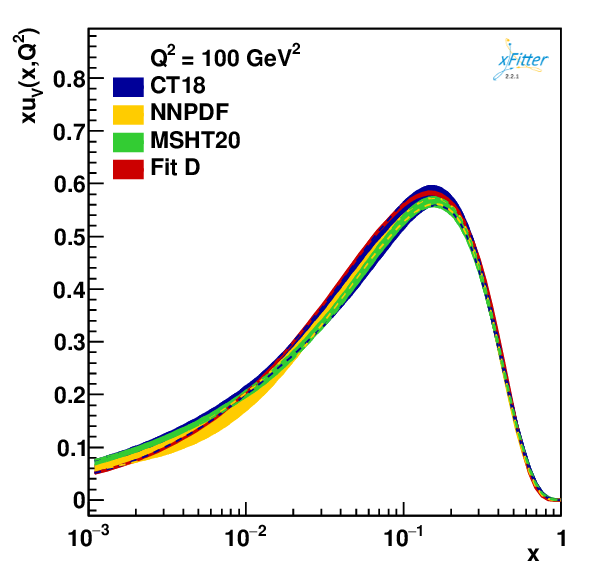}	
\includegraphics[scale = 0.4]{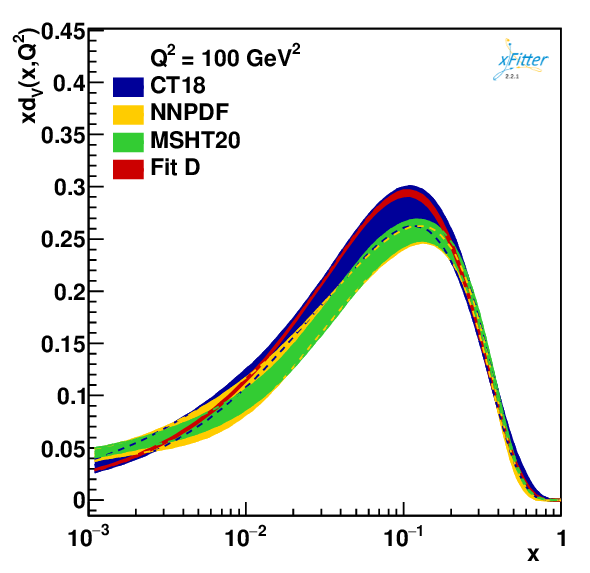}		
\includegraphics[scale = 0.4]{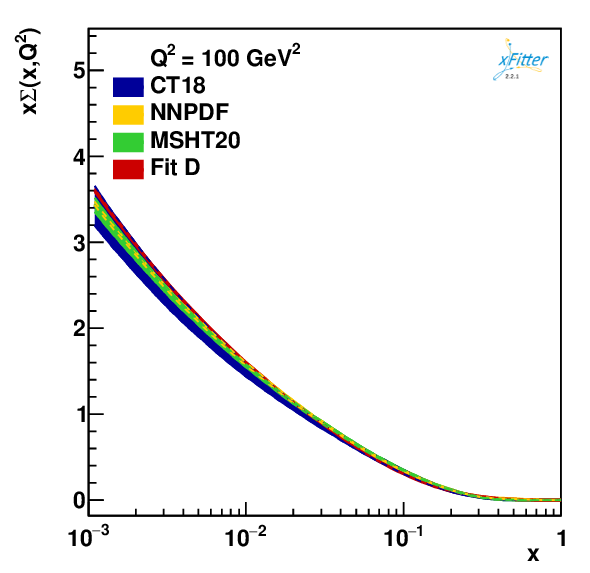}        
\includegraphics[scale = 0.4]{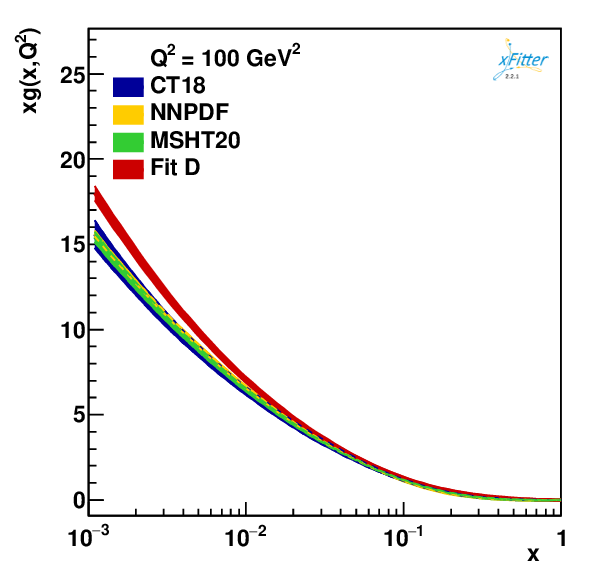}  	
\includegraphics[scale = 0.4]{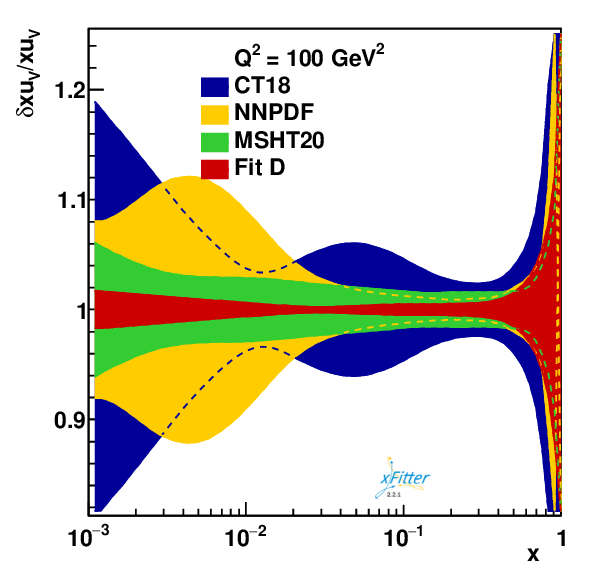}	
\includegraphics[scale = 0.4]{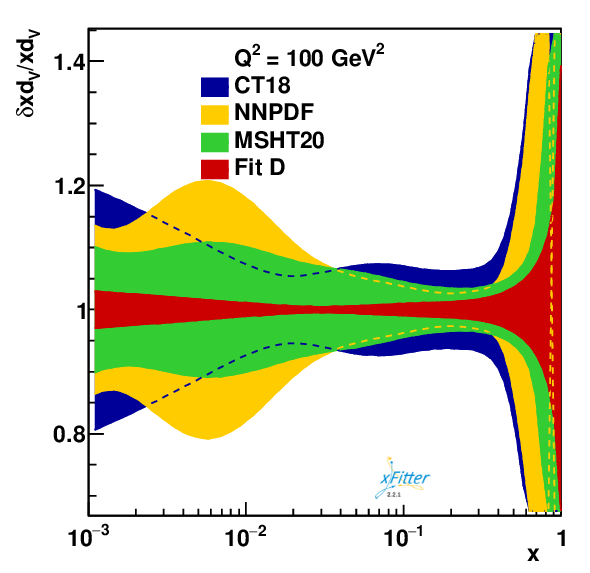}		
\includegraphics[scale = 0.4]{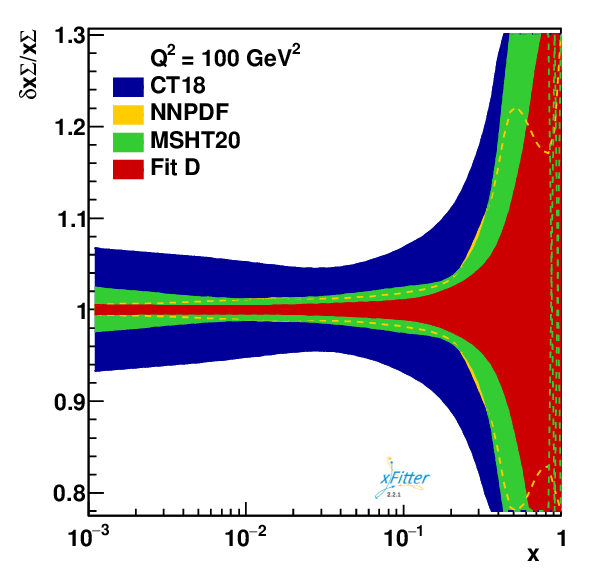}        
\includegraphics[scale = 0.4]{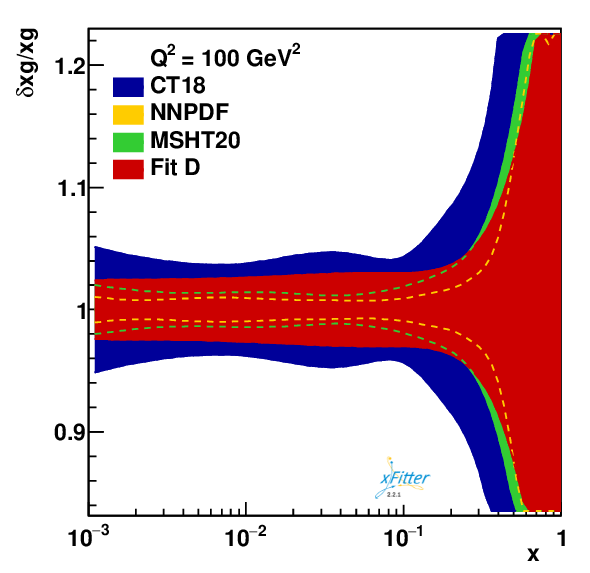}  \\		
\caption{Same as Fig.~\ref{fig:ModernPDF-Q0} but this time for higher  $Q^2$ value of 100~GeV$^2$.}
\label{fig:ModernPDF-100}
\end{center}
\end{figure*}

The same findings hold for higher values of \(Q^2 = 10\) and 100~GeV\(^2\), as shown in 
Figs.~\ref{fig:ModernPDF-10} and \ref{fig:ModernPDF-100}. As can be seen, the gluon PDF 
from all other groups is slightly suppressed in the smaller \(x\) region (\(x < 0.01\)) 
compared to our \texttt{Fit~D}. Due to the lack of data directly constraining the gluon PDF, 
our uncertainty is larger than those of NNPDF4.0 and MSHT20. Once again, 
CT18 generally exhibits larger uncertainties for all parton densities across the entire \(x\) region, 
as shown in Figs.~\ref{fig:ModernPDF-10} and \ref{fig:ModernPDF-100}.

%---------------Log-------- NNPDF and CTEQ gluon PDF NNLO at Q_0^2 = Figure 10-------------------------
%--------------------------------
%--------------------------------

%---------------Log-------- NNPDF and CTEQ gluon PDF NNLO at Q_0^2 = Figure 100-------------------------
%--------------------------------
%--------------------------------

\clearpage 
\newpage

%=================================================  
\subsection{Comparison to experimental data}  
%=================================================  

Here, we illustrate the ability of our NNLO QCD fit to describe the individual experiments 
included in our QCD analysis, with particular attention paid to the Drell-Yan data and W/Z boson production. 
We organize our data/theory discussions according to the specific physical processes.

\begin{figure*}[htb]
\vspace{0.5cm}
\begin{center}	
\includegraphics[scale = 0.6]{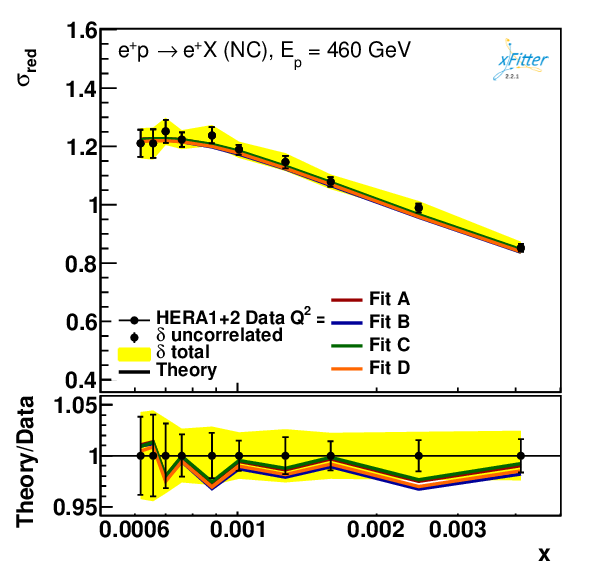}
\includegraphics[scale = 0.6]{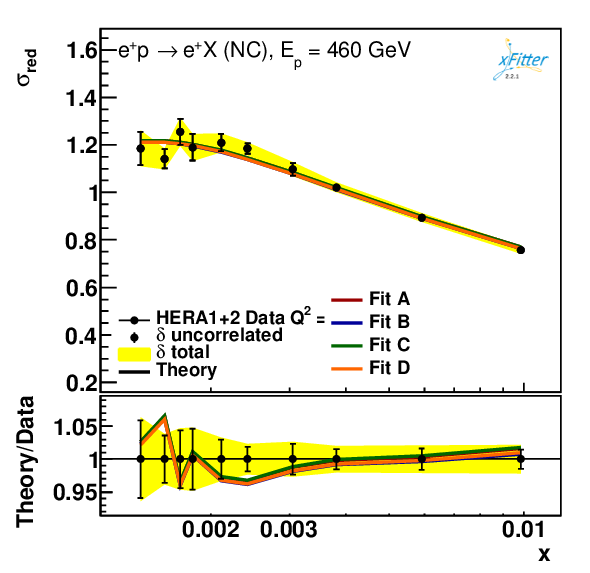}
\includegraphics[scale = 0.6]{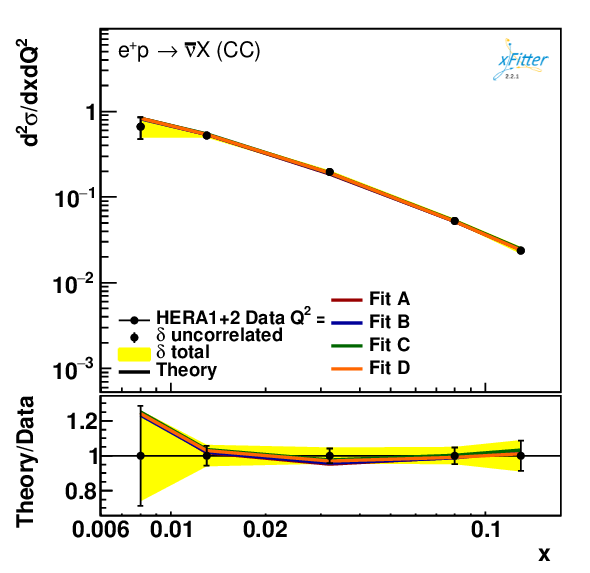}
\includegraphics[scale = 0.6]{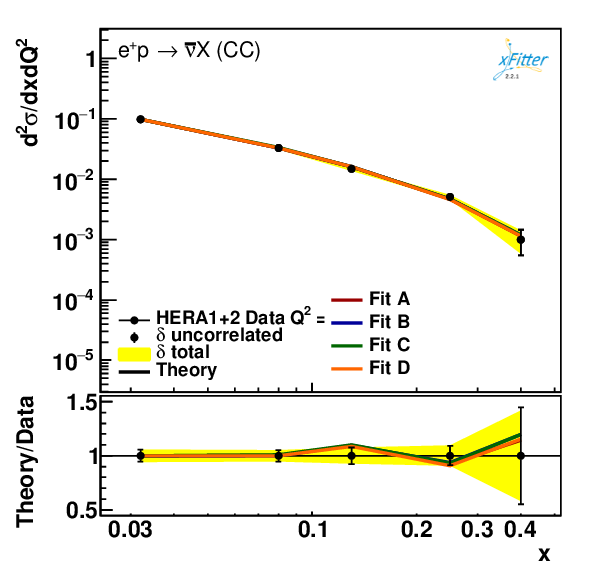}  
\caption{
Comparison between experimental data points and the theory prediction based on our 
NNLO best-fit results for a selection of fitted data points from H1/ZEUS combined DIS 
data sets (see text). The experimental uncertainty shown in the plots is the sum in 
quadrature of all statistical and systematic uncertainties.}
\label{fig:QCDfit1}
\end{center}
\end{figure*}

\begin{figure*}[htb]
\vspace{0.5cm}
\begin{center}	
\includegraphics[scale = 0.6]{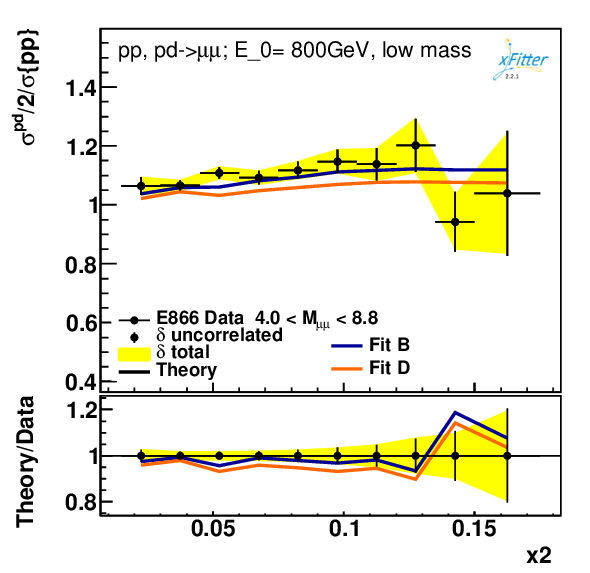}
\includegraphics[scale = 0.6]{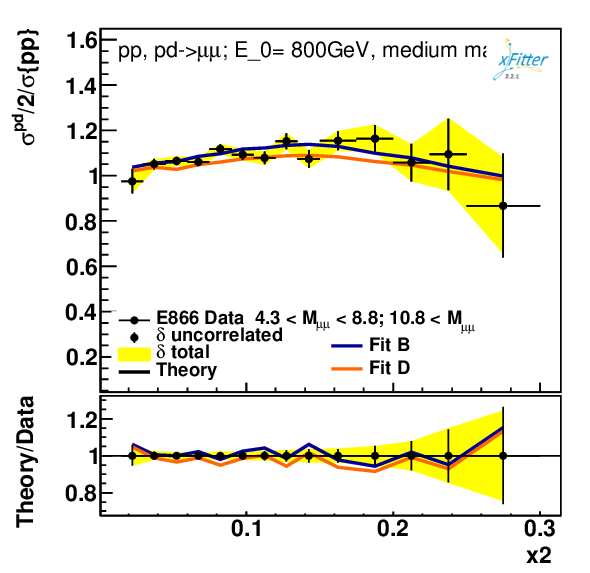}
\includegraphics[scale = 0.6]{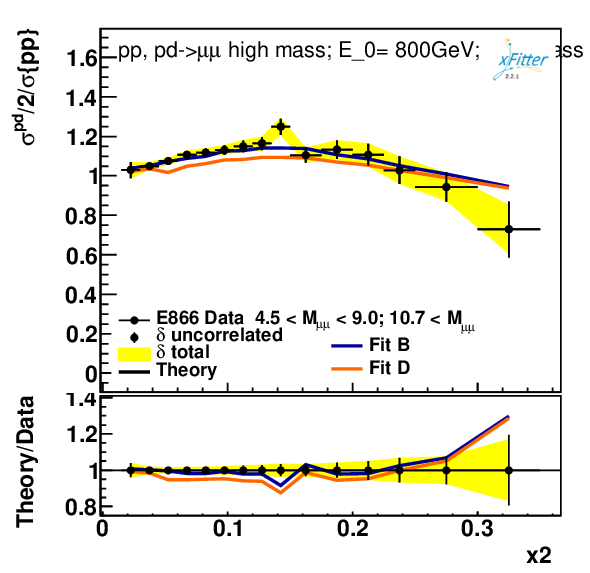}
\includegraphics[scale = 0.6]{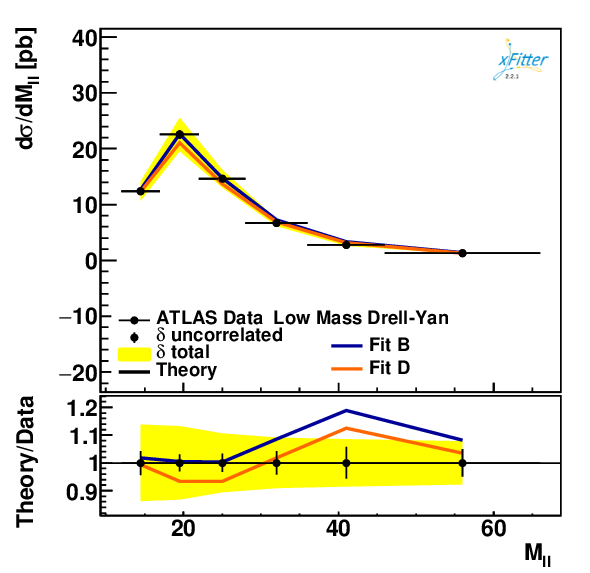}  
\caption{Comparison between experimental data points and the theory prediction based on our 
NNLO best-fit results for a selection of fitted data points for Drell-Yan datasets (see text). } 
\label{fig:QCDfit2}
\end{center}
\end{figure*}

\begin{figure*}[htb]
\vspace{0.5cm}
\begin{center}	
\includegraphics[scale = 0.4]{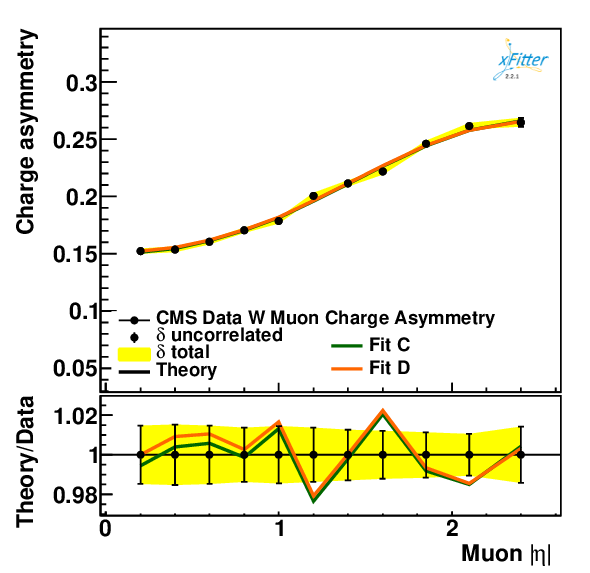}
\includegraphics[scale = 0.4]{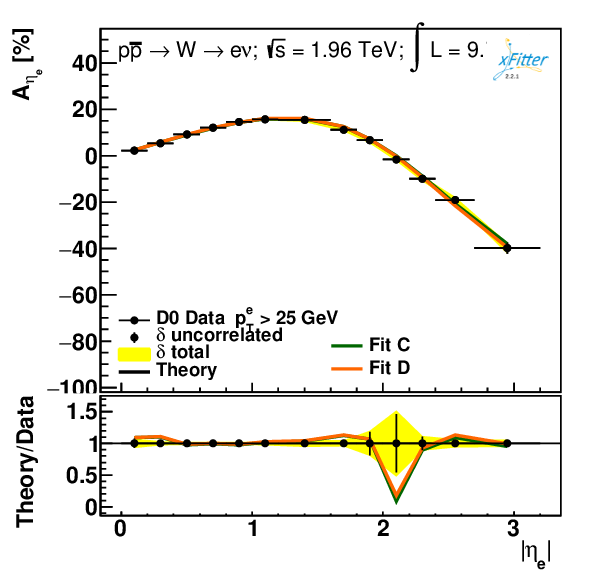}
\includegraphics[scale = 0.4]{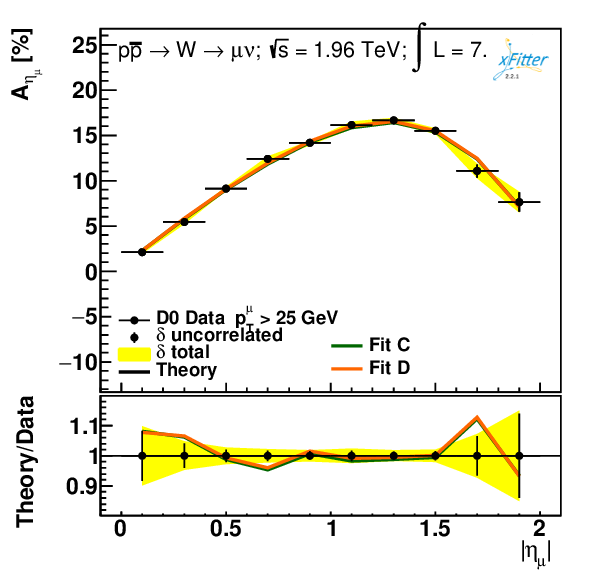}
\includegraphics[scale = 0.4]{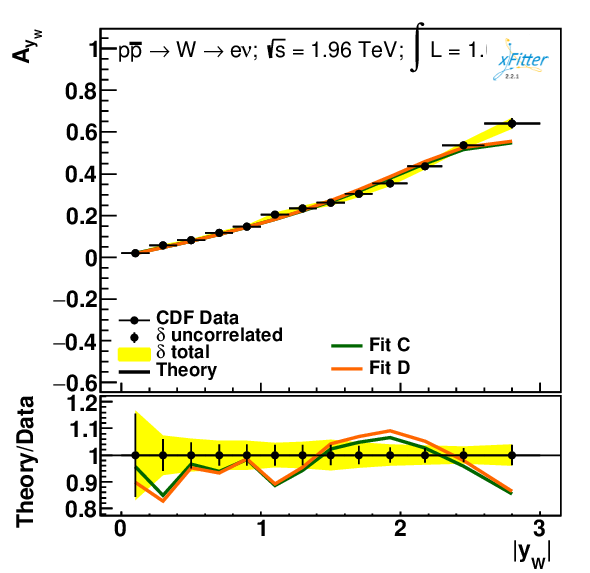}\\
%%%%%%%%%%%%%%%%%%%%%%	
\includegraphics[scale = 0.4]{w-data_3832877-1.eps}
\includegraphics[scale = 0.4]{w-data_2157868-1.eps}
\includegraphics[scale = 0.4]{w-data_9271105-1.eps}
\includegraphics[scale = 0.4]{w-data_5862587-1.eps}\\
%%%%%%%%%%%%%%%%%%%%%%
\includegraphics[scale = 0.4]{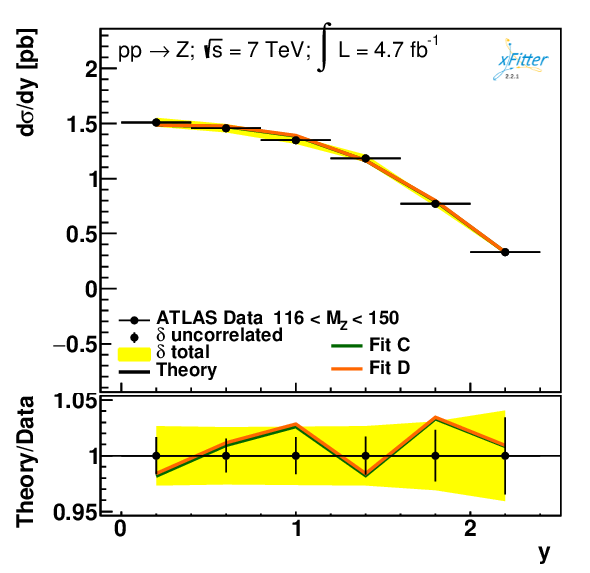}
\includegraphics[scale = 0.4]{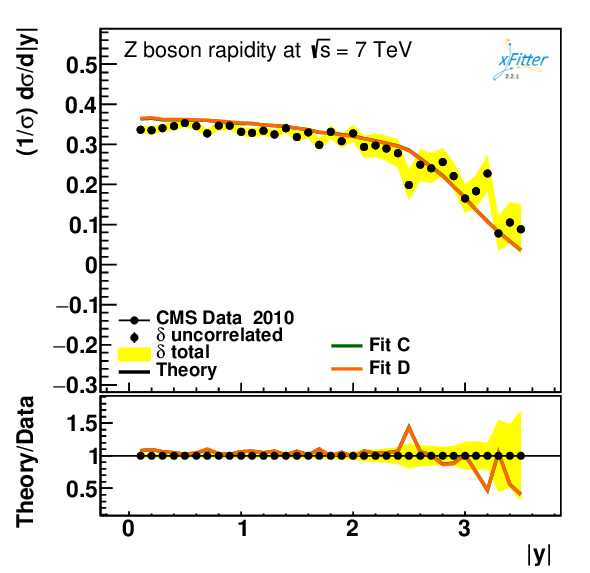}
\includegraphics[scale = 0.4]{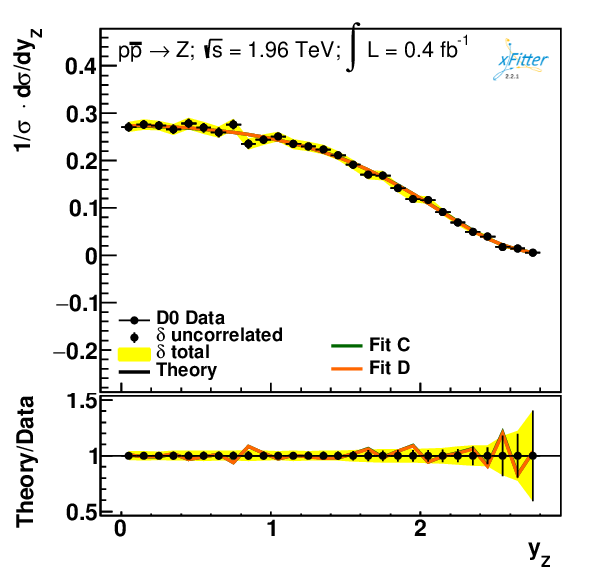}
\includegraphics[scale = 0.4]{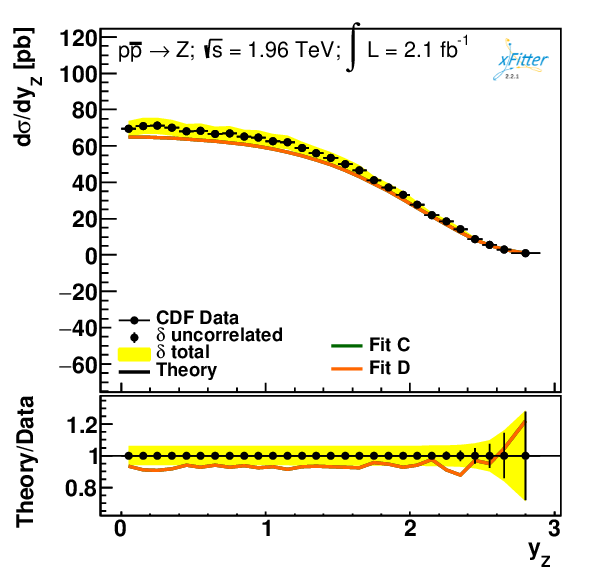}\\
%%%%%%%%%%%%%%%%%%%%%%
\caption{Comparison between experimental data points and the theory prediction based on our 
NNLO best-fit results for a selection of fitted data points for W/Z boson production (see text).}
\label{fig:QCDfit3}
\end{center}
\end{figure*}

In Figs.~\ref{fig:QCDfit1}, \ref{fig:QCDfit2}, and \ref{fig:QCDfit3}, we present a comparison between a 
selection of data included in our QCD fits and the corresponding NNLO best-fit results. 
This comparison aims to visually assess the fit quality and the relative size of data and PDF uncertainties.
The data shown are representative of the global dataset, starting with the H1/ZEUS combined 
data sets, as shown in Fig.~\ref{fig:QCDfit1}. Specifically, we show the results for both NC (top panel) 
and CC (bottom panel) HERA combined DIS data, where the bulk of the sensitivity in our 
fit still arises from HERA data. The data error bars shown in this figure correspond to the 
sum in quadrature of all uncertainties. Quantitative assessments of these comparisons 
are provided by the \(\chi^2\) values presented in Table~\ref{tab:Chi2}.

Since the HERA combined DIS data sets serve as our base data set selection, we present the 
comparison of our NNLO theory predictions for all four sets of PDFs discussed in Sec.~\ref{results}. 
The main finding from Fig.~\ref{fig:QCDfit1} is the very good agreement of our 
NNLO theory predictions with the HERA data for the kinematics shown.

%--------------------------------
%--------------------------------

Next, we discuss the comparisons between our NNLO theory predictions and 
selected Drell-Yan datasets. In Fig.~\ref{fig:QCDfit2}, we show comparisons for the 
E866 Drell-Yan data and the ATLAS low-mass Drell-Yan data. As previously discussed, 
Drell-Yan cross-section measurements have a significant impact on our PDFs compared 
to other datasets. The inclusion of these measurements allows for a clearer separation
of the sea quark distributions, resulting in a notable reduction in the error bands as well.

%--------------------------------
%--------------------------------

Finally, in Fig.~\ref{fig:QCDfit3}, we present a detailed comparison of our NNLO theory predictions 
with selected data sets from W/Z boson production measurements by CMS, ATLAS, D0, and CDF. 
Overall, the fit quality is satisfactory, and our nominal \texttt{Fit~D} provides a generally 
good description of the data. However, some disagreements are observed in specific kinematic 
regions of certain datasets. As indicated by the \(\chi^2\) values in Table~\ref{tab:Chi2}, 
there are evident tensions between specific data sets. These tensions are particularly notable 
for the CDF $W$ asymmetry data~\cite{CDF:2009cjw}, the ATLAS low-mass $Z$ data~\cite{ATLAS:2016nqi}, 
and the HERA1+2 NC \(e^+ p\) 920 dataset~\cite{Abramowicz:2015mha}. Such 
inconsistencies suggest that these data sets impose conflicting constraints on 
the PDFs, potentially affecting the overall quality of the fit.

%--------------------------------
%--------------------------------

%=================================================  
\section{Impact of simulated EIC DIS data on proton PDF determination}\label{EIC} 
%=================================================  

In this section, we explore the impact of simulated inclusive DIS data from the future Electron-Ion Collider (EIC) on 
the determination of proton PDFs at NLO and NNLO accuracy in QCD. 
Comparisons are made relative to our nominal global fit, \texttt{Fit~D}, as discussed in the previous sections. 
We also estimate the expected experimental uncertainty in the strong coupling constant, \(\alpha_s(M_Z^2)\), when simulated EIC 
inclusive data are incorporated into analyses similar to those performed using HERA data.

The EIC, currently being developed at Brookhaven National Laboratory (BNL) in collaboration with the 
Thomas Jefferson National Accelerator Facility, is expected to start collecting data 
around 2030~\cite{AbdulKhalek:2021gbh,Burkert:2022hjz,Accardi:2012qut}. 
The EIC will collide highly polarized electrons with highly polarized protons, as well as light or heavy nuclei. 
In electron-proton mode, the anticipated luminosity will range from \(10^{33} - 10^{34}\) cm\(^{-2}\) s\(^{-1}\), 
with a center-of-mass energy \(\sqrt{s}\) spanning from 29~GeV to 141~GeV. The broad physics program of the EIC includes 
high-precision measurements of inclusive DIS cross sections, with a particular focus on the 
large Bjorken-\(x\) kinematic region, complementing the measurements from H1 and ZEUS at HERA. 
Earlier investigations of the impact of inclusive EIC data on \(\alpha_s\) precision 
and proton PDFs can be found in Refs.~\cite{Cerci:2023uhu,Armesto:2023hnw}. 

Table~\ref{tab:EIC_data} shows the different beam energy configurations and their corresponding center-of-mass energies. 
The EIC will operate with different beam configurations, involving both electron and proton beams at a range of energies. 
The main data sets we include correspond to NC DIS pseudodata generated for five different beam energy combinations, 
with center-of-mass energies ranging from 29 GeV to 141 GeV. The most important electron-proton beam configurations 
included in this study are 10$\times$275~GeV, 18$\times$275~GeV, 5$\times$41~GeV, 5$\times$100~GeV, 
and 10$\times$100~GeV~\cite{AbdulKhalek:2021gbh,Burkert:2022hjz,Accardi:2012qut}. 
Each data set represents an integrated luminosity anticipated for one year of data collection at the EIC, 
with simulated measurements performed across a range of \(x\) and \(Q^2\) values, 
logarithmically spaced across six orders of magnitude. 
The kinematic coverage of the EIC is expected to complement that of HERA, 
particularly by filling the high-\(x\) gap that HERA could not reach. 
Although there will be overlap in the kinematic regions covered, the EIC will provide 
much higher precision, offering new opportunities to probe the proton structure in this poorly constrained region. 
The key advantage of the EIC lies in its ability to provide precise data at large \(x\), 
which remained relatively underconstrained by previous HERA DIS experiments.

%--------------------------------
\begin{table*} 
\caption{\label{tab:EIC_data}
EIC beam energy configurations, center-of-mass energies, and integrated luminosity.}
\begin{center}
\begin{tabular}{c | c | c | c}
\hline 	\hline 
Electron beam energy (GeV)   ~&~ Proton beam energy (GeV)  ~&~ Center-of-mass energy \(\sqrt{s}\) (GeV) ~&~ Integrated Luminosity (fb\(^{-1}\)) ~\\ 	\hline         \hline	
 18  & 275 & 141 & 15.4 \\
 10  & 275 & 105 & 100 \\
 10  & 100 & 63  & 79 \\
 5   & 100 & 45 & 61 \\
 5   & 41  & 29 & 4.4 \\
\hline 	\hline   
\end{tabular}
\end{center}
\end{table*}
%--------------------------------

To improve our understanding of the impact of simulated EIC data on proton PDF determination, 
we have used NC and CC DIS cross-sections at various center-of-mass energies, as anticipated for  
the EIC. These simulated data sets were generated based on predictions using {\tt HERAPDF2.0NLO} 
and {\tt HERAPDF2.0NNLO} as the baselines for NLO and NNLO accuracy, respectively~\cite{Cerci:2023uhu,Armesto:2023hnw}. 
For other inputs, such as heavy quark masses and the value of the strong coupling constant, we used the nominal values from {\tt HERAPDF2.0}.

The pseudodata include statistical uncertainties and systematic errors modeled based on experience from 
HERA, conservatively adjusted to reflect the potential precision of the EIC. Specifically, 
an uncorrelated systematic uncertainty of 1.9\% is applied across most data points, 
extending to 2.75\% for the lowest inelasticity values (i.e., \(y < 0.01\)), along with a 
normalization uncertainty of 3.4\%, fully correlated for each energy configuration. 
These uncertainties encompass both statistical and systematic components, designed to 
capture the challenges expected in high-precision EIC measurements.

The pseudodata points were generated by applying random smearing based on Gaussian 
distributions, with standard deviations determined by the projected uncertainties 
estimated by the ATHENA collaboration~\cite{ATHENA:2022hxb}. 
 This approach allows us to simulate realistic EIC datasets capable of 
significantly constraining the proton PDFs, particularly in the region of $x$  
where existing data from HERA and other sources provide limited coverage.  
For the purposes of our QCD fits, the point-to-point systematic uncertainties were combined in 
quadrature with the statistical uncertainties, while the normalization uncertainties were treated 
as nuisance parameters, following the approach used in Ref.~\cite{H1:2015ubc}.

\begin{figure*}[!htb]
%\vspace{0.5cm}
\begin{center}
\includegraphics[scale = 0.90]{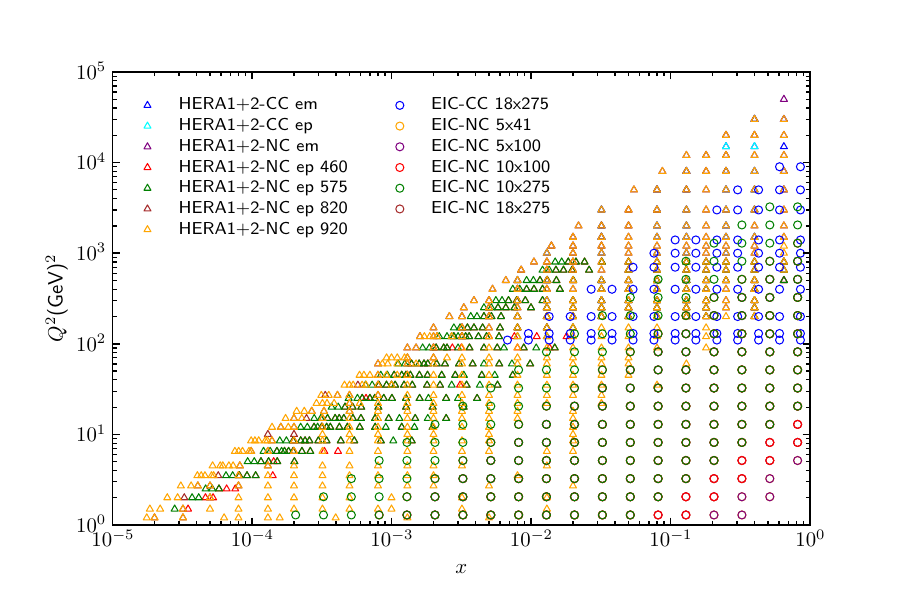}
\caption{The kinematic coverage in the \((x, Q^2)\) plane for the HERA DIS data 
and the EIC simulated NC and CC inclusive DIS pseudodata used in this analysis. 
The same kinematic cuts applied to the HERA DIS datasets have also been used for the EIC pseudodata. } 
\label{fig:EIC_Data}
\end{center}
\end{figure*}

Figure~\ref{fig:EIC_Data} shows the locations of the HERA DIS datasets and EIC 
simulated NC and CC inclusive DIS data points in the \((x, Q^2)\) kinematic plane used in this QCD analysis. 
As illustrated in the figure, the EIC pseudodata overlap with the HERA DIS data in 
some regions while also extending the kinematic reach to higher values of \(x\) in the intermediate \(Q^2\) range. 
Additionally, the same cuts imposed on the HERA datasets were also applied to the EIC pseudodata, 
ensuring consistency in the treatment of kinematic constraints and 
thereby avoiding the need for additional higher twist corrections in our analysis. 

%--------------------------------
%--------------------------------

\begin{figure*}[!htb]
%\vspace{0.5cm}
\begin{center}
  
\includegraphics[scale = 0.48]{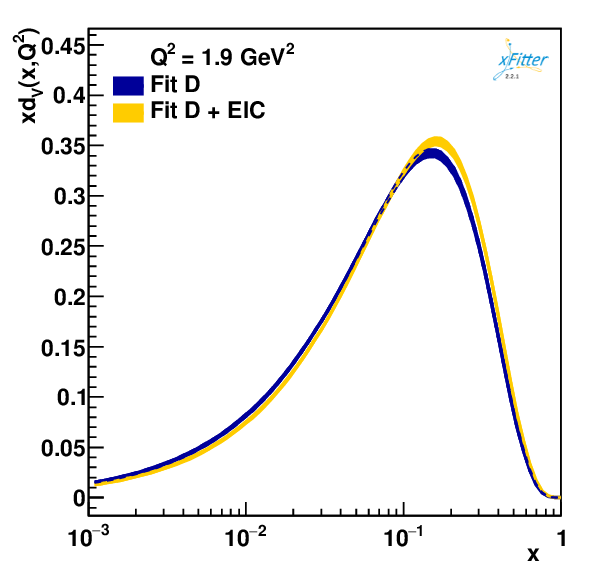}	
\includegraphics[scale = 0.48]{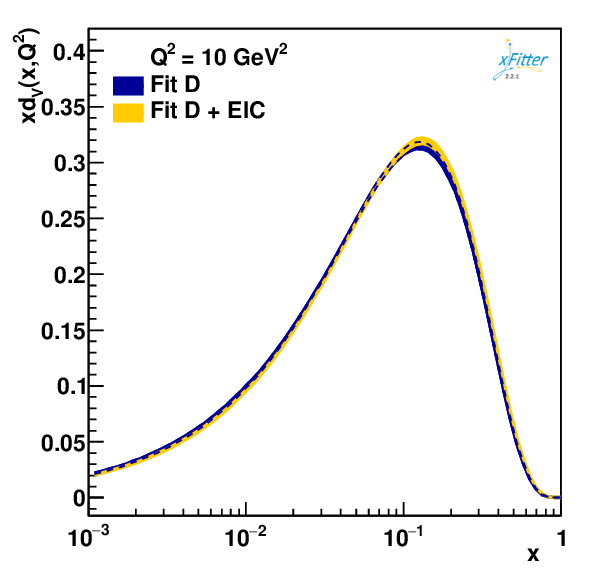}		
\includegraphics[scale = 0.48]{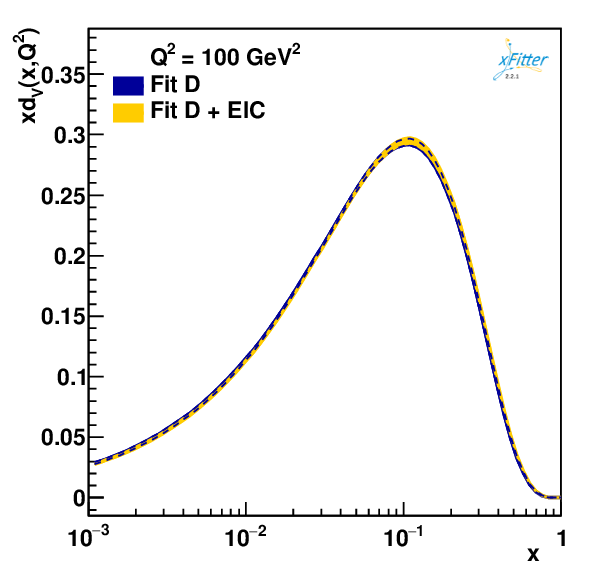}\\        
\includegraphics[scale = 0.48]{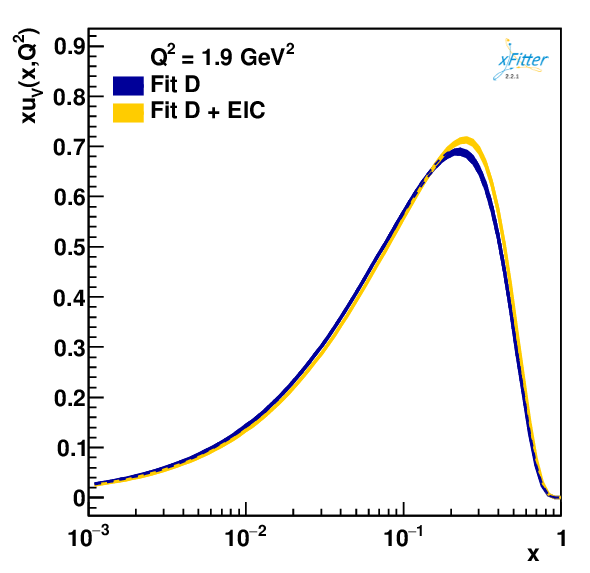}  
\includegraphics[scale = 0.48]{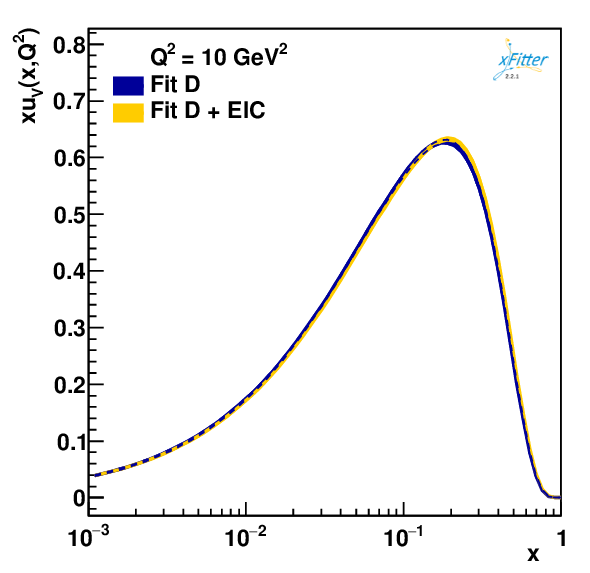}	
\includegraphics[scale = 0.48]{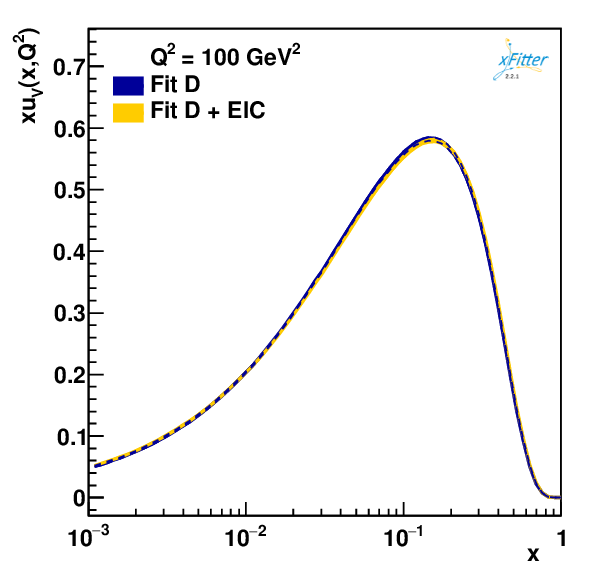} \\		
\includegraphics[scale = 0.48]{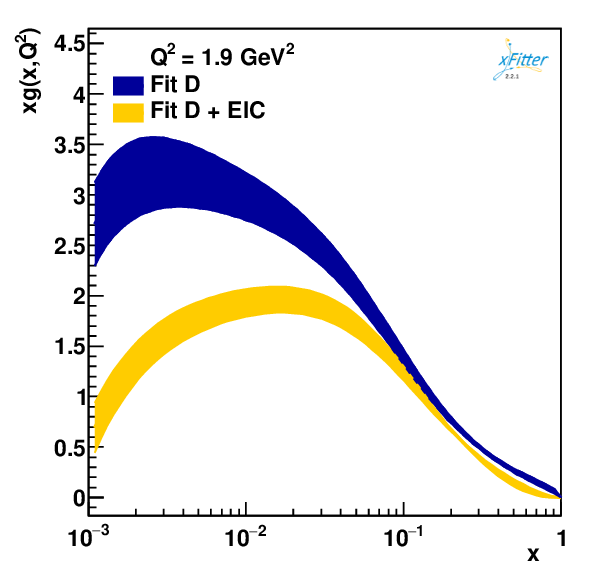}        
\includegraphics[scale = 0.48]{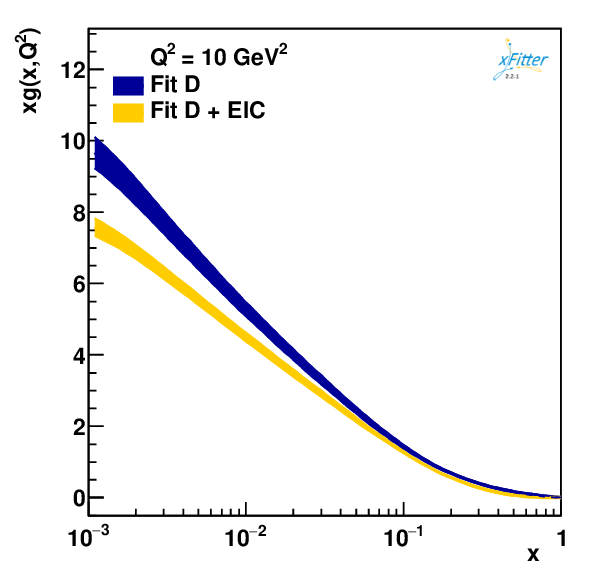}  
\includegraphics[scale = 0.48]{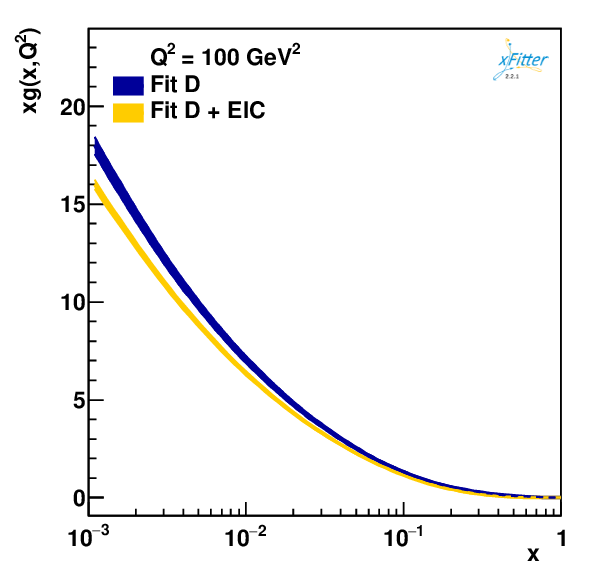} \\
\caption{Impact of EIC pseudo data on $xu_v$, $xd_v$ and $xg$ distribution functions at $Q^2 =$ 1.9, 10, 100~GeV$^2$. } 
\label{fig:EIC}
\end{center}
\end{figure*}

\begin{figure*}[!htb]
%\vspace{0.5cm}
\begin{center}
  
\includegraphics[scale = 0.48]{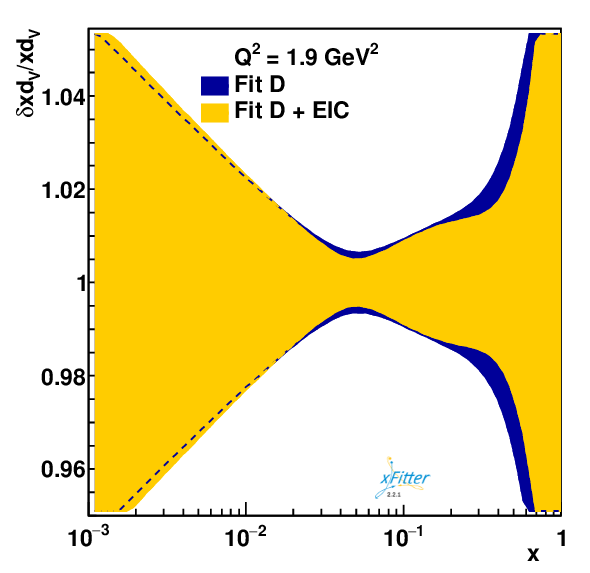}	
\includegraphics[scale = 0.48]{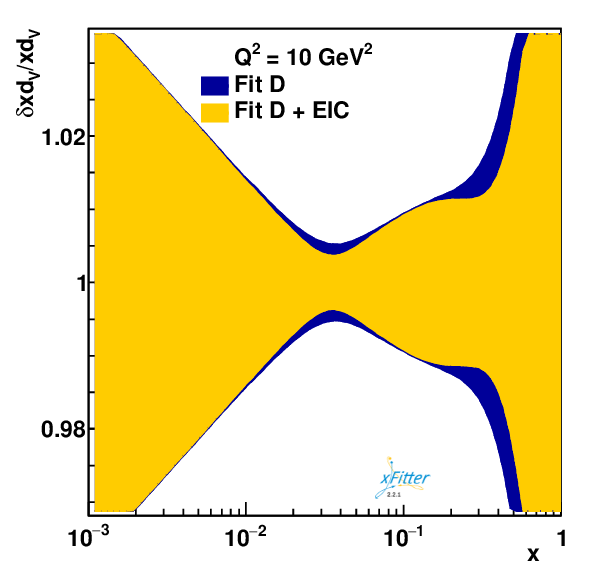}		
\includegraphics[scale = 0.48]{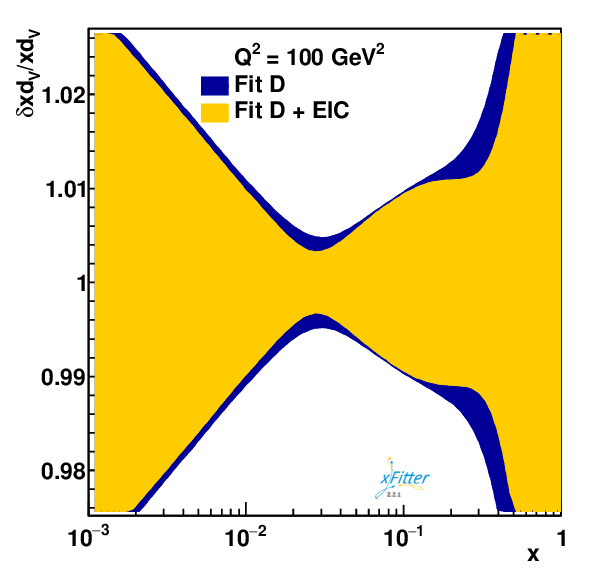}\\        
\includegraphics[scale = 0.48]{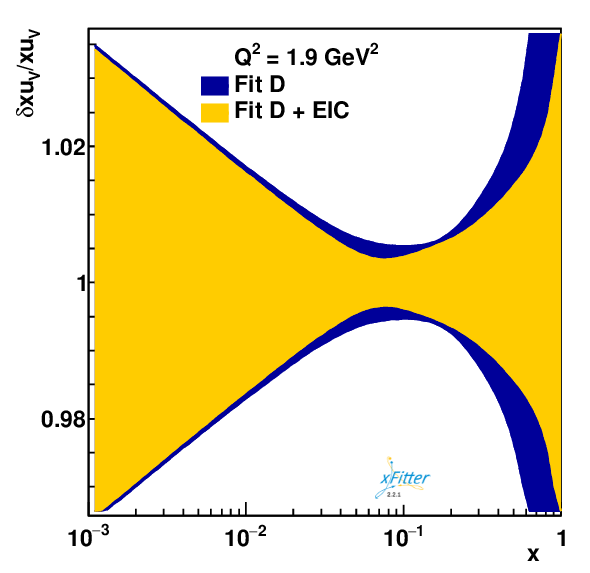}  
\includegraphics[scale = 0.48]{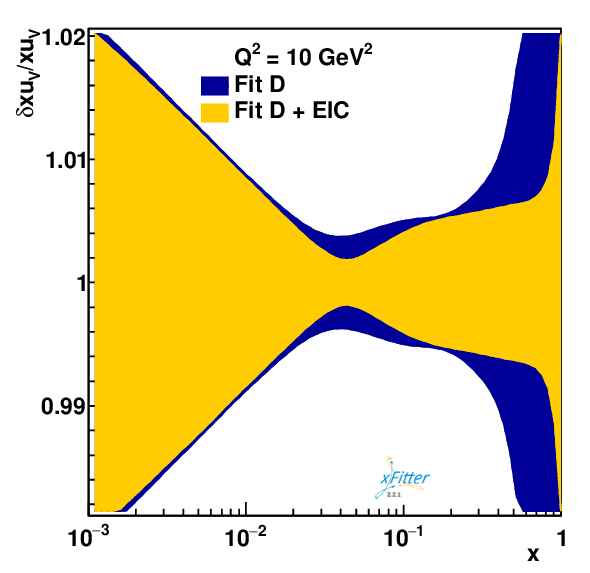}	
\includegraphics[scale = 0.48]{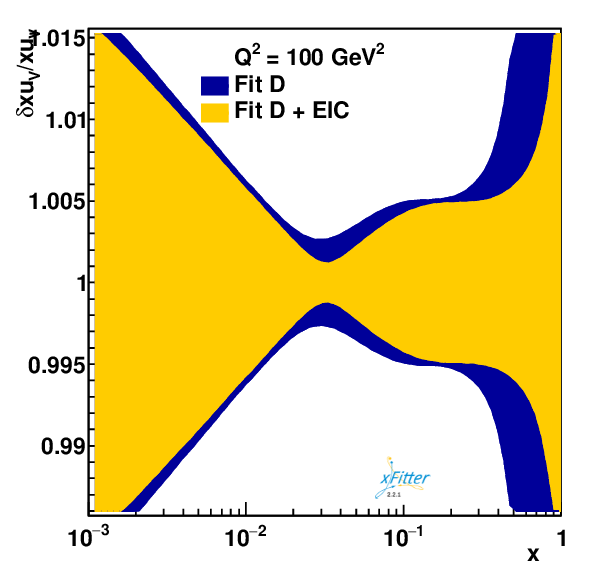} \\		
\includegraphics[scale = 0.48]{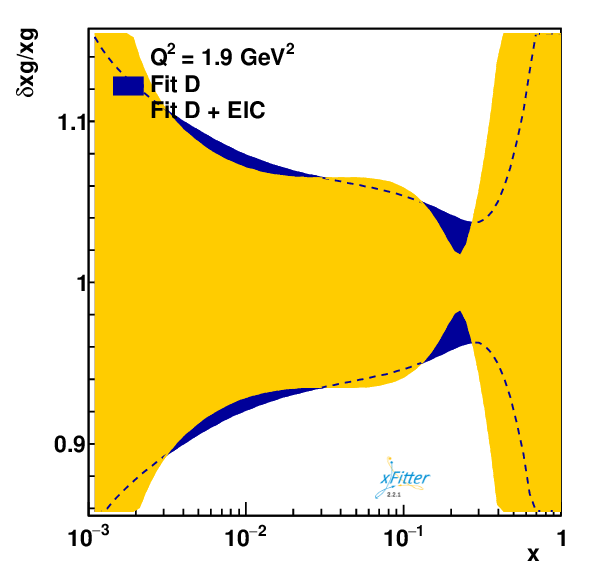}        
\includegraphics[scale = 0.48]{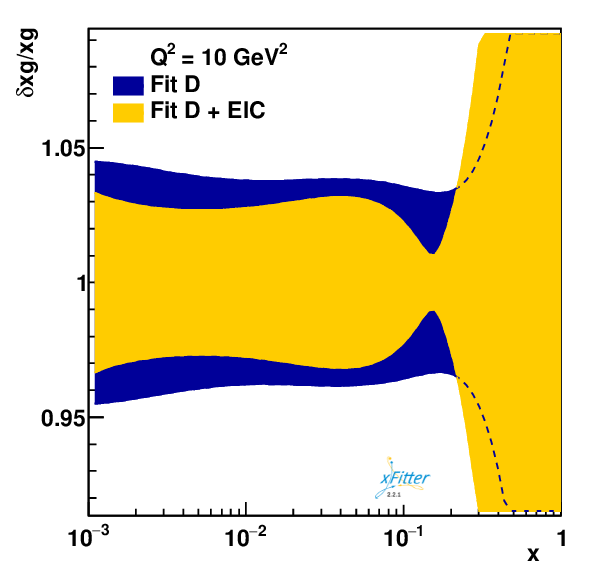}  
\includegraphics[scale = 0.48]{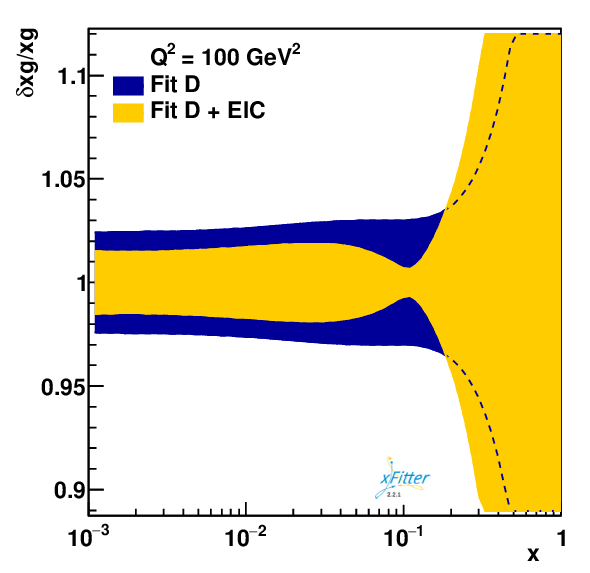} \\
\caption{Same as Fig.~\ref{fig:EIC} but this time for the relative uncertainties $\delta xq(x,Q^2)/xq(x,Q^2)$.} 
\label{fig:EIC_ratio}
\end{center}
\end{figure*}

The inclusion of the simulated EIC pseudodata in our global QCD fits provides 
substantial improvements in constraining the gluon and sea quark distributions, 
especially at medium to large values of \(x\), where the precision added by the 
EIC data is unprecedented. 
As shown in Figs.~\ref{fig:EIC} and \ref{fig:EIC_ratio}, incorporating these 
data sets in our global fit (\texttt{Fit~D + EIC}) reduces the uncertainties 
of gluon PDFs at \(x \leq 0.2\) for moderate-to-high \(Q^2\).
Furthermore, the impact of EIC  on the determination of the strong coupling  
constant \(\alpha_s(M_Z^2)\) is also significant, which we explore in detail in the next section.

%---------------Log--------EIC-------------------------
%--------------------------------
%--------------------------------

%---------------Log--------EIC-Ratio-------------------------
%--------------------------------
%--------------------------------

In Table~\ref{tab:FitD-EIC}, we present the extracted \(\chi^2\) values from our global 
fits, comparing the baseline analysis (\texttt{Fit~D}) with the extended analysis that 
includes the simulated EIC data (\texttt{Fit~D + EIC}). These \(\chi^2\) values, 
corresponding to different experimental data sets, provide a quantitative measure of the 
fit quality and illustrate the impact of incorporating EIC pseudodata in the QCD analysis.

%--------------------------------
\begin{table*}[h]
\begin{ruledtabular}
\begin{center}	
\caption{The values of extracted $\chi^2$ from the global fits with and without EIC data.}
\begin{tabular}{c|r|c}
Experiment & {\tt Fit D} & {\tt Fit D + EIC}   \\       \hline         \hline	
 HERA I+II &  1151/ 1016 & 1130/1016\\
 CMS $W /Z$  & 106.8 / 90 & 101.8 / 90 \\  
 ATLAS $W /Z$  & 104.8 / 91 & 111.3 / 91\\  
 ATLAS Drell-Yan  & 25.3 / 27 & 24.3 / 27\\  
 D0 $W /Z$  & 67 / 51 & 77 / 51\\ 
 CDF $W /Z$  & 75 / 41 & 69 / 41\\
 E866 Drell-Yan  & 50 / 39 & 70 / 39\\ 
 EIC NCep 10x275  & - & 117 / 103  \\ 
 EIC NCep 18x275  & - & 113 / 116  \\
 EIC NCep 5x41  & - & 31 / 36  \\ 
 EIC CCep 18x275  & - & 112 / 88  \\ 
 EIC NCep 5x100  & - & 33 / 55  \\ 
 EIC NCep 10x100  & - & 64 / 66  \\ 
\hline  
\hline
Correlated $\chi^2$  & 113& 149  \\ 
Log penalty $\chi^2$  & -13.22&  -40.97  \\ 
Total $\chi^2$ / dof  & 1678 / 1339 & 2160 / 1803  \\
  & $ = 1.25$ & $= 1.20$  \\ 
\end{tabular}	
\label{tab:FitD-EIC}
\end{center}
\end{ruledtabular}
\end{table*}
%--------------------------------

The addition of the EIC data results in a significant improvement in the overall fit quality, 
with the total \(\chi^2\) per degree of freedom (dof) decreasing from 1.25 in \texttt{Fit~D} 
to 1.20 in \texttt{Fit D + EIC}. 
The fit quality for HERA~I+II slightly improves with the addition of EIC data, 
as indicated by the reduction in \(\chi^2\) from 1151/1016 in \texttt{Fit~D} 
to 1130/1016 in \texttt{Fit~D + EIC}. This suggests that the EIC data are consistent with 
the HERA measurements and help refine the global fit. However, the \(\chi^2\) values 
for the W/Z production data from CMS, ATLAS, D0, and CDF exhibit mixed results, 
with minimal changes in the \(\chi^2\) for the ATLAS and E866 Drell-Yan data sets when EIC data are added.

The simulated EIC data sets show good agreement with theoretical predictions, 
with reasonable \(\chi^2\) values across all beam energy configurations. The 
NC and CC EIC data sets exhibit \(\chi^2/\text{dof}\) values close to 1, indicating 
a good fit within the global QCD analysis. These data sets considerably improve 
the constraints on PDFs, particularly in regions of medium to large \(x\) and 
moderate \(Q^2\), where previous data provided limited constraints.

Our findings clearly demonstrate that the inclusion of EIC pseudodata in the 
global QCD fit significantly enhances the fit quality, as evidenced by the 
improved \(\chi^2/\text{dof}\). 
The EIC data introduce crucial new constraints,  
%particularly at large \(x\) and moderate \(Q^2\), 
resulting in a more precise determination of the proton PDFs.  
Despite minor tensions with certain data sets, 
the overall improvement underscores the importance of the EIC 
for future QCD analyses~\cite{Cerci:2023uhu,Armesto:2023hnw}.

\newpage
\clearpage

%--------------------------------
\begin{table*} 
\begin{ruledtabular}
\begin{center}	
\caption{The \(\chi^2\) values from global QCD fits comparing results with and 
without the inclusion of jet and dijet production data.} 
\begin{tabular}{c|l|c|c|c}
Experiment ~&~ {\tt Fit~D}  ~&~  {\tt Fit~D + jet/dijet}  ~&~    {\tt Fit~D + EIC}   ~&~  {\tt Fit~D + EIC + jet/dijet} \\  \hline \hline 
%            \multicolumn{4}{l}{{\bf CMS Experiment}}\\
HERA I+II &  1151 / 1016 & 1128 / 1016 & 1130 / 1016 & 1127 / 1016\\
 %\hline
CMS $W /Z$  & 106.8 / 90 & 103 / 91 & 101.8 / 90 &99.9 / 90\\  
 %\hline
ATLAS $W /Z$  & 104.8 / 91 & 112 / 91 & 111.3 / 91 & 111/91\\  
%\hline
ATLAS Drell-Yan  & 25.3 / 27 & 24 / 27 & 24.3 / 27& 24.4 / 27\\  
%\hline
D0 $W /Z$  & 67 / 51 & 77 / 51 & 77 / 51&  68 / 51\\ 
%\hline
CDF $W /Z$  & 75 / 41 & 66 / 41 & 69 / 41& 78 / 41\\
%\hline
E866 Drell-Yan  & 50 / 39 & 72 / 39 & 70 / 39& 73 / 39\\ 
%----------------------------------
H1 jet  & - & 30 / 52 & - & 31 / 52\\ 
ZEUS jet  & - & 54 / 60 & -  & 54 / 60\\ 
ZEUS dijet  & - & 18 / 22 & -  & 17 / 22\\ 
%----------------------------------
EIC NCep 5x41  & - & - &  31 / 36  &31 / 36\\ 
EIC NCep 10x275 & - & - & 117 / 103  &117 / 103\\ 
EIC NCep 18x275  & - & - & 113 / 116  &113 / 116\\
EIC CCep 18x275  & - & - & 112 / 88  &112 / 88\\ 
EIC NCep 5x100  & - & - & 33 / 55  & 32 / 55\\ 
EIC NCep 10x100  & - & - & 64 / 66  & 64 / 66\\ 
%----------------------------------
\hline  
\hline
Correlated $\chi^2$  & 113& 142 & 149 &  156\\ 
Log penalty $\chi^2$  & -13.22 & -27 &  -40.97  & -48.72\\ 
Total $\chi^2$ / dof  & 1678 / 1339 &1799 / 1473& 2160 / 1803  & 2259 / 1937\\
  & $ = 1.25$& $= 1.22$ & $= 1.20$ & $= 1.17$ \\ 
%\hline			 
\end{tabular}	
\label{tab:FitD-EIC-jet}
\end{center}
\end{ruledtabular}
\end{table*}
%--------------------------------

%=================================================  
\section{Impact of jet and dijet production data on proton PDF determination}\label{jet} 
%=================================================  

This section discusses the significant role that jet and dijet production data play in 
determining PDFs and reducing their associated uncertainties, particularly for the gluon distribution.  
We also explore the effect of these data on the precision of  the determination  of the strong coupling constant, 
\(\alpha_s\). Historically,  inclusive jet and dijet production measurements have been crucial for constraining 
the gluon density, \(g(x, Q^2)\), due to the high sensitivity of these processes to the gluon PDF, 
especially at high-energy colliders. Early jet data from HERA and Tevatron Run-II 
have had a noticeable impact on global PDF fits, as demonstrated in numerous QCD analyses.

The HERA jet data used in our work include measurements of inclusive jet production in 
DIS at high \(Q^2\) values, ranging from 150 to 15,000~GeV\(^2\), recorded by the 
H1 Collaboration~\cite{H1:2007xjj}. Additionally, we include data from H1 on jet 
production in the lower \(Q^2\) range, below 100~GeV\(^2\), based on an integrated 
luminosity of 43.5~pb\(^{-1}\)~\cite{H1:2010mgp}. The ZEUS Collaboration inclusive jet differential 
cross-section measurements in DIS for \(Q^2 > 125\)~GeV\(^2\) with 38.6~pb\(^{-1}\)~\cite{ZEUS:2002nms} 
and an expanded dataset with 82~pb\(^{-1}\)~\cite{ZEUS:2006xvn} are also included. Furthermore, 
inclusive dijet cross-sections in DIS measured by ZEUS with an integrated 
luminosity of 374~pb\(^{-1}\)~\cite{ZEUS:2010vyw} are considered as well.

\begin{figure}[!htb]
%\vspace{0.5cm}
\begin{center}
\includegraphics[scale = 0.6]{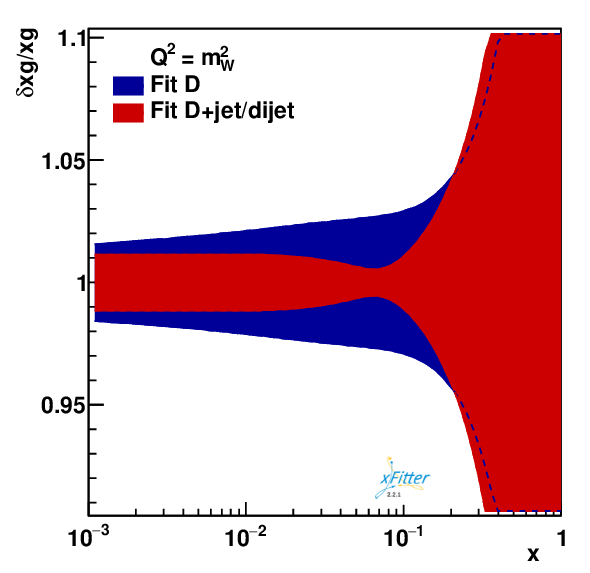}	
\includegraphics[scale = 0.6]{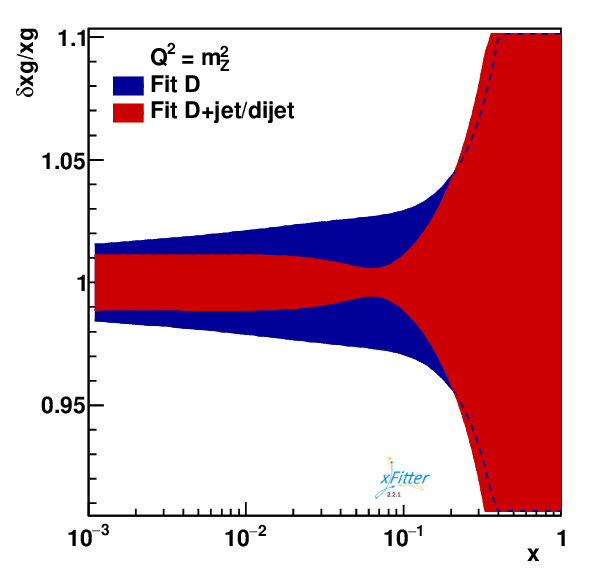}	 
\caption{Comparison of our nominal fit ({\tt Fit~D}) at NNLO accuracy with the extended fit ({\tt Fit~D + jet/dijet}) that includes 
the HERA jet and dijet production data. 
The comparisons are shown for the relative uncertainties of the gluon PDFs, \(\delta xg(x,Q^2)/xg(x,Q^2)\).} 
\label{fig:FitD_j_g_ratio}
\end{center}
\end{figure}

In Fig.~\ref{fig:FitD_j_g_ratio}, we compare our nominal fit ({\tt Fit~D}) at NNLO accuracy with the 
extended fit ({\tt Fit~D + jet/dijet}), which includes the aforementioned HERA jet and dijet production data. 
The comparisons are presented for the relative uncertainties of 
the gluon PDFs, \(\delta xg(x,Q^2) / xg(x,Q^2)\), at $Q^2 = M_W^2$ and $M_Z^2$. 
As can clearly be seen, the inclusion of the gluon-sensitive jet and dijet data significantly reduces 
the error bands, especially at moderate to very small values of $x$. 
This reduction indicates the crucial role of jet and dijet production data in constraining the gluon PDF, 
which is inherently difficult to determine with precision in the absence of such data. 
The enhanced constraints provided by the HERA jet and dijet datasets are particularly impactful in 
reducing uncertainties in the low \(x\) region, where the gluon contribution is less well-determined.
In terms of the overall quality of the fit, as summarized in Table~\ref{tab:FitD-EIC-jet}, the \(\chi^2/\text{dof}\) value 
improves from \(1.25\) in the nominal fit ({\tt Fit~D}) to \(1.22\) when the jet and dijet data are 
included ({\tt Fit~D + jet/dijet}).  
The reduction in \(\chi^2/\text{dof}\) demonstrates the importance of these datasets in 
achieving more precise proton PDF determinations.

%---------------Log--------FitD_j_g_ratio-------------------------
%--------------------------------
%--------------------------------

\begin{figure*}[!htb]
%\vspace{0.5cm}
\begin{center}
  
\includegraphics[scale = 0.48]{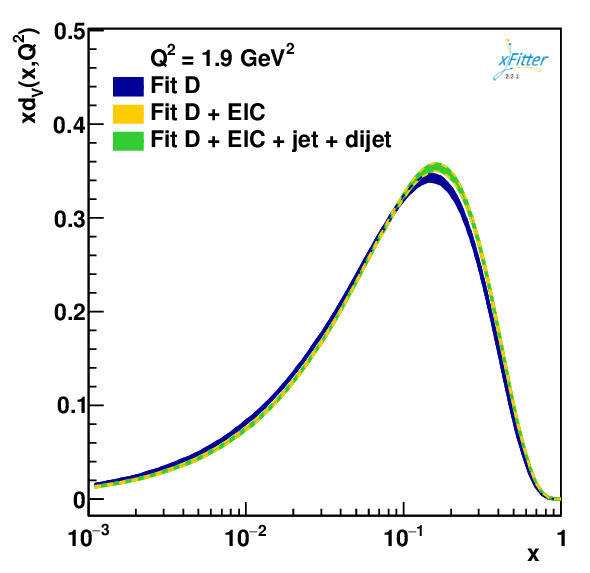}	
\includegraphics[scale = 0.48]{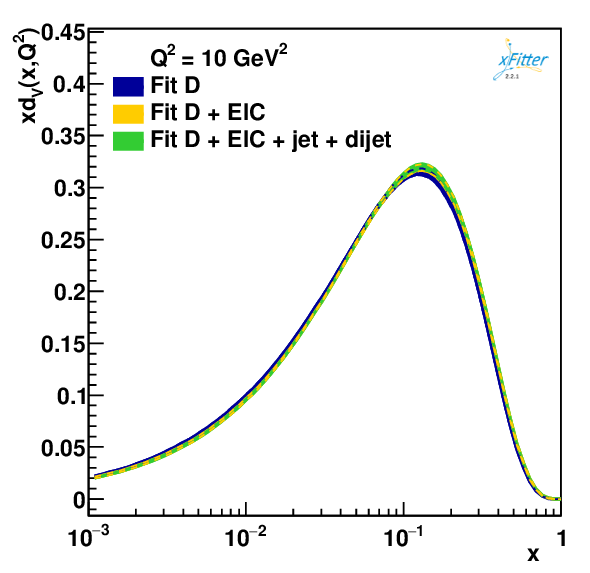}		
\includegraphics[scale = 0.48]{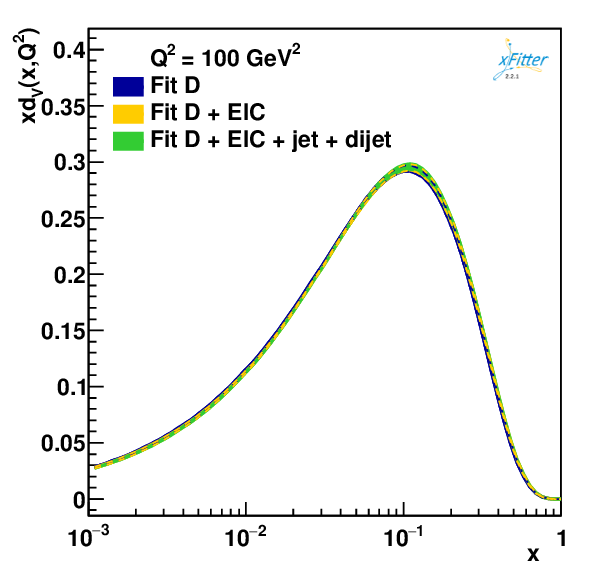}\\        
\includegraphics[scale = 0.48]{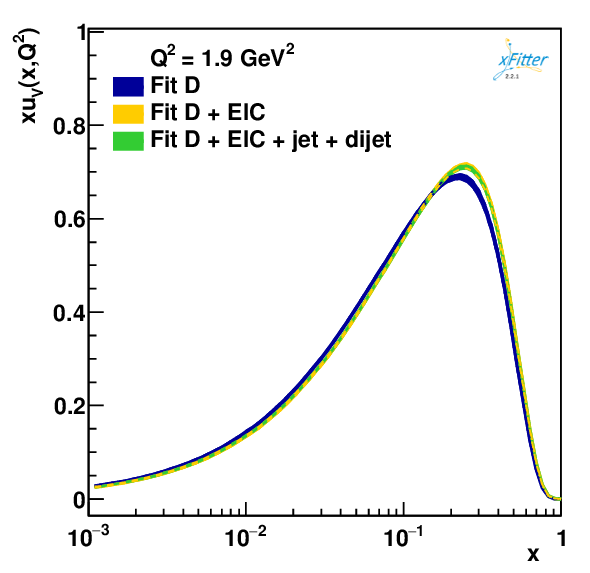}  
\includegraphics[scale = 0.48]{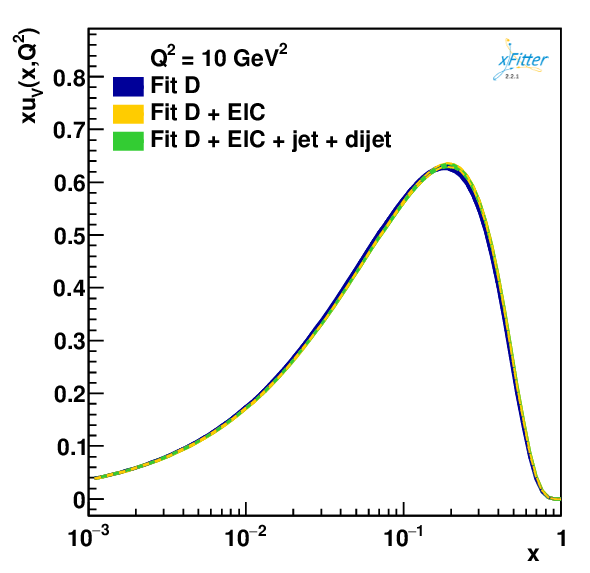}	
\includegraphics[scale = 0.48]{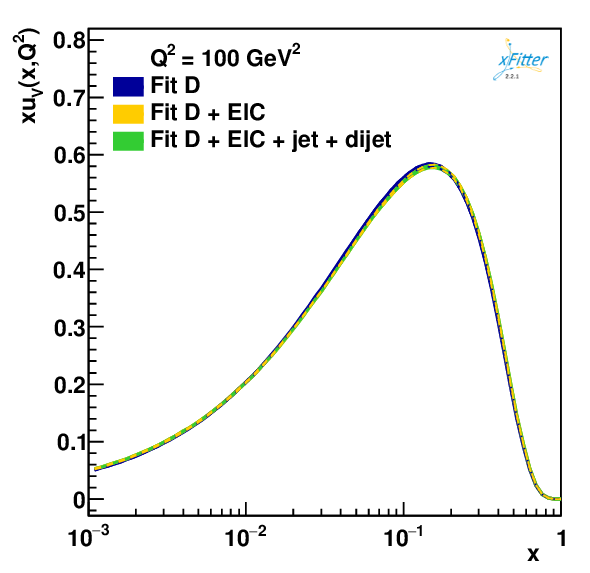} \\		
\includegraphics[scale = 0.48]{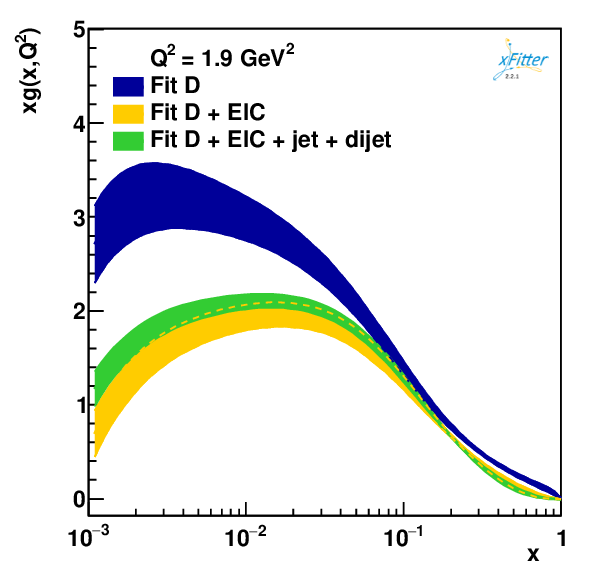}        
\includegraphics[scale = 0.48]{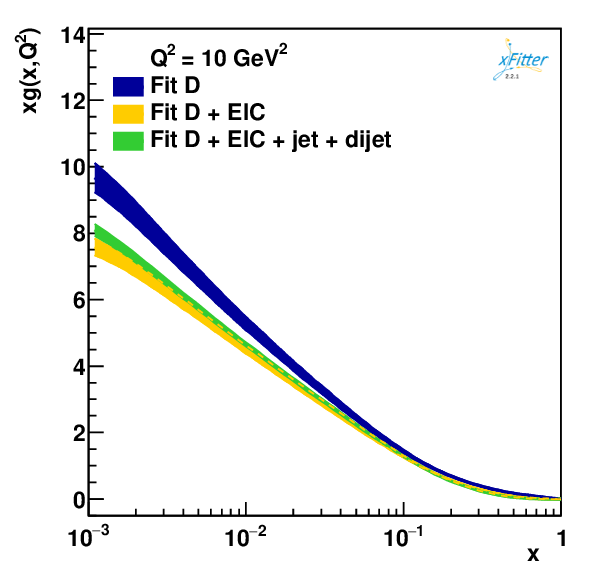}  
\includegraphics[scale = 0.48]{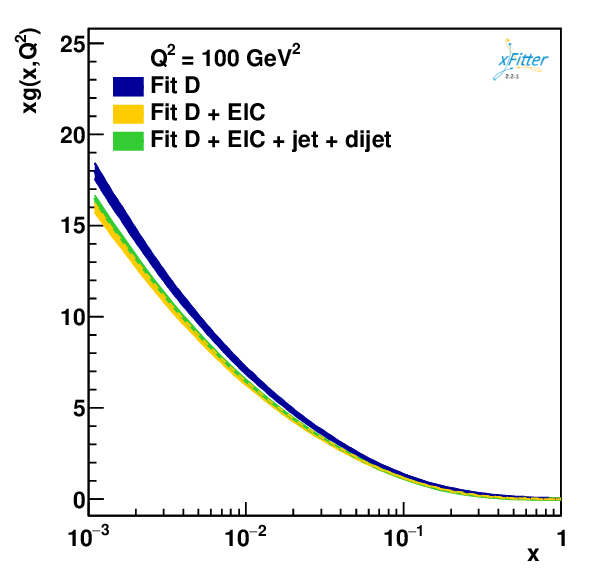} \\
\caption{Comparison of our nominal fit ({\tt Fit~D})  at NNLO accuracy with 
({\tt Fit~D + EIC}) in which include the simulated EIC data, and ({\tt Fit~D + EIC+ jet/dijet}) which include 
the HERA jet data. The comparisons are shown 
for $xu_v$, $xd_v$, and $xg$ distribution as a function of $x$ and at $Q^2 =1.9$, 10 and 100~GeV$^2$. } 
\label{fig:EIC_j}
\end{center}
\end{figure*}

\begin{figure*}[!htb]
%\vspace{0.5cm}
\begin{center}
  
\includegraphics[scale = 0.48]{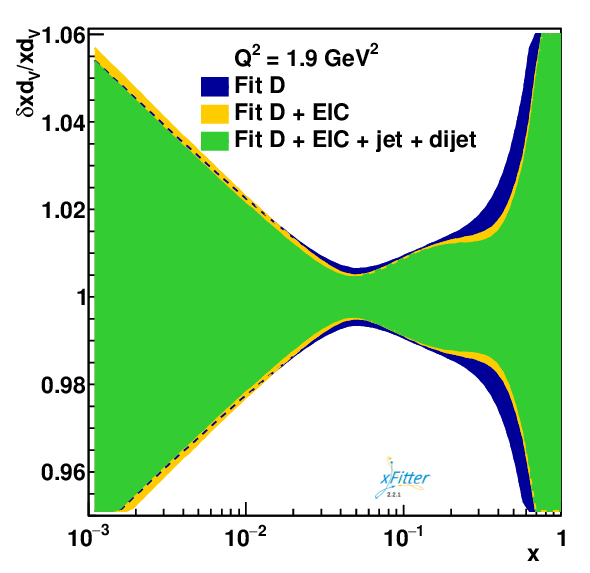}	
\includegraphics[scale = 0.48]{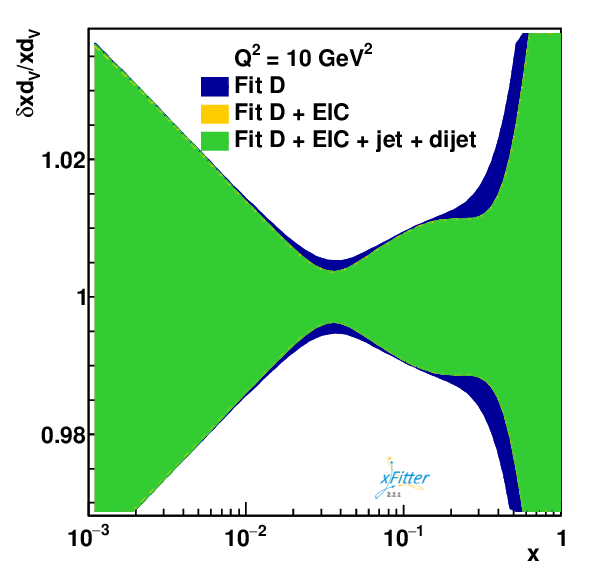}		
\includegraphics[scale = 0.48]{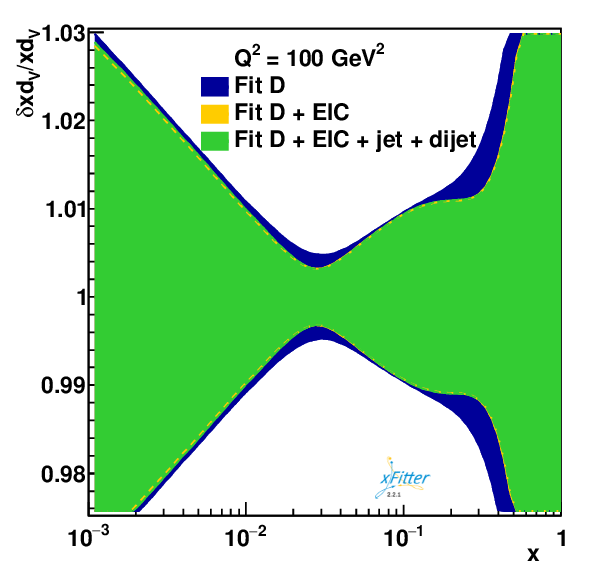}\\        
\includegraphics[scale = 0.48]{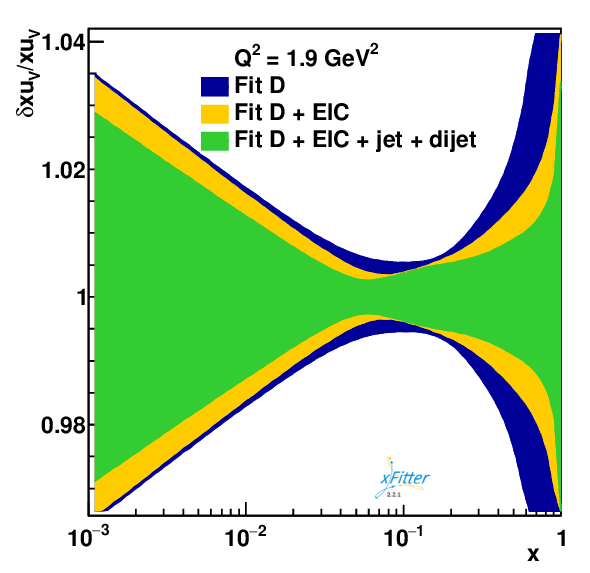}  
\includegraphics[scale = 0.48]{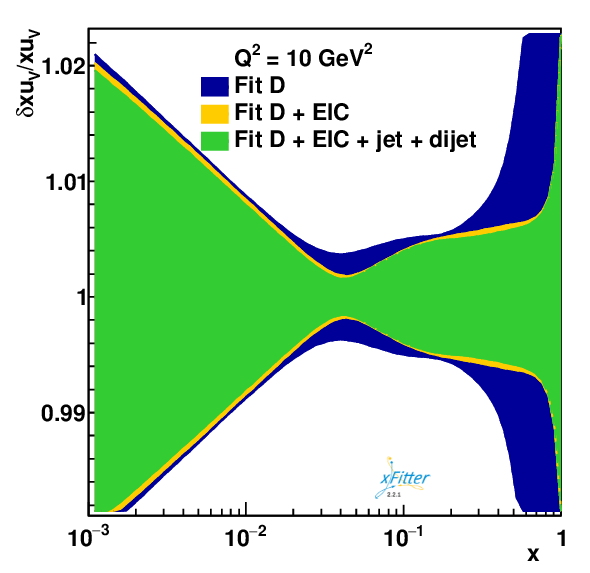}	
\includegraphics[scale = 0.48]{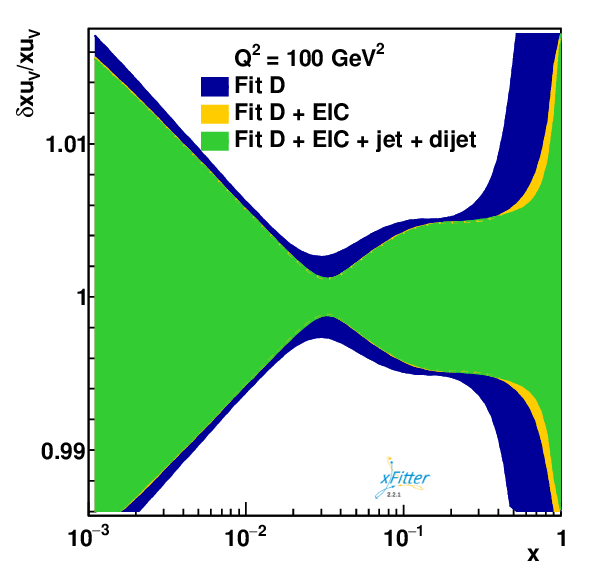} \\		
\includegraphics[scale = 0.48]{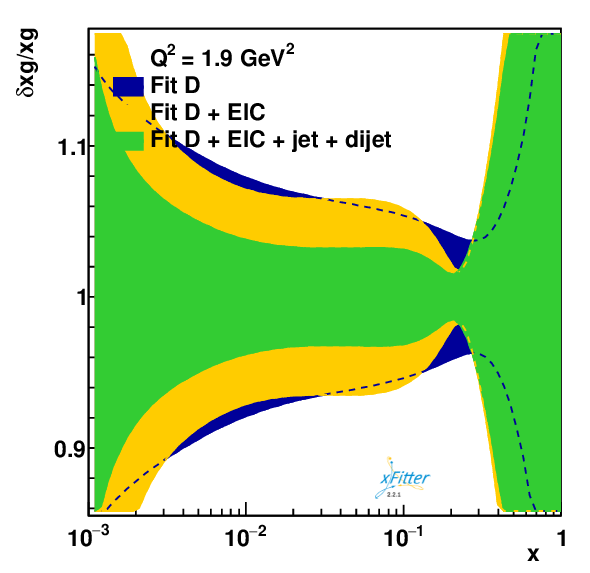}        
\includegraphics[scale = 0.48]{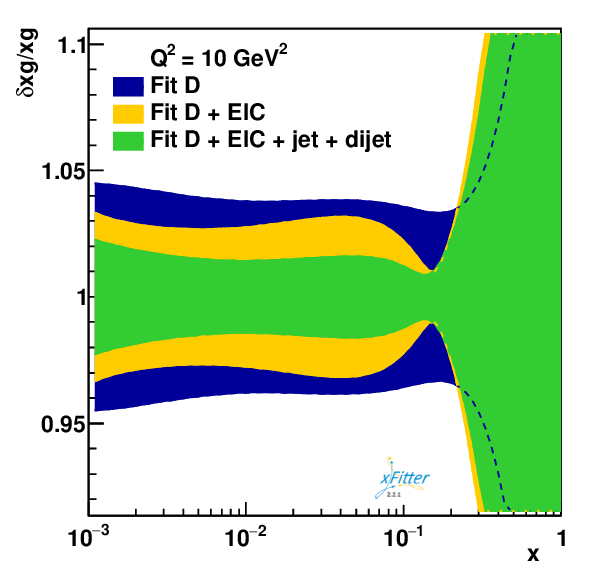}  
\includegraphics[scale = 0.48]{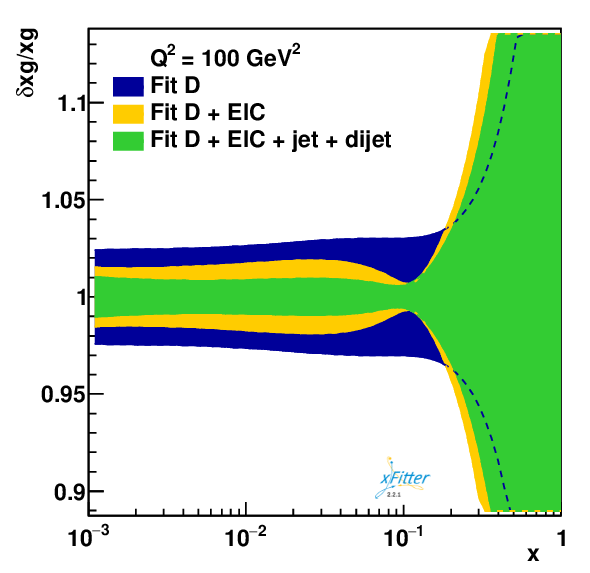} \\
\caption{Same as Fig,~\ref{fig:EIC_j} but this time the comparisons are shown for the relative uncertainties $\delta xf(x,Q^2)/xf(x,Q^2)$.}  
\label{fig:EIC_j_ratio}
\end{center}
\end{figure*}

In Figs.~\ref{fig:EIC_j} and \ref{fig:EIC_j_ratio}, we compare the results of our 
nominal fit ({\tt Fit~D}) at NNLO accuracy with fits that include simulated EIC 
data ({\tt Fit~D + EIC}) and those that incorporate both EIC and HERA jet/dijet 
data ({\tt Fit~D + EIC + jet/dijet}). The comparisons are presented for the 
distributions \(xu_v\), \(xd_v\), and \(xg\) as functions of \(x\), at \(Q^2 = 1.9\), 10, and 100~GeV\(^2\).

%---------------Log--------EIC-------------------------
%--------------------------------
%--------------------------------

%---------------Log--------EIC-Ratio-j-------------------------
%--------------------------------
%--------------------------------

As observed in Figs.~\ref{fig:EIC_j} and \ref{fig:EIC_j_ratio}, the inclusion of HERA jet and dijet 
data significantly constrains the gluon distribution, resulting in reduced uncertainties, 
particularly at medium and very small values of \(x\). This is because jet production 
cross-sections in DIS are directly sensitive to the gluon density. 

Moreover, jet production data are instrumental in determining the strong 
coupling constant, \(\alpha_s(M_Z^2)\). Due to the sensitivity of jet production 
to higher-order QCD corrections, incorporating jet and dijet datasets allows 
for a more precise extraction of \(\alpha_s\), thereby enhancing the overall fit quality of global PDF analyses.

In Table~\ref{tab:FitD-EIC-jet}, we present the \(\chi^2\) values for the global 
QCD fit both with and without the inclusion of jet and dijet production data. 
The \(\chi^2\) for the HERA~I+II data set remains stable, with a slight 
improvement from 1151/1016 to 1130/1016. A significant improvement is observed in 
the \(\chi^2\) values for both CMS and ATLAS W/Z production datasets. The \(\chi^2\) 
values for the ATLAS and E866 Drell-Yan data sets show minor changes, 
with the ATLAS \(\chi^2\) improving slightly.

Finally, the addition of jet and dijet production data from HERA provides  
critical information about the gluon distribution. The \(\chi^2\) values for these data sets 
reflect the strong constraints that these data provide, particularly in the high \(Q^2\) 
region, where gluon-initiated processes dominate. Including these data leads to a significant 
reduction in the overall uncertainty of the gluon PDF, which in turn 
improves the precision of predictions for processes sensitive to gluons. 
The improved \(\chi^2/\text{dof}\) reflects enhanced consistency between theoretical 
predictions and experimental measurements, further supporting the inclusion of 
jet data from the LHC in global QCD fits to improve the precision of proton PDFs and \(\alpha_s\) determinations.

Further improvements in PDF uncertainties, particularly for the gluon distribution, 
are anticipated when inclusive jet and dijet data from the EIC are added to the QCD analysis. 
These data are also expected to significantly contribute to the precision of \(\alpha_s\). 
Future studies could employ theory grids for various EIC energies using the 
fastNLO framework~\cite{Kluge:2006xs}, although such an analysis is beyond the scope of this work. 
Additionally, incorporating a broader range of hadron collider jet and dijet data sets 
could further constrain the gluon PDFs and improve the precision of \(\alpha_s\).

%--------------------------------
\begin{figure*}[!htb]
%\vspace{0.5cm}
\begin{center}
\includegraphics[scale = 0.55]{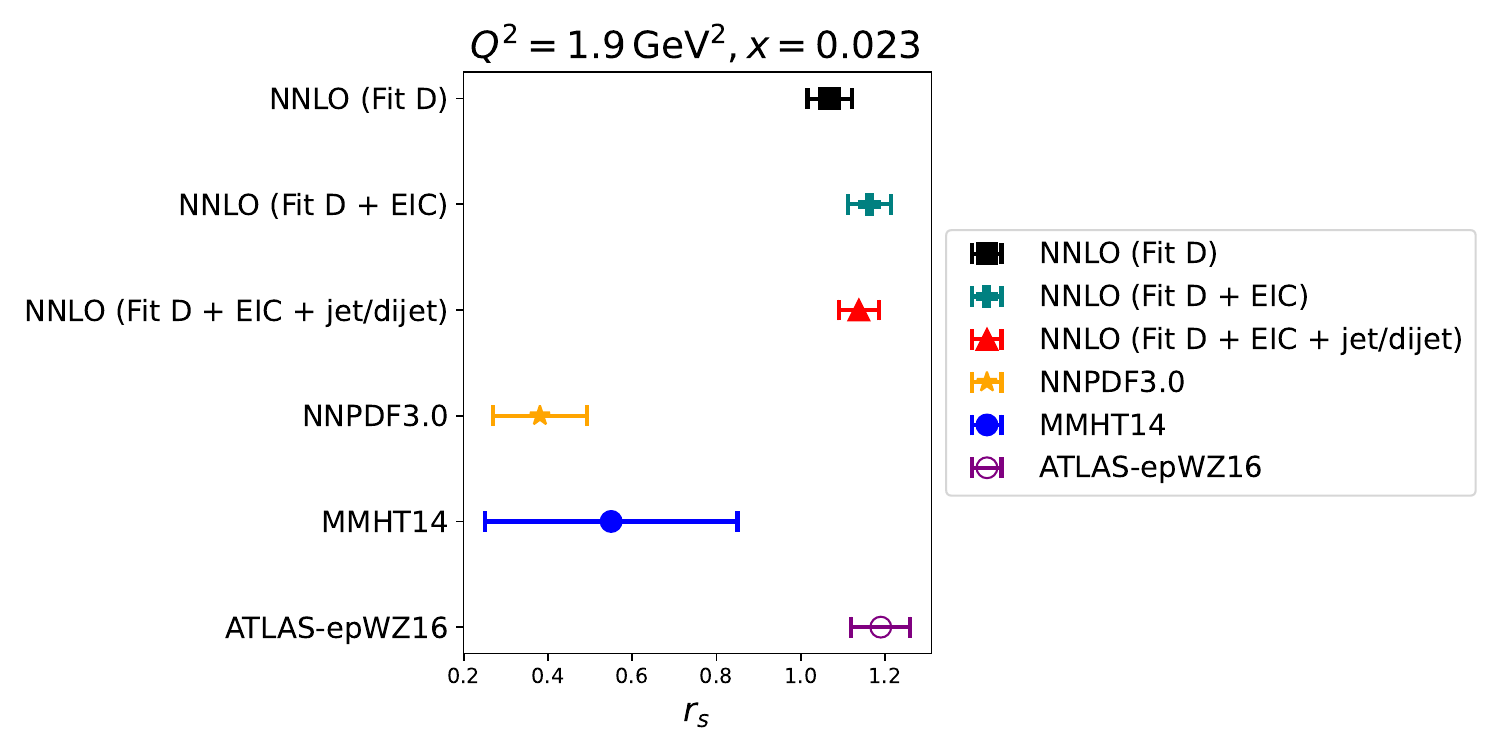} 
\caption{Determination of the relative strange-to-down sea 
quark fraction \(r_s\) calculated at the initial scale \(Q_0^2 = 1.9 \, 
\text{GeV}^2\) and at \(x = 0.023\), which corresponds to the point of 
greatest sensitivity at central rapidity in the ATLAS data.
The results are also compared with other global fit analyses, 
specifically ATLAS-epWZ16~\cite{ATLAS:2016nqi}, NNPDF3.0~\cite{NNPDF:2014otw}, and MMHT14~\cite{Harland-Lang:2014zoa}.} 
\label{fig:r_s}
\end{center}
\end{figure*}
%--------------------------------

%=================================================  
\section{Strange-quark density}\label{Strange-quark}
%=================================================  

The strange quark distribution in the proton plays a crucial role in 
understanding the quark structure of the proton. One of the key quantities 
characterizing the strange-quark 
fraction is the ratio \(r_s\), which is defined as the ratio of the strange-quark 
distribution to the down sea-quark distribution:
\begin{equation}
r_s = \frac{s + \bar{s}}{2 \bar{d}}~. %\nonumber
\end{equation}
This ratio offers insight into the balance between strange and down quarks in the proton sea and is 
important for various QCD and electroweak processes, particularly in high-energy collider experiments.

In this analysis, we determine \( r_s \) at \( Q^2 = 1.9 \, \text{GeV}^2 \) and \( x = 0.023 \), a value 
selected because it lies within the region of maximum sensitivity in the ATLAS data 
for central rapidity at \( \sqrt{s} = 7 \, \text{TeV} \)~\cite{ATLAS:2016nqi}. 

We explore the strange-quark fraction further by examining the values of \(r_s\) obtained under 
different fit scenarios. These are summarized in Table~\ref{tab:par_rs_alpha_s}, which 
shows the extracted values of \(r_s\) from various fits.

%--------------------------------
\begin{table*}
%--------------------------------
\caption{
\label{tab:par_rs_alpha_s} 
The numerical values of $\alpha_s(M_Z)$ and $r_s$ extracted from our {\tt Fit~D}, {\tt Fit~D + jet/dijet}, {\tt Fit~D + EIC}, 
and {\tt Fit~D + EIC + jet/dijet} QCD analyses.} 
\begin{center}
\begin{tabular}{l | c c c c }
\hline 	\hline 
Parameter   ~~&~~   {\tt Fit D} ~~&~~   {\tt Fit D + jet/dijet}  ~~&~~   {\tt Fit D + EIC}    ~~&~~   {\tt Fit D + EIC + jet/dijet}         \\ 	\hline 
$\alpha_s(M_Z)$  &   $0.1128 \pm 0.0014$  &$0.1192 \pm 0.0012$ &   $0.1188 \pm 0.0008$  &   $0.1183 \pm 0.0003$  \\ 
$r_s$            &   $1.069 \pm 0.053$    &$1.148 \pm 0.053$ &   $1.164 \pm 0.051$    &   $1.138 \pm 0.048$                \\
\hline 	\hline   
\end{tabular}
\end{center}
%--------------------------------
\end{table*}
%--------------------------------

The baseline fit ({\tt Fit~D}) yields \(r_s = 1.069 \pm 0.053\), indicating a nearly equal 
strange-to-down sea-quark density, consistent with previous findings of an 
unsuppressed strange quark content at small \(x\). 
For our {\tt Fit~D + EIC} when projected data from the EIC are included in the fit, \(r_s\) increases to \(1.164 \pm 0.051\),  
suggesting that the EIC will significantly enhance the precision of strange-quark density.   
Finally {\tt Fit~D + EIC + jet/dijet} in which the jet and dijet 
production data included to the fit, \(r_s\) is slightly reduced to \(1.138 \pm 0.048\), 
while the uncertainty is reduced even further. 

The ATLAS data, combined with future measurements from the EIC and precision jet experiments, 
will continue to play a crucial role in reducing the uncertainties associated with the 
strange quark distribution. The extracted values of \(r_s\) in this analysis suggest that 
strange quarks contribute significantly to the proton sea. 
These findings will have a broad impact on QCD phenomenology 
and future high-energy physics experiments, particularly in processes involving 
electroweak boson production and DIS.

Our determinations of \(r_s\) for the three different analyses 
mentioned above are illustrated in Fig.~\ref{fig:r_s}. 
The measurements are presented with both the experimental and PDF-fit related uncertainties. 
The results are also compared with recent global fit analyses, 
specifically ATLAS-epWZ16~\cite{ATLAS:2016nqi}, NNPDF3.0~\cite{NNPDF:2014otw}, and MMHT14~\cite{Harland-Lang:2014zoa}. 
As can be seen, both NNPDF3.0 and MMHT14 analyses predict \(r_s\) to 
be significantly lower than unity, with values ranging between approximately 0.3 and 0.6.  
However, our result from the {\tt Fit~D + EIC} analysis shows better 
agreement with the ATLAS-epWZ16 analysis, as both analyses use the same data set.

%=================================================  
\section{Strong coupling constant}\label{Coupling}
%=================================================  

The precise determination of the strong coupling constant, \(\alpha_s\), relies on 
various experimental data sets that are sensitive to different aspects of the proton PDFs. 
In this section, we examine how the EIC and jet/dijet production data contribute to constraining \(\alpha_s\), 
focusing on the complementary information these data sets provide about quark and gluon interactions. 
This sensitivity arises from the unique kinematic coverage and specific processes probed by the EIC and jet production measurements, 
making them highly valuable for reducing uncertainties in \(\alpha_s\) and enhancing the precision of global QCD fits.
The EIC will significantly enhance the determination of the strong coupling constant, \(\alpha_s(M_Z^2)\), 
through its extensive kinematic reach, particularly in regions where current data sets, 
such as those from HERA, are less effective. 

The strong coupling constant, \(\alpha_s\), governs the evolution of quark densities as the energy scale \(Q^2\) increases. 
In the large \(x\) regime, quark distributions are particularly sensitive to \(\alpha_s\) because parton-parton interactions 
dominate in this region. By probing large \(x\), the EIC directly measures the evolution of quark PDFs, 
thereby providing precise constraints on the behavior of \(\alpha_s\) in this regime.

The sensitivity of EIC to \(\alpha_s\) also arises from the specific processes it probes and the unique 
kinematic overlap it offers with previous data. In DIS processes, the interaction between quarks and gluons is 
directly affected by \(\alpha_s\). The structure functions measured in EIC experiments, such as \(F_2(x, Q^2)\), 
depend on quark-gluon interactions, which are controlled by \(\alpha_s\). 
The quantity \(dF_2/d\ln Q^2\), which measures scaling violations, is particularly sensitive to \(\alpha_s\), especially at large \(x\). 
In this region, the evolution is dominated by the \(q \rightarrow qg\) splitting, which involves the product of \(\alpha_s\) and 
the large-\(x\) quark densities. The high precision of EIC measurements leads to more accurate constraints on 
the quark-gluon coupling and, consequently, on \(\alpha_s\), 
with a reduced dependence on the gluon distribution compared to lower \(x\) values.

This decoupling reduces the correlation between \(\alpha_s\) 
and the gluon distribution, resulting in a more precise extraction of \(\alpha_s(M_Z^2)\). 
The inclusion of simulated EIC data in the fits provides additional coverage of the phase space, particularly in regions where 
HERA data alone were limited. This unique contribution from the EIC highlights its role in providing world-leading precision for 
the determination of \(\alpha_s\), supporting future QCD analyses and experimental research~\cite{Cerci:2023uhu,Armesto:2023hnw}. 

\begin{figure*}[!htb]
%\vspace{0.5cm}
\begin{center}
\includegraphics[scale = 0.55]{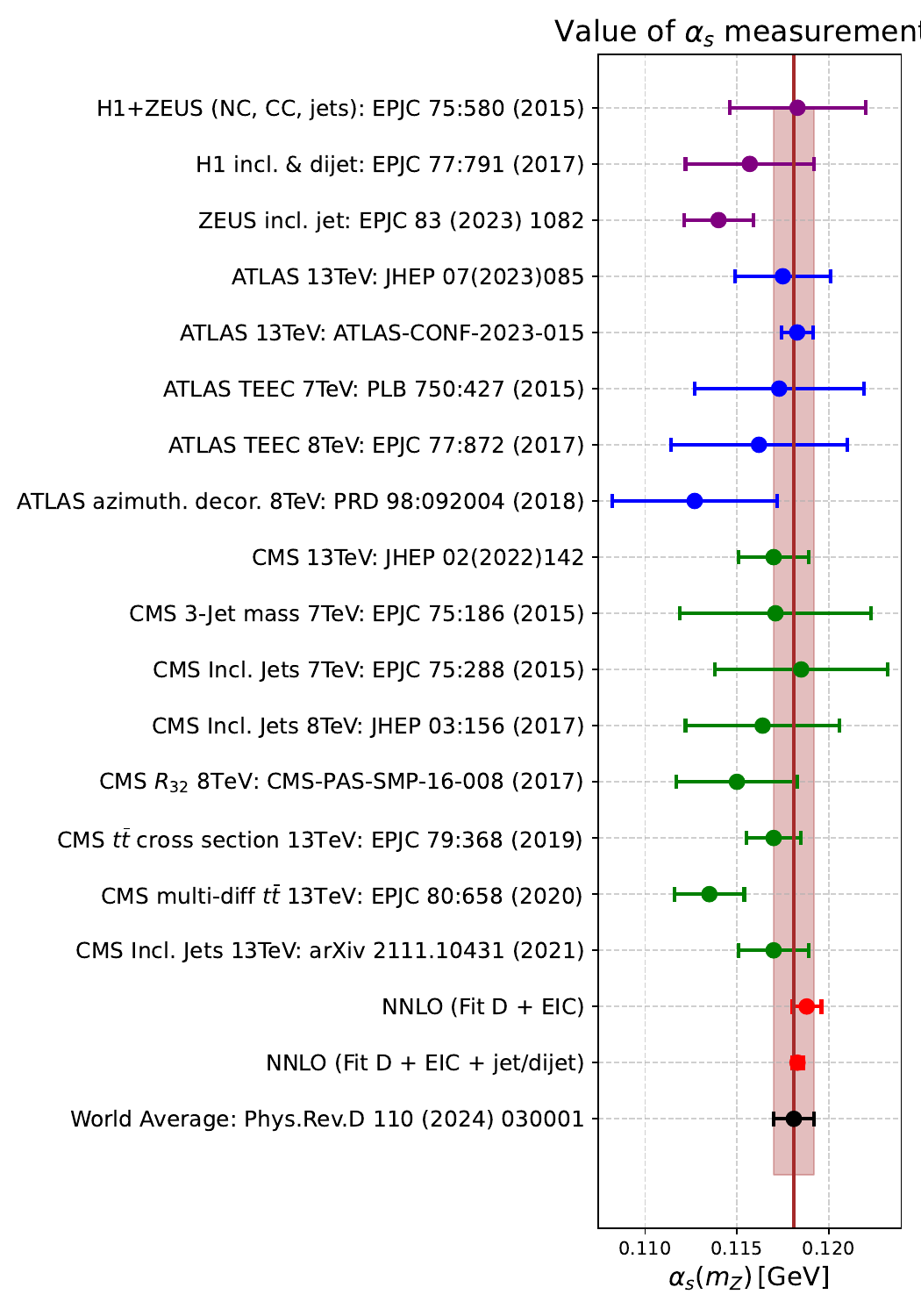}	
\caption{Values of $\alpha_s(M_Z)$ resulting from different measurements, including H1,  ZEUS, 
ATLAS and CMS. \cite{H1:2015ubc,H1:2017bml,ZEUS:2023zie,ATLAS:2023tgo,ATLAS:2023bax,ATLAS:2015yaa,ATLAS:2017qir,ATLAS:2018sjf,CMS:2021yzl,CMS:2014mna,CMS:2014qtp,CMS:2016lna,CMS:2017tvp,CMS:2018fks,CMS:2019esx}.  
Our analyses performed at NNLO are also indicated separately for comparisons.} 
\label{fig:alpha_s}
\end{center}
\end{figure*}

As illustrated in Table~\ref{tab:par_rs_alpha_s}, the inclusion of EIC pseudodata significantly reduces 
the uncertainty on \(\alpha_s(M_Z^2)\) from 0.0014 (in the baseline Fit~D) to 0.0008. 
This reduction is primarily due to the enhanced sensitivity to quark-gluon interactions and 
the broader kinematic coverage provided by the EIC. A summary of the \(\alpha_s(M_Z)\) 
measurements at different collaborations, including the H1,  ZEUS, 
ATLAS and CMS. \cite{H1:2015ubc,H1:2017bml,ZEUS:2023zie,ATLAS:2023tgo,ATLAS:2023bax,ATLAS:2015yaa,ATLAS:2017qir,ATLAS:2018sjf,CMS:2021yzl,CMS:2014mna,CMS:2014qtp,CMS:2016lna,CMS:2017tvp,CMS:2018fks,CMS:2019esx}, as well as the world average~\cite{ParticleDataGroup:2024cfk}, 
is also presented in Fig.~\ref{fig:alpha_s}. Our analyses performed at NNLO are indicated separately for comparison.

%--------------------------------
%--------------------------------

Jet and dijet production measurements offer complementary sensitivity to \(\alpha_s\), primarily through 
their ability to probe the gluon distribution in the proton. Jet production is particularly 
sensitive to gluon-gluon and quark-gluon interactions, both of which are controlled by \(\alpha_s\). 
The strong coupling constant directly enters the cross-sections for jet production, 
making these measurements a powerful tool for constraining \(\alpha_s\). Consequently, 
the inclusion of jet and dijet data in the QCD fit provides direct constraints on the 
gluon distribution and \(\alpha_s\), particularly at high energy scales.

Jet and dijet production data, particularly from the LHC and HERA, cover a broad range of 
high \(Q^2\) values. In this regime, gluon interactions become increasingly important, 
enhancing the sensitivity to \(\alpha_s\). The precise measurement of jet cross-sections 
at large \(Q^2\) significantly reduces the uncertainty in the determination of \(\alpha_s\), 
as evidenced by the improved results in {\tt Fit~D + EIC + jet/dijet}, where the uncertainty on \(\alpha_s(M_Z^2)\) 
decreases to 0.0003.

While the EIC data provide essential constraints on quark distributions at lower energy 
scales, jet and dijet production measurements contribute by refining our understanding of 
gluon distributions at higher energy scales. Together, these datasets offer complementary 
sensitivity to \(\alpha_s\), ensuring that both quark and gluon interactions are accurately 
described in the global fit.

The combination of EIC and jet/dijet data leads to a significant improvement in the 
precision of \(\alpha_s(M_Z^2)\). As seen in Table~\ref{tab:par_rs_alpha_s} and Fig.~\ref{fig:alpha_s}, 
the inclusion of both datasets in the global fit reduces the uncertainty on \(\alpha_s(M_Z^2)\) 
from 0.0008 ({\tt Fit~D + EIC}) to 0.0003 ({\tt Fit~D + EIC + jet/dijet}).

These results suggest that further improvements in the precision of \(\alpha_s(M_Z^2)\) could 
be achieved by incorporating inclusive jet and dijet data from the EIC into the QCD analysis. 
This could be facilitated by using theory grids for EIC energies within the fastNLO framework. 
Additionally, other observables measurable at the EIC could provide further opportunities 
for constraining the strong coupling constant.

Beyond a DIS-only approach, it would be valuable to explore the impact of EIC data on \(\alpha_s\) 
determinations in global QCD fits that integrate data from LHC measurements not included 
in this work, as well as other experimental sources. Before the EIC becomes operational, progress in 
understanding higher-order uncertainties will be crucial. Reaching a consensus on addressing 
these uncertainties in \(\alpha_s(M_Z^2)\) determinations based on EIC data will be essential for 
maximizing the precision of future measurements.

%--------------------------------
\begin{figure*}[!htb]
\begin{center}
\includegraphics[scale = 0.90]{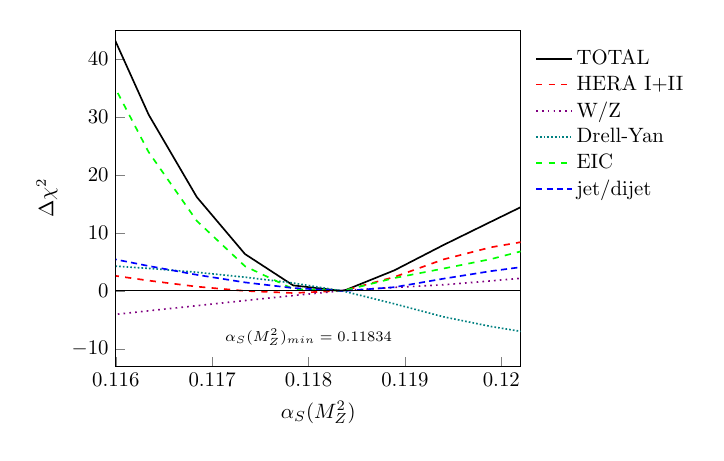} 
\caption{ The curves of \(\Delta\chi^2\) as a function of the strong coupling constant \(\alpha_s(M_Z^2)\) obtained 
from {\tt Fit~D + EIC + jet/dijet} for all types of datasets. 
The \(\Delta\chi^2_{\rm TOTAL}\) curve represents the sensitivity when all experimental data are considered simultaneously.} 
\label{fig:chi-scan}
\end{center}
\end{figure*}
%--------------------------------

%=================================================  
\section{Sensitivity of various datasets to the strong coupling constant}\label{Sensitivity}
%=================================================  

In every proton PDF analysis, the sensitivity of each dataset whether individually or collectively to the 
variation of the strong coupling constant \(\alpha_s(M_Z^2)\) depends on the chosen settings, 
such as the parameterization and the values of applied cuts. One of the most effective approaches for 
assessing this sensitivity is the \(\chi^2\)-scan method. This method allows us to evaluate 
how different experimental datasets influence the fit for \(\alpha_s(M_Z^2)\). The following equation is used for the \(\chi^2\)-scan:
\begin{equation}
\Delta\chi^2_e(\pi) = \chi^2_e(\pi) - \chi^2_e(\pi_{\rm min})~.
\end{equation}
In this equation, \(\Delta\chi^2_e(\pi)\) is a function of the fit 
parameters \(\pi\), which, in this case, is the variation of the QCD coupling constant (\(\pi = \alpha_s(M_Z^2)\)). 
The term \(\chi^2_e(\pi_{\rm min})\) represents the minimum value obtained from the 
final fit, specifically from {\tt Fit~D + EIC + jet/dijet}. This value corresponds to the 
optimal parameter set, denoted as \(\pi_{\rm min}\), which yields the lowest \(\chi^2\). 
In contrast, \(\chi^2_e(\pi)\) refers to the \(\chi^2\) value obtained for each 
variation of \(\alpha_s(M_Z^2)\) for a specific experimental dataset (\(e\)) or for all datasets together.

In the \(\chi^2\)-scan procedure, the value of the parameter under study, \(\alpha_s(M_Z^2)\), 
is fixed at different specific values, while all other fit parameters are 
allowed to vary freely, as described in Section~\ref{method}. The difference between the 
values obtained from these two types of calculations (\(\chi^2_e(\pi)\) and \(\chi^2_e(\pi_{\rm min})\)) 
provides the \(\Delta\chi^2_e(\pi)\) value. This value effectively captures the change 
in the goodness of fit for different fixed values of \(\alpha_s(M_Z^2)\).

Figure~\ref{fig:chi-scan} illustrates the dependence of \(\Delta\chi^2\) for different 
experimental data sets and for the full set of data on variations of \(\alpha_s(M_Z^2)\), 
ranging from 0.116 to 0.120. From the comparison of the different curves in the figure, 
it is evident that \(\Delta\chi^2_{EIC}(\alpha_s(M_Z^2))\) shows a strong sensitivity 
to the variation of \(\alpha_s(M_Z^2)\), especially when \(\alpha_s(M_Z^2)\) is less than 0.11834. 
On the other hand, other data sets, such as HERA I+II, \(W/Z\), Drell-Yan, and jet/dijet 
production data, exhibit lower sensitivity to the variation of the strong coupling constant. 
This suggests that EIC data could play a crucial role in constraining \(\alpha_s(M_Z^2)\) with 
higher precision compared to the other data sets.

The \(\chi^2\) scan method used here is similar to the approach employed in the CT18 analysis, 
which also investigated the sensitivity of different data sets, including DIS and jet production 
data, to \(\alpha_s(M_Z^2)\). Similar to the CT18 results, the \(\Delta\chi^2\) curves for 
individual datasets generally exhibit a parabolic shape, consistent with the central limit theorem, 
indicating their respective pulls on the preferred value of \(\alpha_s(M_Z^2)\). This helps 
identify any tensions or consistencies across different data sets, providing insights into how 
well each data set constrains \(\alpha_s\).

%=================================================
\section{Summary and conclusion}\label{summary}
%=================================================

In this work, we have presented new proton PDFs at both NLO and NNLO accuracy, derived from a comprehensive analysis of 
high-precision data from the LHC, combined HERA DIS dataset, and additional crucial inputs. 
We have explored the impact of including data from Drell-Yan pair production, as well as W and Z boson production, on 
the PDFs and their associated uncertainties.
Our findings demonstrate that the inclusion of these data sets significantly enhances the precision and 
reliability of the proton PDFs, reducing uncertainties across a wide range of kinematic regions. 
Specifically, the Drell-Yan, W, and Z production data provided important sensitivity to 
quark distributions at moderate to high values of Bjorken \(x\), thereby complementing the 
information obtained from DIS data. 
 As have been shown, the inclusion of jet and dijet production data further constrained 
the gluon distribution, reducing the gluon uncertainties for \(x \leq 0.2\).   
The resulting PDFs are accompanied by well-defined error estimates calculated using the Hessian method, 
ensuring a rigorous quantification of uncertainties. 
This approach guarantees that the uncertainties are propagated appropriately, 
making these PDFs highly suitable for precision physics studies.
Our results highlight the importance of continuously updating PDF determinations with the latest experimental data to 
achieve percent-level accuracy in theoretical predictions for high-energy processes. 
This level of precision is crucial for interpreting measurements at the LHC, particularly in processes 
involving electroweak boson production, jet production, and beyond, where a precise 
knowledge of the partonic structure is necessary.
The simulated data from the future Electron-Ion Collider (EIC) also play a significant role in improving our 
understanding of the partonic structure of the proton. We have shown that the inclusion of EIC simulated data, 
in combination with jet and dijet production data from HERA, has the potential to not 
only improve the determination of the gluon distribution but also enhance the precision of 
the determination of the strong coupling constant \(\alpha_s(M_Z^2)\). 
These improvements underscore the critical role of future facilities like the EIC in global QCD analyses, 
particularly for better constraining gluon and sea-quark distributions in 
previously unexplored kinematic regions.
Furthermore, the sensitivity study conducted on \(\alpha_s(M_Z^2)\) demonstrates the complementary nature of different data sets, 
such as DIS, Drell-Yan, W/Z boson, and jet production data, in constraining the strong coupling constant.

%=================================================  
\section*{Availability of proton PDFs sets}\label{LHAPDF} 
%=================================================  

Our proton PDFs, determined at NLO and NNLO in perturbative QCD, have been made available in the {\tt LHAPDF} library to 
facilitate their use in a wide variety of high-energy physics analyses. 
The LHAPDF format allows easy integration of these PDFs with various Monte Carlo event generators, 
QCD analysis tools, and simulation software used in collider experiments. 
These PDF sets, corresponding to different global QCD fit scenarios (e.g., {\tt Fit~D}, {\tt Fit~D + EIC}, {\tt Fit~D + EIC + jet/dijet}), 
are intended to provide flexibility for experimental analyses requiring precise PDF 
information across different kinematic regions.
The uncertainties in the PDFs are quantified using the Hessian method, and each PDF 
set includes eigenvector variations that can be used to assess the uncertainties in observables. 
The Hessian PDF sets are also available in the LHAPDF library, 
allowing users to perform detailed uncertainty propagation in their analyses.
The PDF sets are formatted in compliance with the LHAPDF standard (version~6 and above) to ensure 
maximum compatibility with modern QCD analysis frameworks and simulation-based analyses. 
One can access our PDF sets via email upon request, where they are categorized based on 
the fit scenario (e.g., {\tt Fit~D}, {\tt Fit~D + EIC}).

%=================================================
\section*{Acknowledgement}   
%=================================================

The authors gratefully acknowledge the helpful discussions and insightful 
comments provided by Katarzyna Wichmann and Andrei Kataev, which significantly elevated 
and enriched the quality of this paper. The authors also thank the School of Particles and 
Accelerators at the Institute for Research in Fundamental Sciences (IPM) for their financial 
support of this project. 
Hamzeh Khanpour appreciates the financial support from NAWA under grant 
number BPN/ULM/2023/1/00160, as well as from the IDUB program at AGH University.  
The work of  UGM was also supported in part by the Chinese
Academy of Sciences (CAS) President's International
Fellowship Initiative (PIFI) (Grant No.\ 2025PD0022).

%=================================================

%=================================================

%=================================================  
\end{document}